\documentclass[twocolumn]{aastex631}
\usepackage{graphics}
\usepackage{enumitem}
\usepackage{amsmath}
\usepackage{xcolor}
\usepackage{soul}
\usepackage{url}
\usepackage{color,soul}

\newcommand{\N}[1]{\textcolor{blue}{NS: #1}} 

\begin{document}

\shortauthors{Temim et al.}
\shorttitle{Crab Nebula with JWST}

\title{Dissecting the Crab Nebula with JWST: \\
Pulsar wind, dusty filaments, and Ni/Fe abundance constraints on the explosion mechanism}

\author[0000-0001-7380-3144]{Tea Temim}
\correspondingauthor{Tea Temim} 
\email{temim@astro.princeton.edu}
\affiliation{Department of Astrophysical Sciences, Princeton University, 4 Ivy Ln, Princeton, NJ 08544, USA}

\author[0000-0002-3362-7040]{J. Martin Laming}
\affiliation{Space Science Division, Code 7684, Naval Research Laboratory, Washington DC 20375, USA}

\author[0000-0001-6872-2358]{P.\ J.\ Kavanagh}
\affiliation{Department of Experimental Physics, Maynooth University, Maynooth, Co. Kildare, Ireland}

\author[0000-0001-5510-2424]{Nathan Smith} 
\affiliation{Steward Observatory, University of Arizona, 933 N. Cherry Ave., Tucson, AZ 85721, USA}

\author[0000-0002-6986-6756]{Patrick Slane}
\affiliation{Center for Astrophysics $\vert$ Harvard \& Smithsonian, 60 Garden Street, Cambridge, MA 02138, USA}

\author[0000-0003-2379-6518]{William P. Blair}
\affiliation{Department of Physics and Astronomy, Johns Hopkins University, 3400 N. Charles Street, Baltimore, MD, 21218, USA}

\author[0000-0001-9419-6355]{Ilse De Looze}
\affiliation{Sterrenkundig Observatorium, Ghent University, Krijgslaan 281-S9, B-9000 Gent, Belgium}

\author[0000-0002-8848-1392]{Niccol\`o Bucciantini} 
\affiliation{INAF Osservatorio Astrofisico di Arcetri, Largo Enrico Fermi 5, 50125 Firenze, Italy}

\author[0000-0001-8005-4030]{Anders Jerkstrand} 
\affiliation{The Oskar Klein Centre, Department of Astronomy, Stockholm University, Albanova University Center, 106 91, Stockholm, Sweden}

\author[0009-0008-8797-0865]{Nicole Marcelina Gountanis}
\affiliation{Department of Astronomy, The Ohio State University, Columbus, Ohio 43210, USA}

\author[0000-0001-8858-1943]{Ravi Sankrit}
\affiliation{Space Telescope Science Institute,
3700 San Martin Dr., Baltimore MD 21218, USA}

\author[0000-0002-0763-3885]{Dan Milisavljevic}
\affiliation{Purdue University, Department of Physics and Astronomy, 525 Northwestern Ave, West Lafayette, IN 47907, USA}
\affiliation{Integrative Data Science Initiative, Purdue University, West Lafayette, IN 47907, USA}

\author[0000-0002-4410-5387]{Armin Rest}
\affiliation{Space Telescope Science Institute,
3700 San Martin Dr., Baltimore MD 21218, USA}

\author[0000-0001-6436-8304]{Maxim Lyutikov}
\affiliation{Purdue University, Department of Physics and Astronomy, 525 Northwestern Ave, West Lafayette, IN 47907, USA}

\author{Joseph DePasquale}
\affiliation{Space Telescope Science Institute,
3700 San Martin Dr., Baltimore MD 21218, USA}

\author[0000-0002-3074-9608]{Thomas Martin}
\affiliation{D\'epartement de physique, de g\'enie physique et d\'optique, Universit{\'e} Laval, 2325, rue de l'universit{\'e}, Qu{\'e}bec, (Qu{\'e}bec), G1V 0A6, Canada}

\author[0000-0003-1278-2591]{Laurent Drissen}
\affiliation{D\'epartement de physique, de g\'enie physique et d\'optique, Universit{\'e} Laval, 2325, rue de l'universit{\'e}, Qu{\'e}bec, (Qu{\'e}bec), G1V 0A6, Canada} 

\author[0000-0002-7868-1622]{John Raymond}
\affiliation{Center for Astrophysics $\vert$ Harvard \& Smithsonian, 60 Garden Street, Cambridge, MA 02138, USA}

\author[0000-0003-2238-1572]{Ori D. Fox}
\affiliation{Space Telescope Science Institute, 3700 San Martin Drive, Baltimore, MD 21218, USA}

\author[0000-0001-7132-0333]{Maryam Modjaz}
\affiliation{Department of Astronomy, University of Virginia, Charlottesville, VA22904, USA}

\author[0000-0001-9179-9054]{Anatoly Spitkovsky}
\affiliation{Department of Astrophysical Sciences, Princeton University, 4 Ivy Ln, Princeton, NJ 08544, USA}

\author[0000-0002-7756-4440]{Lou Strolger}
\affiliation{Space Telescope Science Institute,
3700 San Martin Dr., Baltimore MD 21218, USA}



\begin{abstract}

We present JWST observations of the Crab Nebula, the iconic remnant of the historical SN~1054. The observations include NIRCam and MIRI imaging mosaics, plus MIRI/MRS IFU spectra that probe two select locations within the ejecta filaments. We derive a high-resolution map of dust emission and show that the grains are concentrated in the innermost, high-density filaments. These dense filaments coincide with multiple synchrotron bays around the periphery of the Crab’s pulsar wind nebula (PWN). We measure synchrotron spectral index changes in small-scale features within the PWN's torus region, including the well-known knot and wisp structures. The index variations are consistent with Doppler boosting of emission from particles with a broken power-law distribution, providing the first direct evidence that the curvature in the particle injection spectrum is tied to the acceleration mechanism at the termination shock. We detect multiple nickel and iron lines in the ejecta filaments and use photoionization models to derive nickel-to-iron abundance ratios that are a factor of 3--8 higher than the solar ratio. We also find that the previously reported order-of-magnitude higher Ni/Fe values from optical data are consistent with the lower values from JWST when we reanalyze the optical emission using updated atomic data and account for local extinction from dust. We discuss the implications of our results for understanding the nature of the explosion that produced the Crab Nebula and conclude that the observational properties are most consistent with a low-mass iron-core-collapse supernova, even though an electron-capture explosion cannot be ruled out.

\end{abstract}




\section{Introduction}

The Crab Nebula is the first astronomical object identified to have resulted from a supernova (SN) explosion \citep{mayall39}, and since its discovery almost 300 years ago, it remains one of the most studied objects in the sky \citep[e.g.][]{davidson85,hester08}. It serves as a benchmark for understanding neutron stars and pulsars \citep{Ostriker69}, and its well-established physical properties provide a critical anchor for calibrating models of neutrino-powered explosions \citep[e.g.][]{sukhbold16}.

We know the Crab Nebula resulted from a core-collapse explosion because it left behind a rapidly rotating pulsar. The pulsar converts its spin-down energy into a wind of relativistic particles that, in turn, produce a synchrotron-emitting pulsar wind nebula (PWN). 
The ejecta in the Crab Nebula form a network of complex filaments consisting of ejected material that has been swept up and photoionized by the PWN. Studies of filament abundances and photoionization calculations support a relatively low mass progenitor \citep[e.g.][]{macalpine08}.

The most recent estimate for the total mass of gas in the filaments is around $7\:M_{\odot}$, consistent with a $\sim$ 9~$M_{\odot}$ progenitor star and with most of the ejected mass accounted for in the filaments \citep{owen15}. The typical ejecta velocities of $\sim$1200 $\rm km\: s^{-1}$ \citep{temim09} account for less than $10^{50}$~ergs of kinetic energy, and yet the explosion was an order of magnitude more luminous than a normal Type II SN \citep[see][]{smith13crab}.
\citet{chevalier77} argued that there must be a more extended supernova remnant (SNR) beyond the visible Crab that would account for the missing energy, but despite intensive searching, no signatures of such a component have been found \citep{frail95,seward06}. While \citet{sollerman00} did detect a \ion{C}{4} $\lambda$1550 absorption feature in the far-UV spectrum of the Crab's pulsar, the inferred mass and velocity are only 0.3~$M_{\odot}$ and 2500~$km\:s^{-1}$, respectively, accounting for an additional energy of only $1.5\times10^{49}$ ergs.

Historically, such a weak explosion was theorized to be produced by an electron-capture SN (ECSN) resulting from the collapse of a super asymptotic giant branch (AGB) star with a degenerate O-Ne-Mg core \citep{miyaji80,nomoto84,nomoto87, janka08}.
An ECSN was thus proposed to be the most likely culprit to produce the observed properties of the Crab \citep[e.g.][]{nomoto82,kitaura06,smith13crab,yang15}. More recent works, however,  show that the Crab's properties are consistent with a wider range of models, including iron core collapse explosions \citep[e.g.][]{hitomi18, gessner18, stockinger20}. Precisely what type of star and SN explosion produced the Crab Nebula remains unknown.

Here we present JWST \citep{gardner23} imaging and spectroscopic observations on the Crab Nebula that offer a uniquely detailed view of its PWN, ejecta filaments, and dust formation. With its unprecedented sensitivity and spatial resolution at infrared wavelengths \citep{rigby23}, the JWST observations allow us to probe the spatial variations in the Crab's synchrotron emission that reveal insights about the acceleration of particles injected by the pulsar, to isolate emission from dust to determine where the grains formed and to more accurately measure ejecta abundances that constrain the Crab's origin. 

The paper is organized as follows. In \S\ref{redux}, we describe the observations and data reduction, followed by a description of the imaging mosaics and the method for producing various emission maps in \S\ref{mosaics_section}. We discuss the large-scale morphology of the ejecta filaments in \S\ref{large}, the dust distribution map in \S\ref{dust}, large and small-scale synchrotron emission in \S\ref{synch}, and the determination of the nickel-to-iron abundance ratios using photoionization models in \S\ref{ratios}. In \S\ref{explosion}, we discuss the implications of our results for the SN explosion and progenitor type and summarize our conclusions in \S\ref{conclusion}.

\begin{figure*}
\center
\includegraphics[width=1.0\textwidth]{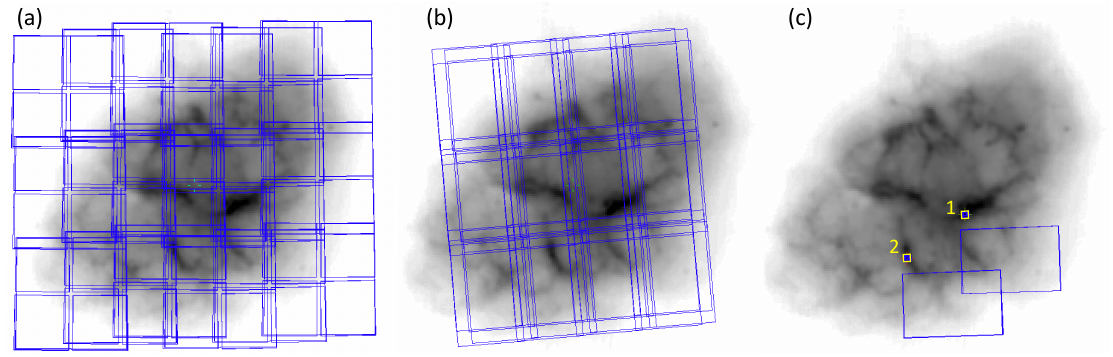}
\caption{\label{coverage} The NIRCam (panel a) and MIRI (panels b) imaging mosaic footprints are overlaid in blue on the \textit{Spitzer}~24~\micron\ image of the Crab Nebula. Panel c shows the footprints of the MIRI MRS observations (small yellow squares), along with the fields-of-view of the two MIRI imaging observations taken simultaneously with the MRS (blue).}
\end{figure*}

\begin{deluxetable*}{lrccccl}[tp]
  \tablecaption{JWST Imaging Observation \label{obs_table} }
\tablehead{\colhead{}  &  \colhead{Filter}  & \colhead{Wave.}  & \colhead{FWHM}  & \colhead{Exp.} & \colhead{Extinction} & \colhead{Dominant Emission Components} \\  
                                   & & ($\mu$m)  & (\arcsec) & Time (s) & Correction & }
                                   \startdata
				NIRCam & F162M     & 1.626 & 0.055 & 870 & 1.184$\pm$0.068 & [\ion{Fe}{2}] 1.644; [\ion{Si}{1}]  1.645; synchrotron\\
				               & F480M    &   4.834               &  0.164  & 870 & 1.030$\pm$0.010 & synchrotron \\
                   \\
\hline
				 MIRI       & F560W   &  5.589       &        0.207   & 555     & 1.027$\pm$0.010 & [\ion{Fe}{2}]  5.34; synchrotron\\
				                & F770W   & 7.528         & 0.269 & 1099    & 1.030$\pm$0.010 & [\ion{Ar}{2}] 6.99; synchrotron  \\
				                & F1130W & 11.298      & 0.375  & 422   & 1.049$\pm$0.018 &  [\ion{Ni}{2}] 10.68; [\ion{Ni}{3}] 11.00;
                    [\ion{He}{1}] 11.24; [\ion{Ni}{4}] 11.73;  synchrotron \\
				                & F1500W & 14.932       & 0.420 & 1055    & 1.032$\pm$0.012 &[\ion{Ne}{3}] 15.555; synchrotron \\
				                & F1800W   & 17.875       & 0.591 & 111   & 1.041$\pm$0.015 & [\ion{S}{3}] 18.713; dust; synchrotron \\
				                & F2100W  & 20.563     & 0.674  & 344   & 1.040$\pm$0.014 & [\ion{S}{3}] 18.713; dust; synchrotron \\
				                & F2550W  & 25.147    & 0.803  & 1099   & 1.035$\pm$0.013 & [\ion{O}{4}] 25.89; [\ion{Fe}{2}] 25.99;  dust; synchrotron \\
				  \enddata
\tablecomments{Values for filter wavelengths and full-width-half-maximum of the point spread function (PSF) have been adopted from {\sl JWST} User Documentation (\url{https://jwst-docs.stsci.edu/}). Derivation of the extinction correction factors is outlined in Section~\ref{method}. }
\end{deluxetable*}




\begin{figure*}
\center
\includegraphics[width=0.9\textwidth]{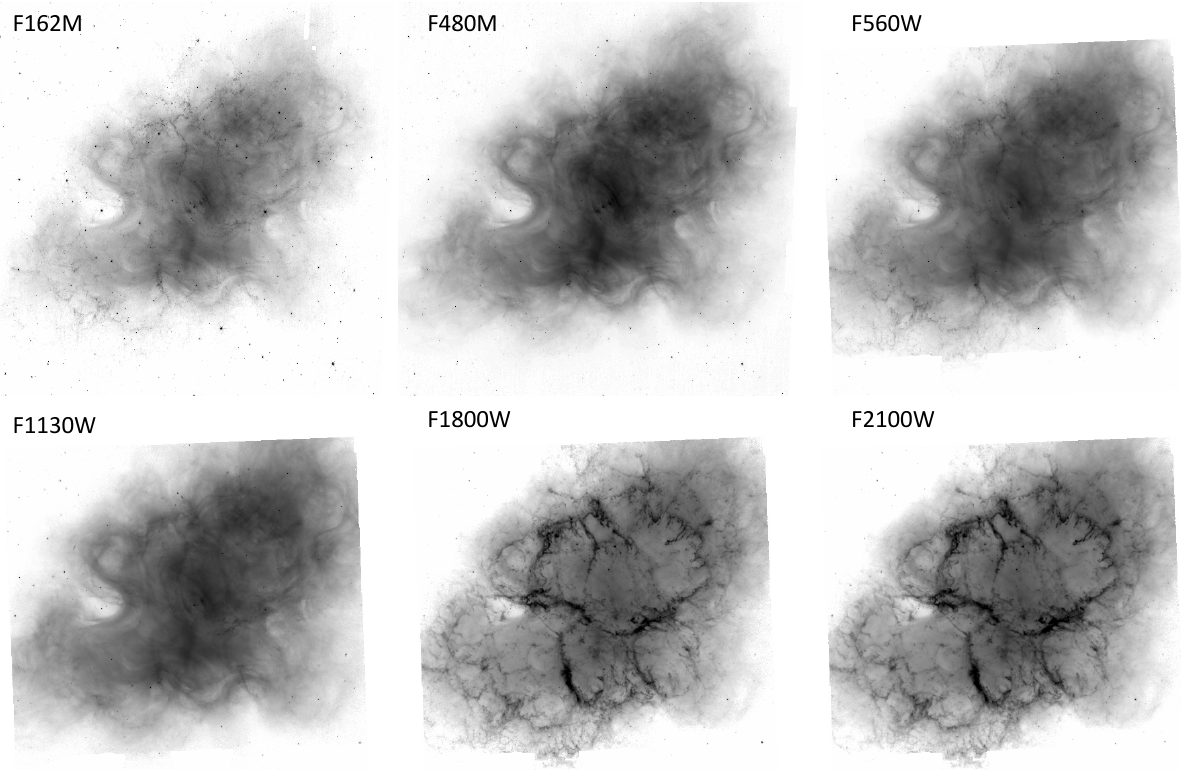}
\caption{\label{mosaics} The NIRCam and MIRI imaging mosaics of the Crab Nebula with dominant emission components listed in Table \ref{obs_table}. The size of each mosaic is 4\farcm9 $\times$ 5\farcm5 and they are oriented with north up.}
\end{figure*}


\begin{figure*}
\center
\includegraphics[width=0.75\textwidth]{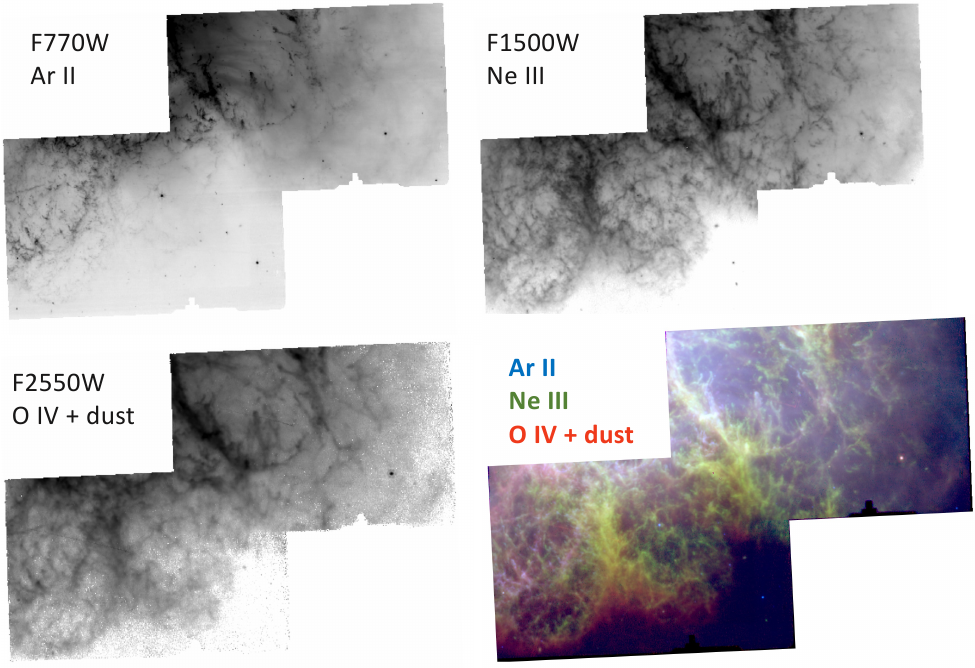}
\caption{\label{simul} MIRI two-tile mosaics composed of images observed simultaneously with the MRS and the three-color image that combines all three filters. The position of the footprint relative to the entire Crab Nebula is shown in Fig.~\ref{coverage}c. The size of each of the two MIRI mosaic tiles is 1\farcm2 $\times$ 1\farcm9.}
\end{figure*}

\section{Observations and Data Reduction} \label{redux}

The Crab Nebula was observed with JWST in the General Observer (GO) Cycle 1 under Program 1714 (PI: Temim). The majority of the SNR was imaged with NIRCam \citep{rieke23} and MIRI \citep{wright23} instruments, while Medium-Resolution Spectrometer (MRS) spectroscopy \citep{argyriou23} was acquired in two chosen locations centered on the ejecta filaments. The footprints of the imaging and spectroscopic observations are shown in Fig.~\ref{coverage} and the imaging filter parameters are summarized in Table~\ref{obs_table}, including the central wavelengths of the filters, the full-width at half-maximum (FWHM) of the PSF. All the data were downloaded from the Mikulski Archive for Space Telescopes (MAST) at the Space Telescope Science Institute and can be accessed via \dataset[DOI: 10.17909/6264-w578]{https://doi.org/10.17909/6264-w578}.

\subsection{Imaging with NIRCam and MIRI}
The NIRCam observations were carried out on 2022 Oct 31 using the F162M and F480M filters. The Crab was mapped by a 3$\times$1 mosaic, shown in panel (a) of  Fig.~\ref{coverage}), using the default row and column overlap region of 10\%. The size of the complete mosaic is 7$\farcm$0$\times$6$\farcm$3. The chosen observing configuration was the FULL array mode utilizing the FULL 3TIGHT primary dither pattern and 3 STANDARD sub-pixel dithers, for a total number of nine dithers per mosaic tile position. We used a BRIGHT1 readout pattern with 5 groups per integration for each dither position, for a total exposure time of 870 s per tile. 

The MIRI observations were performed on 2023 Feb 24 using the F560W, F1130W, F1800W, and F2100W filters. The imaging mosaic consists of 4$\times$3 tile positions, shown in Fig.~\ref{coverage}b, and has an approximate size of 5$\farcm$0$\times$5$\farcm$6. It was carried out with a 4-point extended source dither pattern, a 10\% row and column overlap region, and the FASTR1 readout pattern. The exposure parameters for the F560W, F1130W, and F1800W filters all include a single integration with 50, 38, and 10 groups, respectively. The F2100W observations were carried out with 2 groups of 15 integrations. The corresponding exposure time for each tile position is listed in Table~\ref{obs_table}. A dedicated background observation was taken at a blank-sky position off of the SNR, centered on coordinates $\alpha (\rm J2000.0)$ = 05$^{\rm h}$34$^{\rm m}$38\fs3095, $\delta (\rm J2000.0)$ = +21$^{\rm o}$55$\arcmin$53\farcs53, and using the same exposure, dither, and filter parameters as the source observations. 

Additionally, during the MIRI-MRS observations (see next section), simultaneous imaging observations were carried out in adjacent fields of view with the positions and orientation set by the telescope roll angle on the date of each observation. The footprint of these simultaneous imaging observations is shown in Fig.~\ref{coverage}c. We chose a different imaging filter for each MRS grating change, leading to a total of three filters: F770W, F1500W, and F2550W. The exposure time for each of the filters is comparable to the total MRS exposure time and listed in Table~\ref{obs_table}. 

All imaging observations were processed using the JWST calibration pipeline version 1.12.0, the calibration reference data system (CRDS) version 11.17.2, and the CRDS context file $\mathtt{jwst\_1130.pmap}$. We used the WCS alignment tool JHAT \citep{rest23} to align the NIRCam images to Gaia DR2. The $\mathtt{skymatch}$ step was turned off in the $\mathtt{calwebb\_image2}$ step of the pipeline since the extended emission from the Crab Nebula fills the entire field of view for some tiles. We used the MIRI dedicated sky observations to construct background images for each filter by co-adding and sigma clipping the four individual dithers. This step is particularly important for removing the thermal telescope background that dominates at wavelengths longer than 15~\micron \citep{rigby23}. The background images were then subtracted from each of the level 2 images before they were combined into the final mosaics in the $\mathtt{calwebb\_image3}$ step. The final pixel scales of the NIRCam F162M, NIRCam F480M, and MIRI images are 0.031$^{\prime\prime}$, 0.063$^{\prime\prime}$, and 0.11$^{\prime\prime}$ per pixel, respectively. All of the mosaics are shown in Fig.~\ref{mosaics} and the additional two-tile coverage obtained from the simultaneous imaging is shown in Fig.~\ref{simul}.

\subsection{Medium Resolution Spectroscopy} \label{mrs}

The MIRI~MRS spectroscopic observations were carried out on 2023~Mar~17 using all four channels (1-4) and the three grating settings (SHORT, MEDIUM, and LONG) to cover the entire wavelength range from 5-28~$\micron$ split into 12 sub-bands, one for each channel/band combination. The MIRI~MRS observations at the two filament positions shown in Fig.~\ref{coverage}c) used the 4-point extended source dither pattern, with one group of 100 integrations for each dither in the FASTR1 readout mode, for a total exposure time of 1110 s per channel/band combination. The same configuration was used to observe a background region located to the south of the Crab Nebula, except the 4-point POINT~SOURCE dither pattern was used.

\begin{figure*}
\center
\includegraphics[width=0.85\textwidth]{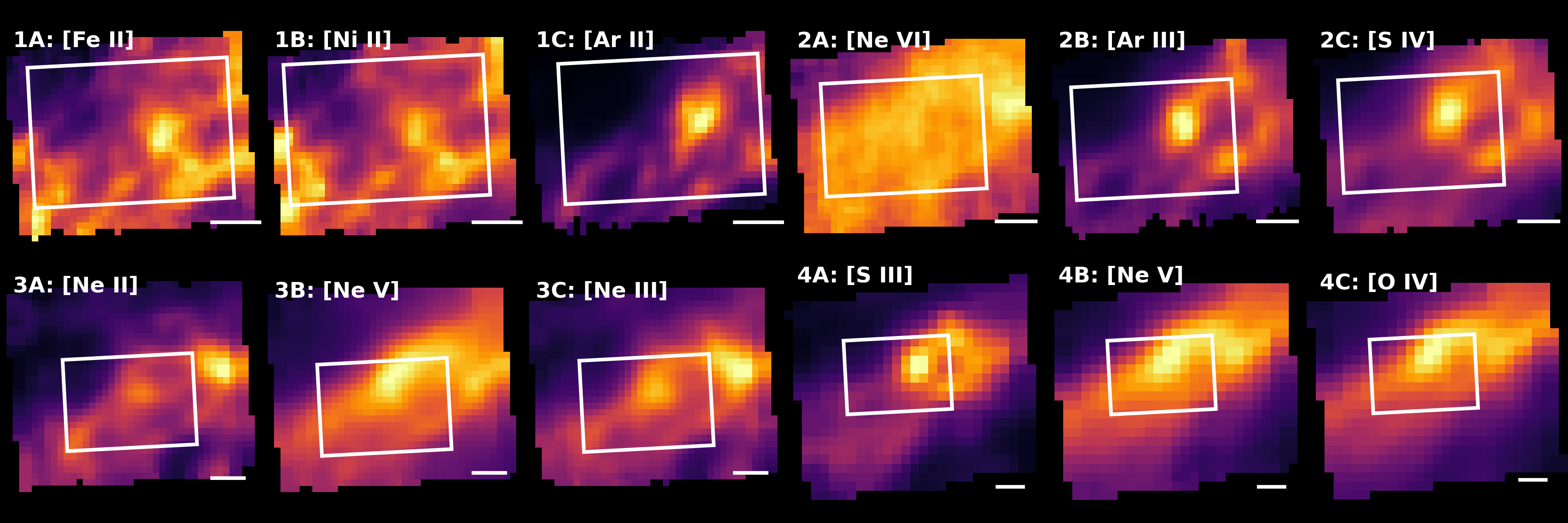}
\caption{\label{mrs_apertures} The extraction aperture for Position 1 (white rectangle) plotted on images of emission lines in each of the 12 MRS sub-band cubes. The sub-band labels and species are shown in the top left-hand corner of each panel. The white lines in the bottom right of each panel represent $1^{\prime\prime}$ to
highlight the increasing size of the FOV from sub-bands 1A-4C. The decreased spatial resolution from channels 1–4 due to the diffraction limit is also evident. North is up, east is left.}
\end{figure*}

We reduced the MRS data using version 1.12.0 of the JWST Calibration Pipeline, with versions 11.17.4 and `jwst\_1154.pmap' of the CRDS and CRDS context, respectively. We processed all raw level 1B files through \texttt{calwebb\_detector1} to produce level 2A rate detector images. We create `master' detector background files for each MRS channel/band combination by median combining the rate images, before subtracting these from the corresponding detector images for each of the filaments. We then used \texttt{calwebb\_spec2} to produce calibrated level 2B detector images, which included the non-default detector level residual fringe correction. We constructed cubes for each of the 12 MRS sub-bands using \texttt{Spec3Pipeline} which implements the cube-building algorithm described in \citet{Law2023}. As described in \citet{argyriou23}, each of the sub-bands has differing spatial dimensions, with band 4C having a larger FOV than 1A. This is illustrated in Fig.~\ref{mrs_apertures}.

\begin{figure*}
\center
\includegraphics[width=0.9\textwidth]{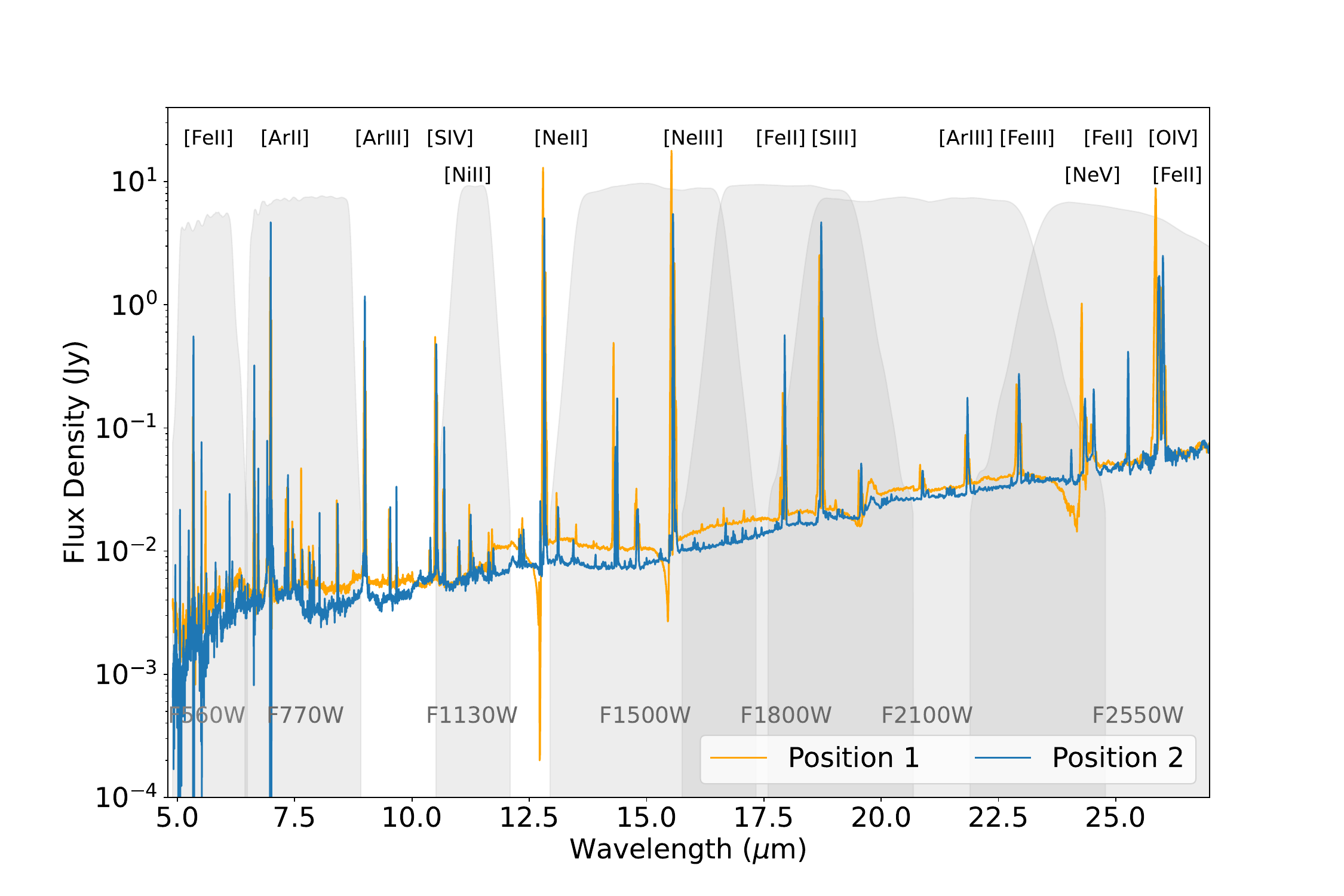}
\caption{\label{spectrum} MIRI MRS spectra from the two observed positions centered on the ejecta filament in the equatorial region (Position 1 in orange) and the southern filament (Position 2 in blue). The two positions are shown in Fig.~\ref{coverage}. The detector artifacts that affect the shape of the spectrum are discussed in \S\ref{mrs}.}
\end{figure*}

\begin{figure*}
\center
\includegraphics[width=0.9\textwidth]{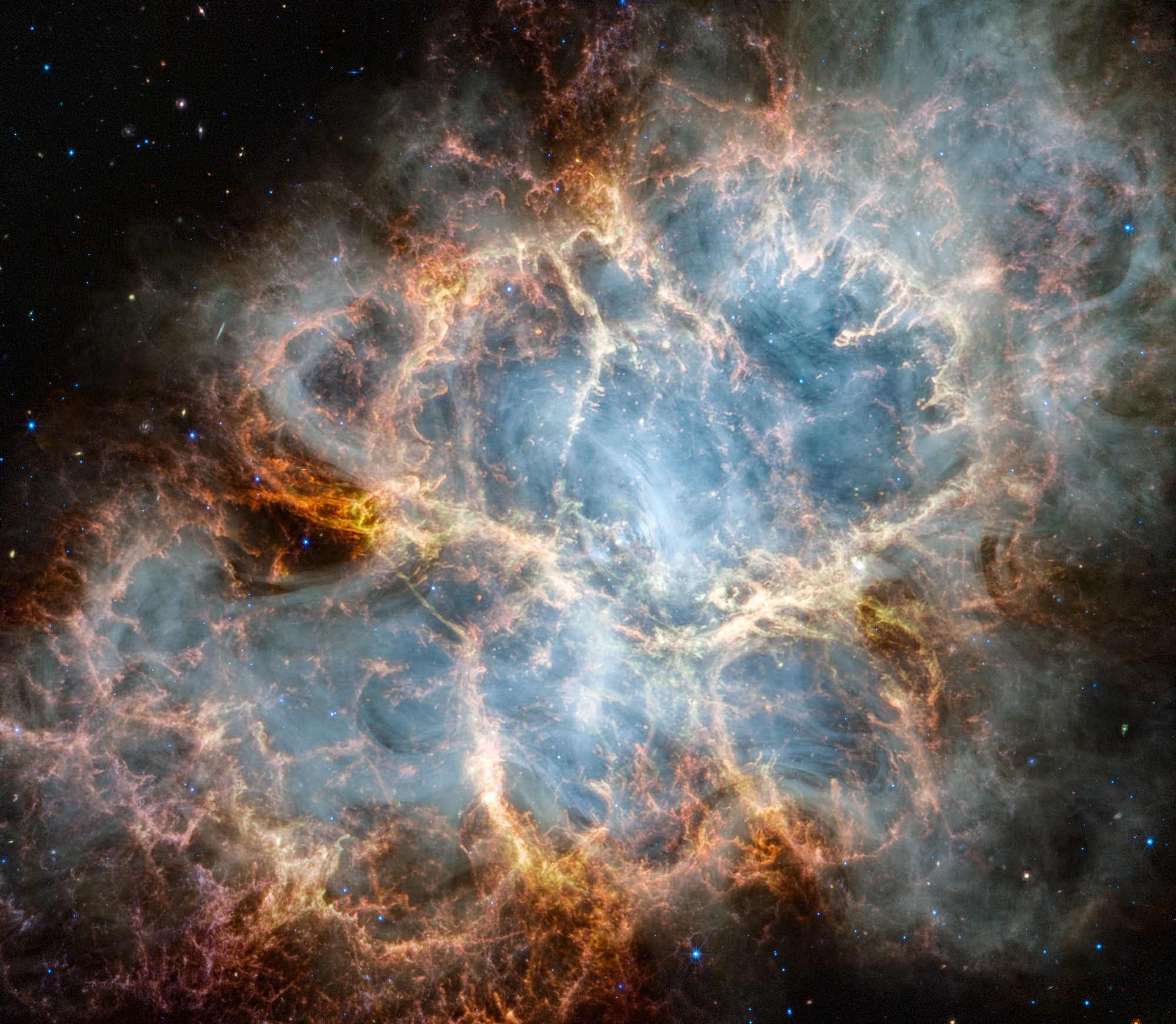}
\caption{\label{composite} STScI and NASA press release image of the Crab Nebula containing all MIRI and NIRCam images. The composite image shows the [\ion{S}{3}] emission in red-orange, [\ion{Fe}{2}] emission in blue, dust emission in yellow-white and green, and synchrotron emission from the PWN in smoky white. The colors assigned to different filters are blue (F162M), light blue (F480M), cyan (F560W), green (F1130W), orange (F1800W), and red (F2100W). Credit: NASA, ESA, CSA, STScI, T. Temim, J. DePasquale.}
\end{figure*}

We extracted spectra from a fixed rectangular aperture that is slightly smaller in size than the band 1A FOV from all the MRS sub-band cubes (see Fig.~\ref{mrs_apertures}). These spectra were processed with the additional spectrum-level residual fringe correction available in the JWST calibration pipeline, to account for any residual fringing, in particular the high-frequency fringes in channels 3 and 4, thought to originate in the MRS dichroics \citep{argyriou23} which is not completely removed by the detector level correction in \texttt{calwebb\_spec2}.

Observing objects with extremely bright emission lines such as supernova remnants can produce several detector effects leading to spurious or misleading features in extracted spectra. These include the ``pull-up/pull-down’’ electronic cross-talk effect \citep{dicken22}, light scattering in the detector across slices and spectral channels, and persistence \citep{argyriou23}. After careful visual inspection of the extracted spectra, we found that all three affect the Crab Nebula spectra to some degree. Persistence and scattered light effects produced false, unidentified broad and narrow emission lines which we flagged and ignored in our analysis. The pull-up/pull-down affected the continuum in spectral channels with saturated emission lines. This manifests as the broad `dips' next to the brightest neon lines in the Position~1 spectrum in Fig.~\ref{spectrum} that then also appear in other channels (e.g. the dip near 24~\micron). There is currently no correction for the pull-up/pull-down effect.

\begin{figure*}
\center
\includegraphics[width=0.89\textwidth]{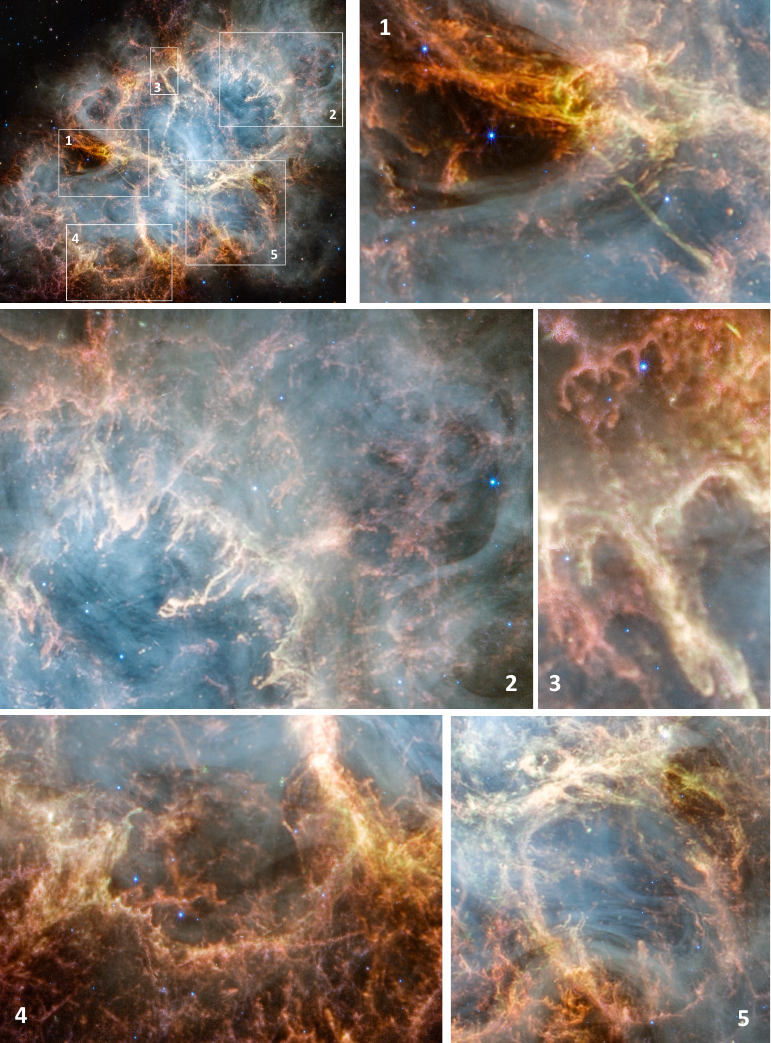}
\caption{\label{composite2} Enlarged regions on the composite image shown in Fig.~\ref{composite}. The [\ion{S}{3}] emission is shown in red-orange, [\ion{Fe}{2}] emission in blue, dust emission in yellow-white and green, and synchrotron emission from the PWN in smoky white. 
}
\end{figure*}

\section{Imaging Mosaics} \label{mosaics_section}

The composite image of the Crab Nebula created from all of the NIRCam and MIRI imaging mosaics is shown in Fig.~\ref{composite}. Fig.~\ref{composite2} enlarges particular regions of the composite image. For reference, at a distance of 2~kpc \citep{trimble68}, one MIRI pixel corresponds to a spatial size of $\sim$ 0.001~pc. These figures highlight the incredible detail present in both the synchrotron emission and the emission filaments while also demonstrating the large-scale structure: an inner `cage' of bright filaments is energized by the bright synchrotron from the pulsar.  One gets the impression that the synchrotron emission then expands outward through gaps in the cage, in some cases extending beyond the outside edge of the ejecta filaments. Because of the combination of filters in this image, the filamentary emission shows a range of ionization levels and samples a range of filament densities. To provide insight into this, the dominant emission components sampled by each of the imaging filters are listed in Table~\ref{obs_table}, and the MIRI imaging filter transmission curves are overlaid on the MRS spectra in Fig.~\ref{spectrum}. Below we discuss how at least some of this complexity can be resolved by processing the images further to separate various components.

For our study, we have chosen the minimum number of filters required to achieve the goal of mapping out the synchrotron emission, determining the distribution of dust grains, and mapping out the Fe emission. All of the images contain a significant contribution from synchrotron emission, but most also include some emission from the ejecta filaments.

The NIRCam F162M image contains synchrotron emission from the PWN as well as the \ion{Fe}{2} 1.644~\micron\ and \ion{Si}{1} 1.645~\micron\ emission from the ejecta filaments. 
The NIRCam F480M image is dominated by synchrotron emission with a minimal contribution from the ejecta filaments. This image is shown in Fig.~\ref{synch_composite} and it provides a high-resolution and high dynamic range view of the PWN structure at IR wavelengths.

All of the MIRI images show synchrotron emission from the PWN, but each also samples different emission lines from the filaments. The dominant emission line in the F560W mosaic is the [\ion{Fe}{2}]~5.34~\micron\ line. While the F1130W mosaic contains mostly synchrotron emission, there is a small contribution from the ejecta filaments traced primarily by three different nickel lines ([\ion{Ni}{2}]~10.68~\micron, [\ion{Ni}{3}]~11.00~\micron, and [\ion{Ni}{4}]~11.73~\micron) and the \ion{He}{1}~11.24~\micron\ line. The ejecta filaments in the F1800W and F2100W mosaics include emission from dust and the [\ion{S}{3}]~18.71~\micron\ line. 

The imaging regions observed simultaneously with the MRS, shown in Fig.~\ref{simul}, are dominated by [\ion{Ar}{2}]~6.99~\micron\ and [\ion{Ne}{3}]~15.56~\micron\ lines for the F770W and F1500W filters, respectively, and by the [\ion{O}{4}]~25.89~\micron\ and [\ion{Fe}{2}]~25.99~\micron\ for the F2550W filter.  Clearly the higher ionization [\ion{Ne}{3}] and [\ion{O}{4}] line emission extends significantly further to the southwest compared with the [\ion{Ar}{2}] emission and is related to the [\ion{O}{3}] `skin' noted in previous HST studies \citep{sankrit97}.  Comparing the morphology of emission in these three images highlights the effect of ionization and density variations, with the low ionization [\ion{Ar}{2}] appearing clumpy and structured while the [\ion{O}{4}] emission appears fluffy and diffuse. Although the [\ion{Ne}{3}] and [\ion{O}{4}] panels look similar in extent in the black and white panels, the color image shows that their relative intensity changes with position as the green filaments are bounded by red toward the lower right.

\subsection{Method for Producing Emission Maps} \label{method}

Here we outline the procedure used to create maps that contain isolated emission from dust and individual emission lines. The background emission for the MIRI filters was subtracted using the off-source blank-sky observations, as outlined in \S\ref{redux}. For the NIRCam images, we measured the background surface brightness in a circular aperture with a 10\arcsec\ radius, centered on $\alpha (\rm J2000.0)$ = 05$^{\rm h}$34$^{\rm m}$40\fs815, $\delta (\rm J2000.0)$ = +22$^{\rm o}$02$\arcmin$48\farcs347 and subtracted the measured value from the mosaics. An extinction correction was applied to all the mosaics using $A_V=1.08\pm0.38$ \citep{delooze19} and the G23 average extinction curve from \citep{gordon23} using $R_V =3.1$. To calculate the extinction correction factors for each wavelength, we used the Python package $\mathtt{dust\_extinction}$ \citep{gordon23a}. The resulting extinction correction factor for each filter is listed in Table~\ref{obs_table}.
Before producing the various emission maps, the images were convolved to the spatial resolution of the lowest resolution image in the set, using convolution kernels produced from PSF models generated by WebbPSF version $\mathtt{1.2.1}$ and the Python package $\mathtt{pypher}$ \citep{boucaud16}. 

\subsubsection{Synchrotron Maps}

Our first step was to produce maps of the synchrotron spectral index and its uncertainty from the synchrotron-dominated F480M and F1130W images and use them to produce maps of the predicted synchrotron emission for each of the other MIRI filters. This process is outlined in \S3.1 of \citet{temim12b}. We have assumed the currently quoted NIRCam photometric calibration uncertainties of $\sim$1\% for the F162M filter and 4\% for the F480M filter\footnote{\url{https://jwst-docs.stsci.edu/jwst-data-calibration-considerations/jwst-calibration-uncertainties}}, and 5\% for each of the MIRI filters \citep{dicken24}. 
The derived synchrotron map was then subtracted from the F560W, F1800W, and F2100W mosaics, leaving mostly emission from the filaments. 
We note here that since the NIRCam and MIRI data were acquired four months apart, temporal changes in the synchrotron wisps (discussed in \S\ref{wisps}) produced artifacts in the F480W to F1130W spectral index map in the central regions around the pulsar. These temporal changes did not affect regions farther from the pulsar and had minimal effect on the emission from the ejecta filaments. We note that to examine the spectral index variations in the regions around the pulsar, we also produce a spectral index map using the contemporaneous MIRI F560W and F1130W images (see \S\ref{indexvar}).

\begin{figure*}
\center
\includegraphics[width=0.8\textwidth]{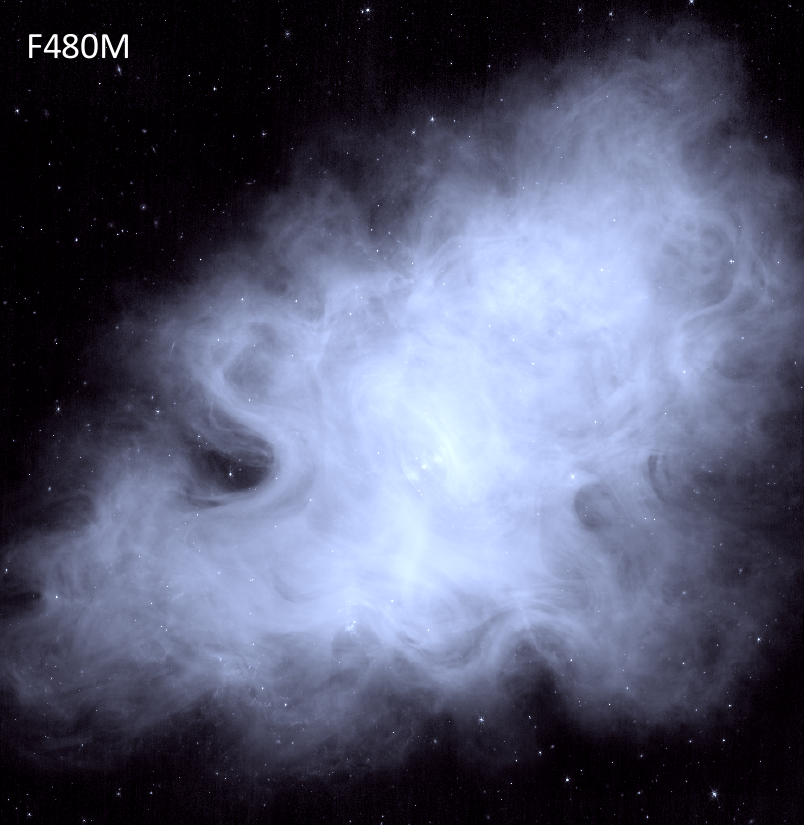}
\caption{\label{synch_composite} The NIRCam F480M image of the Crab Nebula shown here is dominated by synchrotron emission from the PWN and has only a minimal contribution from the ejecta filaments. The image is shown on a log scale to maximize the dynamic range. The size of the mosaicked image is 4\farcm9 $\times$ 5\farcm5 and the orientation is north-up.}
\end{figure*}

\subsubsection{Line and Dust Maps}

\begin{figure}
\center
\includegraphics[width=0.48\textwidth]{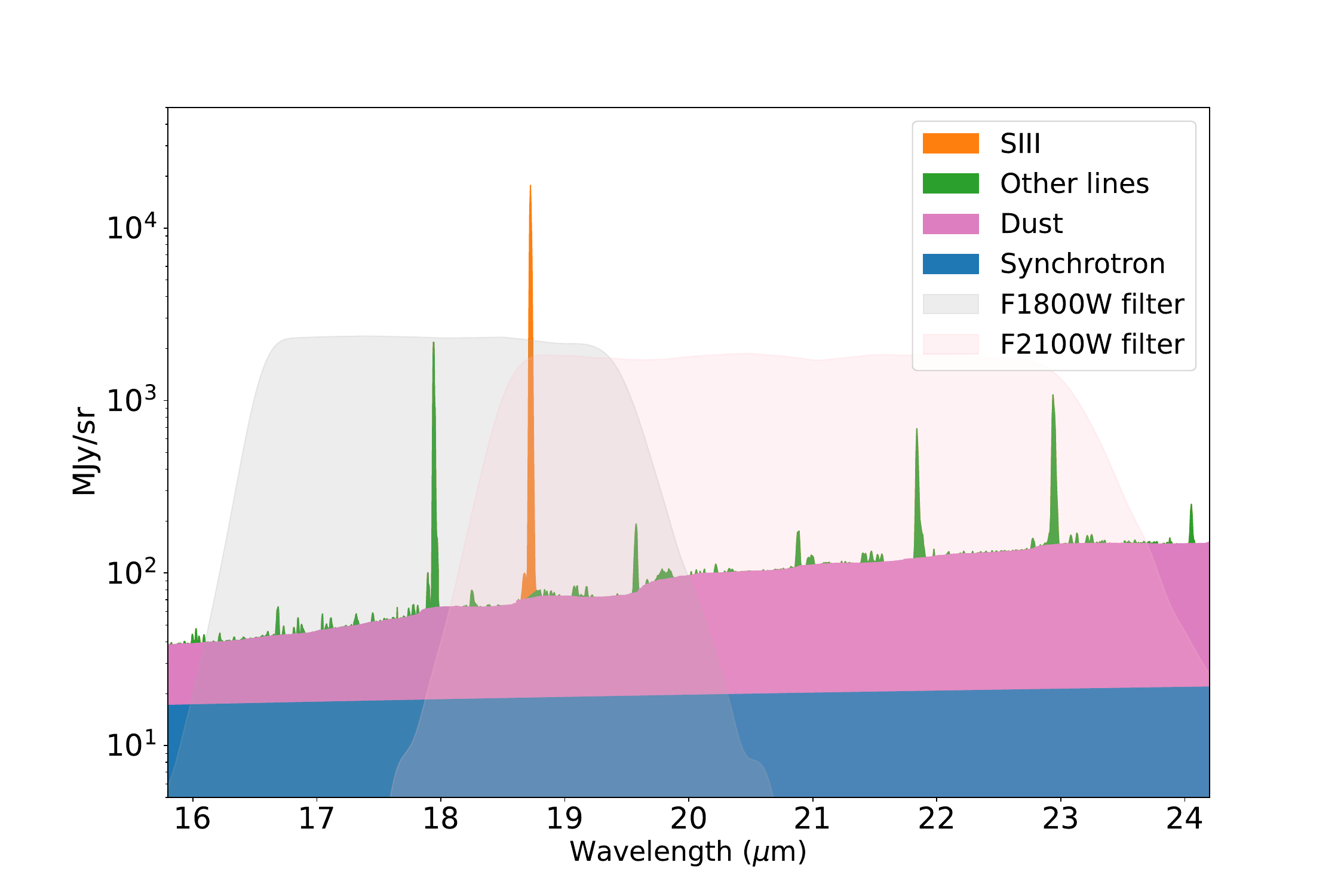}
\includegraphics[width=0.48\textwidth]{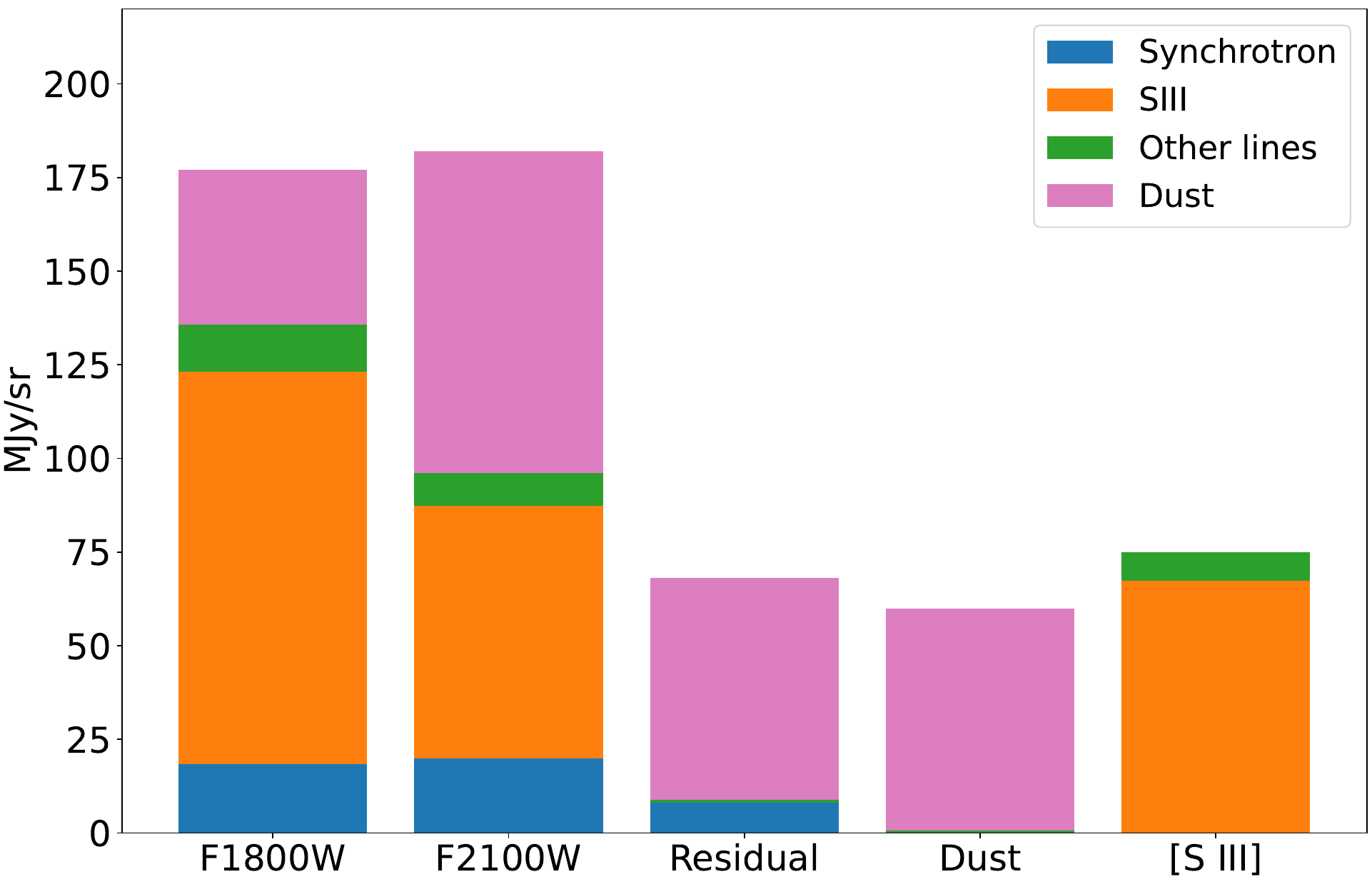}
\caption{\label{dustmrs} Top panel shows the Position 2 MRS spectrum with the F1800W and F2100W filter transmission curves overlaid and the different emission components shaded in different colors. The brightest lines contributing to the two filters are labeled. The bar chart in the bottom panel shows the contribution of the corresponding emission components to the total surface brightness in the F1800W and F2100W filters within the MRS field of view at Position 2 (first two columns). It also shows the contribution of emission components to the weighted residual of the two filters (third column), where Residual = F2100W - 0.65$\times$F1800W. The dust map (fourth column) is given by subtracting the synchrotron component from the residual (Residual - Synchrotron), and the [\ion{S}{3}] image (fifth column) is given by [S III] = F2100W - Synchrotron - 1.65 $\times$ Dust. See \S\ref{method} for details.}
\end{figure}

As summarized in Table~\ref{obs_table}, once the synchrotron component is subtracted from the remaining MIRI images, the residual emission in the F560W mosaic is dominated by the [\ion{Fe}{2}] line emission, and the F1800W and F2100W mosaics by dust and [\ion{S}{3}] line emission. 
To produce separate emission maps for dust and [\ion{S}{3}], we utilized the fact that the filter transmission curves for the F1800W and F2100W images conveniently overlap across a wavelength region that includes the [\ion{S}{3}]~18.713~\micron\ line. This is demonstrated in the top panel of Fig.~\ref{dustmrs} that shows the two filter transmission curves overlaid onto the MRS spectrum from Position~2 and the relative contributions of various emission components shaded in different colors. The synchrotron component in blue has been extrapolated from the F1130W band, using the F480M-to-F1130W spectral index. The remaining continuum emission is assumed to arise from dust and is colored in red. The [\ion{S}{3}] is shown in orange. The bar graph in the bottom panel of Fig.~\ref{dustmrs} depicts the contribution of each emission component to the total surface brightness within the MRS field-of-view at Position~2. Even though the synchrotron component has been subtracted from the images, we show its contribution to the original image for reference. To isolate the dust emission, we subtracted the weighted F1800W image from the F2100W image, with the weighting factor of 0.65 chosen to completely remove the [\ion{S}{3}] emission and isolate the dust emission. Other lines that contribute to the two images are colored green and labeled in the top panel of the figure. While their total contribution appears to partially cancel out in the bar chart, this would only be true if the spatial distribution of the emission lines was the same. Nevertheless, since the contribution of these lines to the total surface brightness is an order of magnitude lower than the contribution from dust, we are confident that the resulting residual image is dominated by dust emission. We note here that while the dust emission map provides the spatial distribution of warm dust grains, the absolute values of the surface brightness are lower than they should be due to the dust emission in the F1800W being subtracted by the procedure we have used here.

\begin{figure*}
\center
\includegraphics[width=1\textwidth]{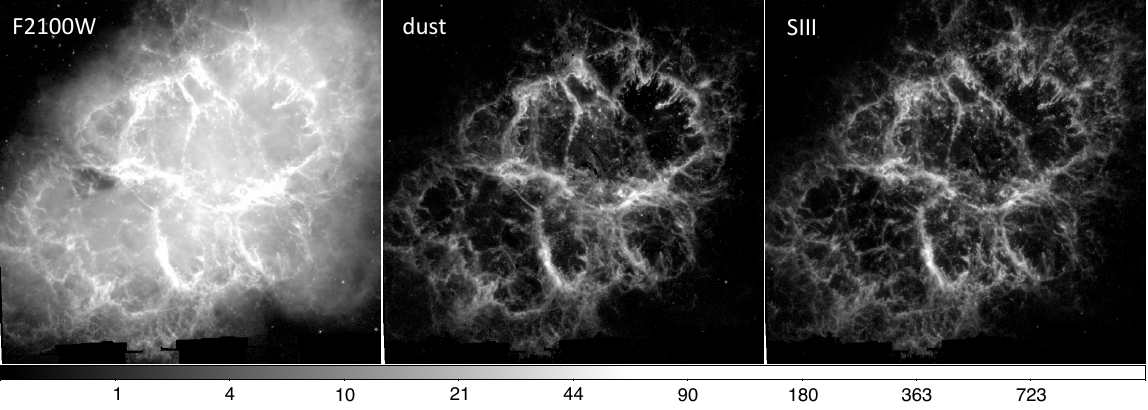} \\
\caption{\label{dustmap} The first panel shows the original F2100W image that contains synchrotron, [\ion{S}{3}], and dust emission. The middle and last panels show the maps of warm dust and [\ion{S}{3}] emission that have been derived from the synchrotron-subtracted F1800W and F2100W images (see \S\ref{method}). }
\end{figure*}

Our last step was to produce a map of the [\ion{S}{3}] emission by scaling the dust emission map (fourth bar in the bottom panel of Fig.~\ref{dustmrs}) up to match the total contribution of dust in the synchrotron-subtracted F2100W image and subtracting it from that image (rightmost bar in the bar chart). This new residual image is then dominated by [\ion{S}{3}] emission with a small contribution from the other lines. Fig.~\ref{dustmap} shows the final resulting maps of the dust emission (middle panel) and the [\ion{S}{3}] emission (right panel). The fourth panel of Fig.~\ref{linemaps} shows the color composite comparison of these two maps, where the dust emission is seen to be concentrated in the innermost filaments while the [\ion{S}{3}] emission extends to larger radii.

\section{Morphology of the Filaments} \label{large}

\begin{figure}
\center
\includegraphics[width=0.42\textwidth]{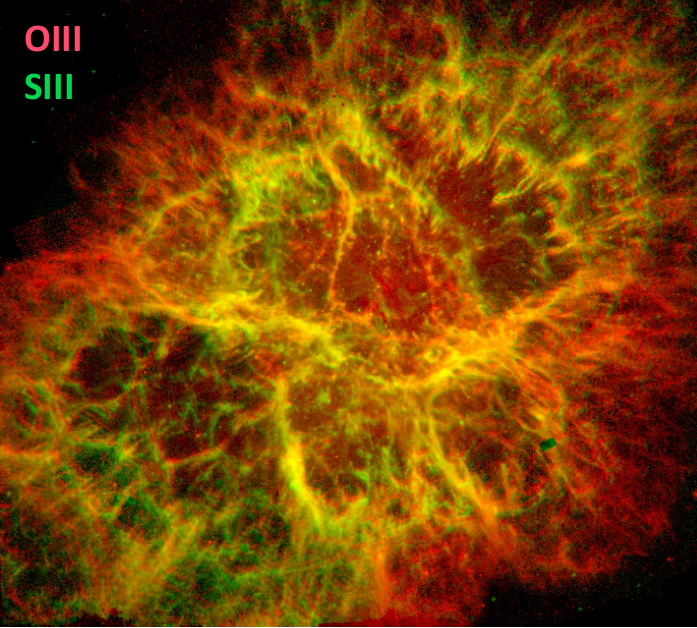}
\includegraphics[width=0.42\textwidth]{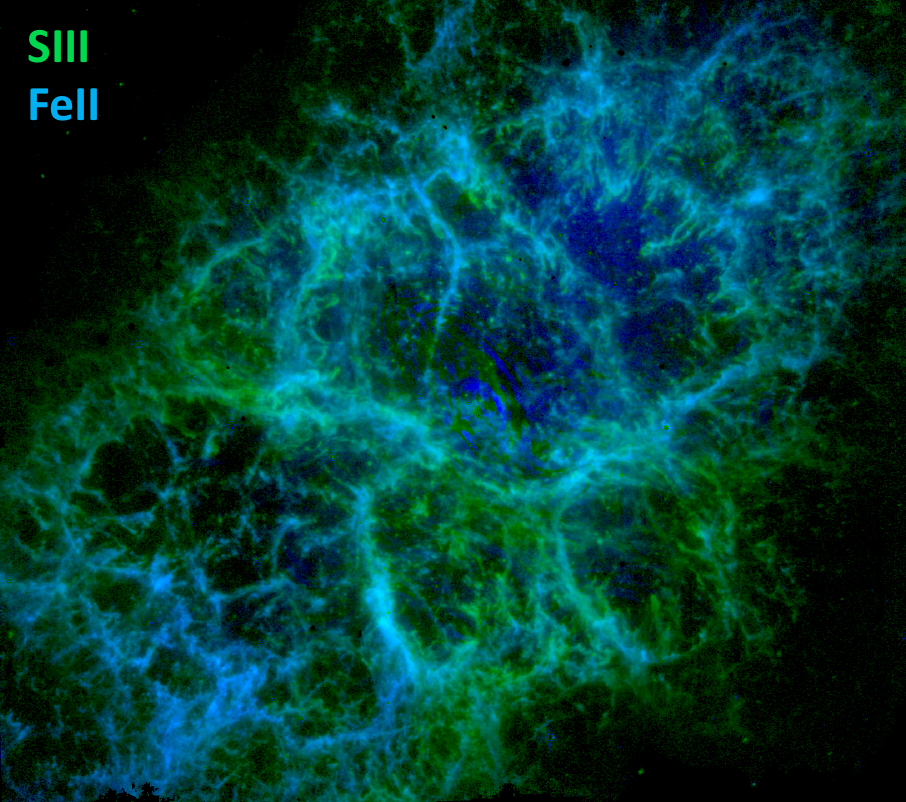}
\includegraphics[width=0.42\textwidth]{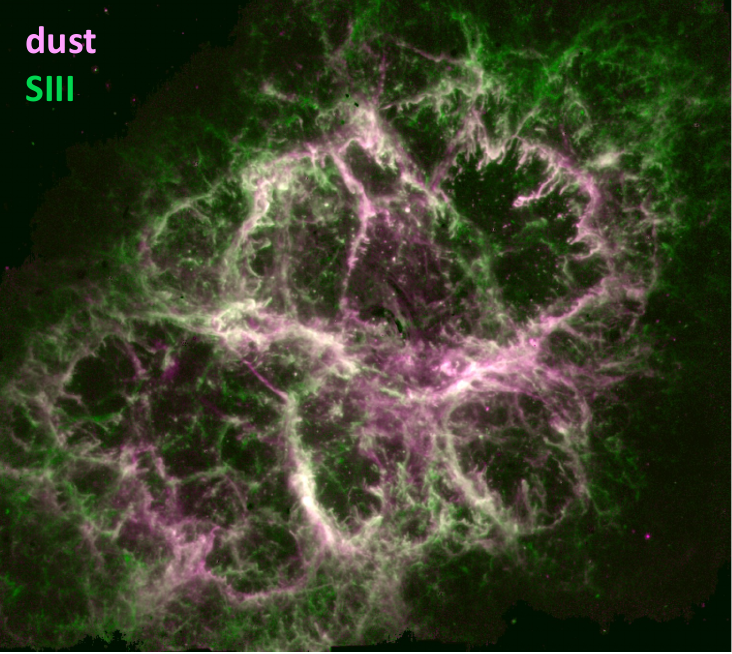}
\caption{\label{linemaps} This figure shows color composites produced from the MIRI line emission images (see \S\ref{method}). The [\ion{O}{3}] emission shown in the top panel was obtained in 1999-2000 with the WFPC2 instrument aboard HST \citep{loll13} and does not align on fine scales with the JWST images. 
See \S\ref{large} for details.}
\end{figure}

In this section, we present a brief assessment of the large-scale spatial distributions of the various maps constructed above in \S\ref{method}.  Fig.~\ref{linemaps} shows color figures highlighting various combinations of these maps.  In all three panels, the [\ion{S}{3}] image is used as a reference and is shown in green. The `cage' of inner filaments stands out from the more extended structure in all three of the comparisons shown.  Interpretation of these comparisons is somewhat complicated because of the differing ionization stages of the various ions and the known correlation of ionization with density variations, especially in the bright filaments.  Here we will concentrate on the larger-scale trends, and leave more detailed smaller-scale filament comparisons to a future paper.

The top panel in Fig.~\ref{linemaps} shows [\ion{O}{3}] from HST \citep{loll13} in red compared with [\ion{S}{3}], so regions bright in both ions appear various shades of yellow. 
Because of the time separation between the existing HST data (circa 2000) and the JWST images, the proper motion of many filaments is obvious, and filament-by-filament comparisons are compromised. On larger scales, both ions display a similar filamentary morphology but their spatial distributions are different. We see [\ion{O}{3}] extending to greater radii, especially in the northeast and southwest regions aligned with the short axis of the nebula, while both ions extend along the major axis to comparable distances.  The ionization potential for $\rm O^{+}$ to $\rm O^{++}$ is 35.1 eV while for $\rm S^{+}$ to $\rm S^{++}$ it is 23.3 eV.  Hence, it is tempting to attribute the different spatial extents to asymmetrical ionization effects. 
However, the ionization balance calculations described in the next section reveal [\ion{O}{3}] and [\ion{S}{3}] emit over similar density and temperature ranges, so we cannot rule out the importance of varying elemental composition in contributing to what is seen in this panel.

The middle panel of Fig.~\ref{linemaps} shows the distribution of [\ion{Fe}{2}] in blue to [\ion{S}{3}] in green, both from JWST.  With a low ionization potential of only 7.9 eV for $\rm Fe^{+}$, this ion is expected to trace the densest cores of filaments which are being ionized from the outside inward by the synchrotron emission. Indeed, it is primarily the cage of bright ejecta that are prominent in blue in this panel although it is interesting that fainter outer filaments along the major axis show up as well.
The bottom panel of Fig.~\ref{linemaps} shows the warm dust map relative to [\ion{S}{3}]. The dust emission primarily traces the bright inner filaments that likely have the highest densities. Along the major axis, some of the fainter dust emission seems to trace [\ion{Fe}{2}] in the SE, but this behavior does not appear to be mirrored in the extended northwest region. The dust emission map is discussed in more detail in \S\ref{dust} and the last paragraph of \S\ref{pwnmorph}.

\section{Dust Emission} \label{dust}

The first evidence for dust in the Crab Nebula came from an observed IR excess above the spatially-integrated synchrotron spectrum \citep[e.g.][]{trimble77,glaccum82}, and later from observed knots of optical extinction of the background synchrotron radiation \citep{woltjer87, fesen90}. Estimates of the total dust mass in the Crab Nebula based on the total integrated spectrum have varied considerably, with a range of 0.02--0.5~$M_{\odot}$ \citep{gomez12a,temim13, owen15}.
The total dust mass estimate was more recently refined to be 0.03-0.05\,M$_{\sun}$ after updates to the corrections for the Crab’s synchrotron spectrum and foreground interstellar dust emission \citep{delooze19}.

Determining the spatial distribution of the emitting dust has been challenging; broadband imaging filters usually also contain bright emission lines and IR spectroscopy only exists for select regions in the Crab.
HST images showed that dust extinction features are concentrated in the cores of the ejecta filaments \citep{blair97,sankrit98}, and spatially resolved \textit{Spitzer} spectroscopy of select positions later confirmed that dust emission does indeed reside in the filaments \citep{temim12b}. 
\citet{gomez12a} used \textit{Herschel} imaging in broadband filters that is likely dominated by dust to spatially associate a cool (28-34\,K) dust component with the ionized ejecta filaments that emit brightly in the optical \citep{fesen92} and that host copious amounts of molecular hydrogen \citep{loh12}.

The new JWST observations have allowed us to isolate the dust emission across the entire Crab Nebula, providing a dust distribution map with unprecedented spatial resolution. This map is shown in the middle panel of Fig.~\ref{dustmap} and the lower-right panel of Fig.~\ref{linemaps}. The temperature of the dust grains emitting in the MIRI F2100W image is approximately 55~K \citep{temim13}, so this map traces warm dust in an intricate network of ejecta filaments within the PWN.  The method for producing this map is outlined in \S\ref{method}. 

\begin{figure}
\center
\includegraphics[width=0.48\textwidth]{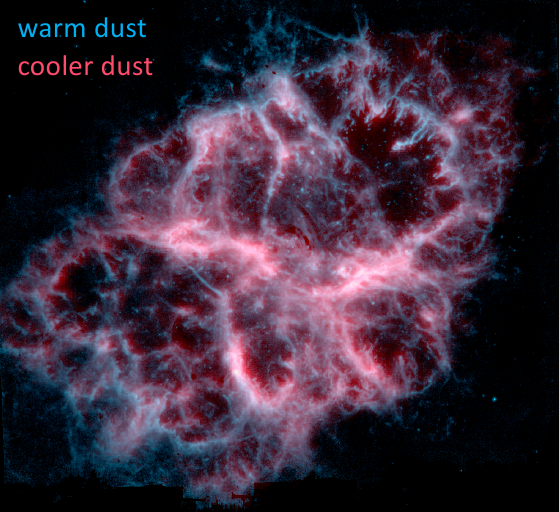} \\
\caption{\label{dusttemp} A color composite image with the warm dust component from JWST shown in blue and the \textit{Herschel}~70~\micron\ image dominated by cooler dust shown in pink. We note that the spatial resolution of the JWST/MIRI image of warm dust has a factor of eight times higher spatial resolution than the \textit{Herschel}/PACS image (PSF FWHM of 0.674\arcsec\ versus 5.6\arcsec)}.
\end{figure}

Fig.~\ref{dusttemp} shows a color-composite comparison of the JWST dust emission and the \textit{Herschel} 70~\micron\ emission (PSF FWHM = 5.6\arcsec) that is likely dominated by the much cooler grains. The brightest emission detected in the \textit{Herschel} image also shows up in the new JWST data, highlighting that the warm dust is mainly distributed along the ejecta filaments that host most of the cool dust material in the Crab. Explaining such a wide dust temperature range may require a scenario in which the cooler dust is located deeper within the filaments, shielded from strong radiation, whereas the warmer dust may be located on the outskirts of the filaments where they are heated to higher temperatures by the PWN's radiation. Fig.~\ref{dusttemp} also shows that the outermost filaments along the long axis of the Crab Nebula appear bluer and contain relatively more warm dust, while the pink, cooler dust emission peaks in the inner filaments. This could also be explained by more shielding of grains in the innermost massive and dense filaments, and less shielding in the outer filaments with lower densities.

The shielding of dust within dense filaments is consistent with recent modeling efforts to explain the observed $^{36}$ArH$^{+}$ emission in the Crab Nebula \citep{barlow13} that require high total hydrogen densities of $>10^{3}$ cm$^{-3}$ \citep{priestley17} and 10$^{4-6}$ cm$^{-3}$ \citep{das19} at the interface of ionized and neutral regions within the Crab.
Fig.~\ref{dusttemp} also shows that the emission from the cooler grains peaks at the tips of the Rayleigh-Taylor fingers where the densities are higher and more shielding from radiation is expected to occur.
Furthermore, the high densities in these regions may have stimulated efficient coagulation of grains, creating a size distribution that preferentially populates the large grain sizes (a~$>$~0.1~$\mu$m), consistent with polarized dust emission in the Crab \citep{chastnet22} and physical dust heating models \citep{temim13, owen15, priestley19}. 

The composition of dust in the Crab Nebula is still not well known. The three brightest ejecta filaments were detected in polarised emission with the SOFIA HAWC+ instrument \citep{chastnet22}, and rather low polarization fractions (4-10$\%$) seem to indicate that at least part of the ejecta dust is composed of carbonaceous material. This is in line with observations of carbon-rich ejecta material found in the Crab \citep{gomez12a,owen15}. 
One of the goals of the MRS observations was to determine the composition of dust grains in the Crab Nebula. However, since the very bright emission lines in the Crab produced various still-uncharacterized detector effects that contaminate the continuum emission detected by MRS (see \S\ref{mrs} and Fig.~\ref{spectrum}), we leave a detailed study of the grain properties to a future paper.

\section{Synchrotron Emission}\label{synch}

\subsection{Overall Morphology and the Synchrotron Bays} \label{pwnmorph}

The synchrotron-emission-dominated F480W NIRCam image of the Crab Nebula is shown in Fig.~\ref{synch_composite}. A remarkable level of detail in the PWN structure is evident in the image. The bright torus surrounds the pulsar in the center, with the outer nebula appearing to have a fibrous structure of intricate ripples and loops, seen in the enlarged images of select regions of the PWN in Fig.~\ref{composite2}, particularly panel 2.
A long-known observational feature of the PWN is its elongation along the SE/NW axis that is aligned with the pulsar's jet. The long axis of the PWN measured $\sim$~7\farcm6 in length, while the equatorial SW/NE axis that is aligned with the pulsar's torus is $\sim$~5\farcm5.

This elongation of the Crab Nebula's PWN was proposed to be caused by the pinching effect of the pulsar's toroidal magnetic field \citep{begelman92}. However, some subsequent magnetohydrodynamic models for the evolution of the PWN showed that the elongated shape does not persist when the models are extended into three dimensions and that the close-to-uniform total pressure produces a more spherical PWN \citep{porth14a,porth14b}. In this case, the elongated shape would more likely be caused by an aspherical SN ejecta distribution or a disk-like circumstellar material (CSM) distribution that confines the PWN in the equatorial plane, as proposed by \citet{fesen92}. On the other hand, 3-D models that assume an anisotropic distribution of the energy flux in the wind do produce an elongation in the PWN along the pulsar's rotational axis \citep{olmi16}.

\citet{michel91} noted an hour-glass structure in polarized light images of the Crab Nebula and a highly organized ``bay'' structure in the east that was interpreted as “scalloping” in the outer nebula due to the PWN’s interaction with the ejecta filaments. In this scenario, the filaments form a conducting cage around the PWN and the wind expands outward between the dense filaments \citep{michel91}. Based on the measurements of the expansion of the synchrotron PWN, \citet{bietenholz91,bietenholz15} also suggested that the relativistic gas is “bursting through” the net of filaments and accelerating into the surrounding low-density medium.

These so-called synchrotron bays, the indentations in the synchrotron emission on the east and west sides of the PWN that give the nebula an hourglass shape, are shown in Fig.~\ref{synch_composite} and enlarged in panels 1 and 5 of Fig.~\ref{composite2}. The misalignment between the axis of the hourglass which is oriented N-S and the axis of the torus oriented in the SW-NE direction is a curiosity and has generated some debate on whether the global elongated morphology of the Crab Nebula is shaped by asymmetries intrinsic to the PWN, by its confinement by asymmetric surrounding material, or some combination of both.

The bays are proposed to be created through the interaction between the PWN and ejecta or CSM in the equatorial plane \citep{michel91, fesen92, li92}. \citet{fesen92} argued that the bays could be formed through the E-W confinement of the PWN by a disk-like CSM structure, reminiscent of the ring in SN 1987A. The nebular magnetic field that wraps around this torus blocks the relativistic particles, causing the indentations in the PWN. The He-rich abundances of the E-W filaments that align with the bay structures \citep{uomoto87,macalpine89} offered additional support for this interpretation.  Furthermore, a bipolar hourglass shape that is oriented N-S is seen in the kinematics of emission lines from the filaments, such as [O~{\sc iii}] $\lambda$5007 emission \citep{smith03}. The Crab's ``chimney" feature also extends directly to the north of the inferred explosion center and is tilted from the SE-NW torus/jet axis \citep{davidson85}.

\begin{figure}
\center
\includegraphics[width=0.48\textwidth]{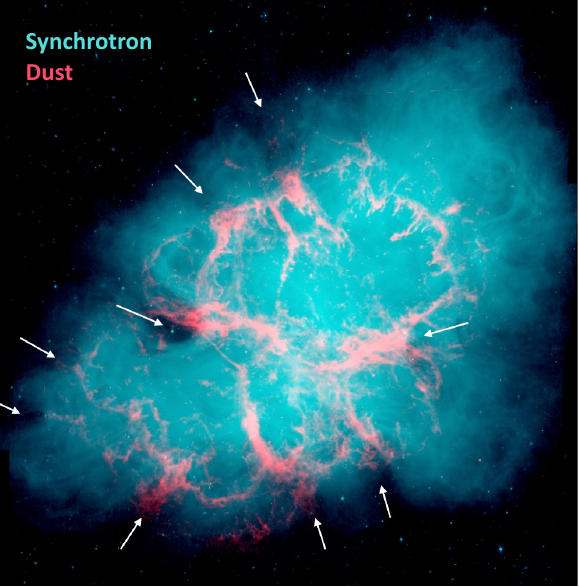}
\caption{\label{baysimage} This image shows the F480M image of the Crab Nebula in blue that traces the PWN's synchrotron emission and the dust emission map in red. The white arrows indicate the locations of indentations in the PWN that resemble the larger ``bays'' in the East and West equatorial regions that have been noted previously.}
\end{figure}

\citet{hester95} pointed out that the E-W band of filaments may not be significantly influencing the structure of the PWN since other filament “rings” are also present in the nebula. They note that the east and west bays may look particularly prominent because the scalloping of the PWN by the filaments lies along our line of sight. 
There are, in fact, other bay-like structures around the perimeter of the PWN. Fig.~\ref{baysimage} shows the F480M synchrotron emission in blue and the dust emission map (\S\ref{dust}) overlaid in red. It can be seen that various other indentations or “bays” also coincide with the location of the densest filaments, as traced by the dust emission. The more prominent thicker filaments appear to be associated with larger indentations in the PWN, and the thinner filaments in the SE and NW with smaller ones. This comparison suggests that the PWN is confined by multiple prominent filaments that lie in the wider NE-SW band in the plane of the torus, and not only by the E-W running filaments, although the East equatorial filament does appear particularly prominent. 
The PWN is less confined in the NW and SE lobes that extend further out from the pulsar and show smaller indentations aligned with less prominent filaments.

Recent 3-D reconstruction of the structure of the Crab Nebula using imaging Fourier transform spectrometer data from the SITELLE instrument showed that the 3D volume follows a “heart-shaped” distribution that is symmetrical about the plane of the pulsar torus, with the two lobes separated by $\sim$120$ ^{\circ}$ \citep{martin21}. This distribution is consistent with the confinement of the SN ejecta along the plane of the PWN torus. While this geometry may be produced by the expansion of a spherical PWN into CSM or ejecta that happen to be distributed in the plane of the torus, the alignment with the torus makes it more plausible that an axisymmetric nature of the PWN produces the asymmetry. 
 
Another example of a very elongated PWN in the SNR 3C~58 \citep{reynolds88} shows the long axis to be aligned with the rotational axis of the pulsar \citep{slane04}, with optical ejecta filaments distributed in a roughly spherical distribution centered on the pulsar and extending to the radius of the PWN's short axis \citep{fesen08}.

It can be seen in Fig.~\ref{baysimage} that the synchrotron emission from the PWN extends well past the innermost dense filaments traced by the dust emission. Three-dimensional simulations of \citet{blondin17} show that when the PWN expands into the outer steep part of the ejecta density profile, Rayleigh-Taylor instabilities can fragment the ejecta shell. The shocked pulsar wind breaks out of the shell and accelerates to larger radii in the freely expanding SN ejecta, leaving an inner shell of swept-up ejecta with about half its total mass enclosed within half of the PWN radius. The distribution of the dense filaments in Fig.~\ref{baysimage} appears consistent with this picture, particularly along the NW direction where the synchrotron emission extends well beyond the detected optical filaments \citep[e.g.][]{loll13}. For the PWN to reach the steep part of the ejecta density profile, its deposited energy needs to exceed the SN explosion energy, consistent with a low-energy explosion for the SN that produced the Crab Nebula \citep{blondin17}. However, this would require that about half of the kinetic energy we observe today in the Crab filaments is deposited by the PWN, and thus, the original SN explosion was even lower by about a factor of two. Since a generous assessment of the current filament kinetic energy is $<\:10^{50}$ \citep{smith13crab}, this would require an extremely low SN explosion kinetic energy of only a few times $10^{49}$ erg and would not account for the high observed luminosity of SN~1054.

A similar filament morphology and PWN ``blowout" will occur \citep{og71} if the PWN had encountered a thin shell produced by the interaction of the SN ejecta with significant CSM, as proposed by \citet{smith13crab}. During the CSM interaction phase, a large fraction of the SN kinetic energy is converted to radiation, making it even easier for the PWN energy to exceed the shell's kinetic energy.  This would also predict that the original SN explosion kinetic energy was actually somewhat more than the currently observed filament kinetic energy (i.e., more plausibly around 10$^{50}$ erg), because some of that explosion kinetic energy would have been lost to radiation to power the bright SN~IIn event \citep{smith13crab}.  This scenario is also consistent with the lack of freely expanding SN ejecta and a blast wave outside the Crab filaments, since in this CSM interaction scenario, the forward shock has been decelerated by CSM.  This latter scenario is discussed further in \S\ref{explosion}.

\begin{figure}
\center
\includegraphics[width=0.48\textwidth]{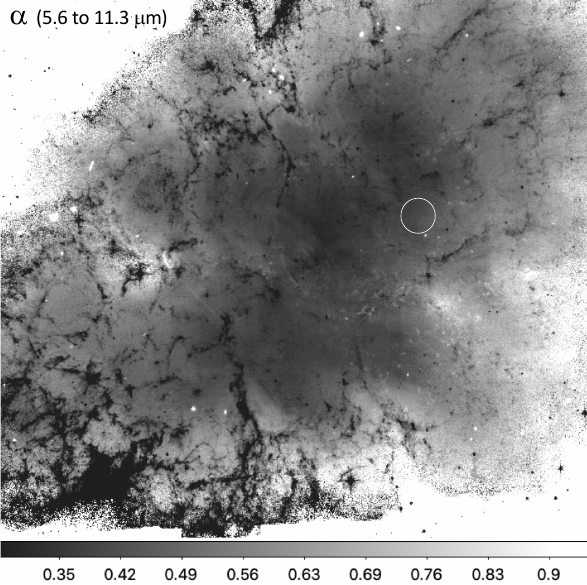} 
\includegraphics[width=0.48\textwidth]{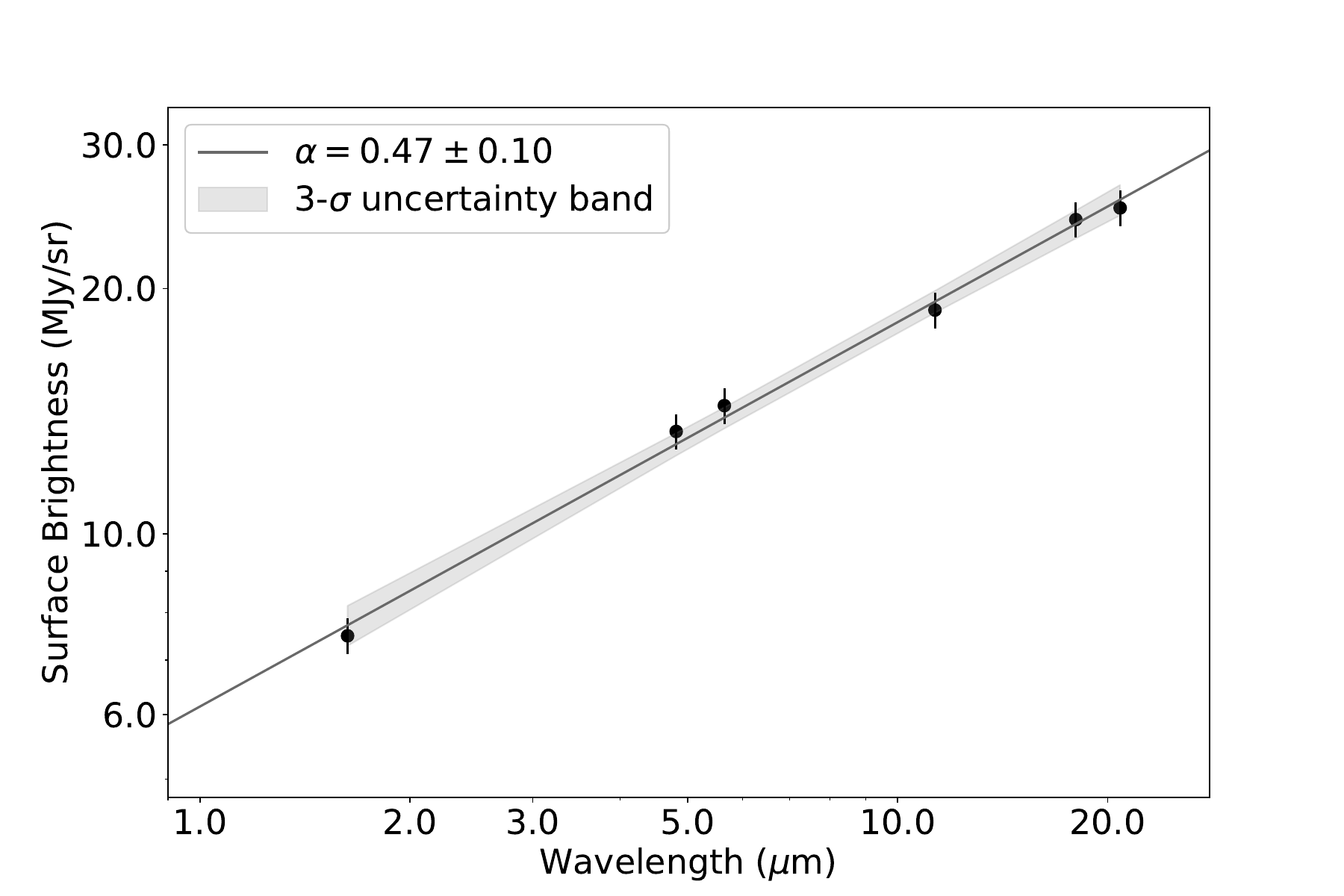}
\caption{\label{indexmap} Top panel shows the synchrotron spectral index map computed from the coeval F560W and F1130W MIRI images. The map is shown on a linear scale with the index scale bar running along the bottom. The ejecta filaments have very flat indices and appear black in the image. The plot in the bottom panel shows the average surface brightness value within the white circular aperture shown in the top panel, a region chosen to avoid filament emission. The uncertainties on the values are dominated by photometric calibration uncertainties. The black line is a power-law model with a best-fit index of $\alpha=0.47\pm0.10$.}
\end{figure}

\begin{figure*}
\center
\includegraphics[width=1.0\textwidth]{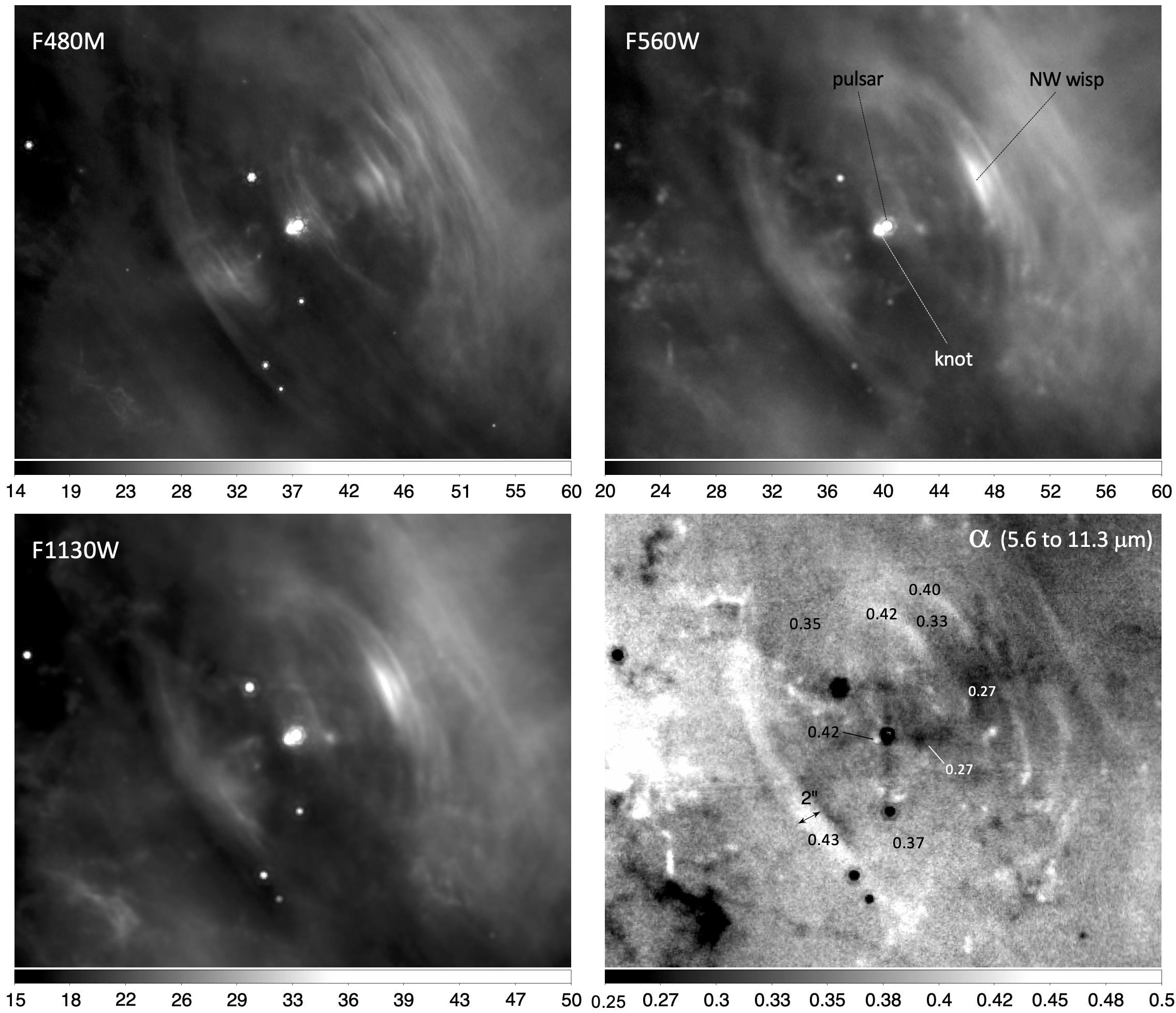}
\caption{\label{wispfigure} The first three panels are images of the inner pulsar region showing the structure of the synchrotron wisps, with a field-of-view size of 40\arcsec\ across and the scale in units of MJy/sr. The F560W and F1130W images were taken on the same day, while the F480M image was obtained four months before. The last panel shows the synchrotron spectral index map between the F560W and F1130W images. The spectral index values are labeled for several regions. While the absolute values of the spectral indices have an average 3-$\sigma$ uncertainty on the order of $\sim$0.15, the relative values in the index map have 3-$\sigma$ uncertainties of only $\sim$0.025.}
\end{figure*}

\subsection{Spectral Index Variations}\label{indexvar}

To explore the spatial variations in the synchrotron spectral index across the PWN, we computed a spectral index map between the coeval F560W and F1130W MIRI images (see \S\ref{method}). This map is shown in the top panel of Fig.~\ref{indexmap}. Since the F560W image has a significant contribution from [\ion{Fe}{2}], the ejecta filaments stand out in the index map as black regions with low index values. The index map of the synchrotron emission itself is fairly smooth and seen to steepen with distance from the pulsar region, as previously seen in the optical \citep{veron-cetty93} and IR \citep{temim06,temim12b} observations. In this MIRI wavelength range, the index values are approximately 0.35 $\pm$ 0.05 in the inner torus region to 0.80 $\pm$ 0.07 in the outermost regions of the PWN. The quoted 1-$\sigma$ uncertainties are dominated by the 5\% photometric calibration uncertainty for the MIRI images and are six times lower than those measured for the spectral index map derived from the 3.6 and 4.5 \micron\ \textit{Spitzer} images \citep{temim06}.

The global spectral indices of the PWN measured at radio and optical wavelengths imply that a break in the spectrum occurs somewhere in the IR wavelength range (please see the broadband spectral energy distribution of the Crab Nebula in Figure~1 of \citet{lyutikov19}).
We searched for evidence of a break in the JWST data by selecting a region in the Crab Nebula that is dominated by the synchrotron component and relatively free of filament emission. This region was also chosen to be away from the inner torus region that exhibits temporal brightness changes on short timescales. It is shown as the white circle in the top panel of Fig.~\ref{indexmap}. The average surface brightness values measured within this region for the NIRCam and MIRI images are plotted in the bottom panel of the figure. The black line is the best-fit power-law model with a 3-$\sigma$ uncertainty band shown in gray. All points are well-fitted by a power law model with a spectral index of 0.47 $\pm$ 0.10, indicating that we cannot confirm a spectral break in the JWST wavelength range with the present calibration uncertainty values.

\begin{figure*}
\center
\includegraphics[width=0.49\textwidth]{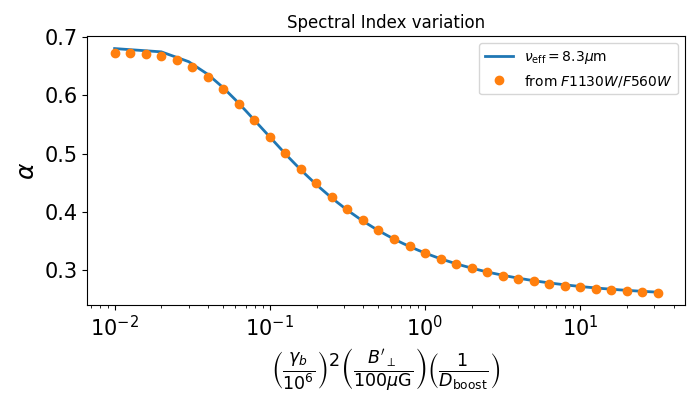}
\includegraphics[width=0.49\textwidth]{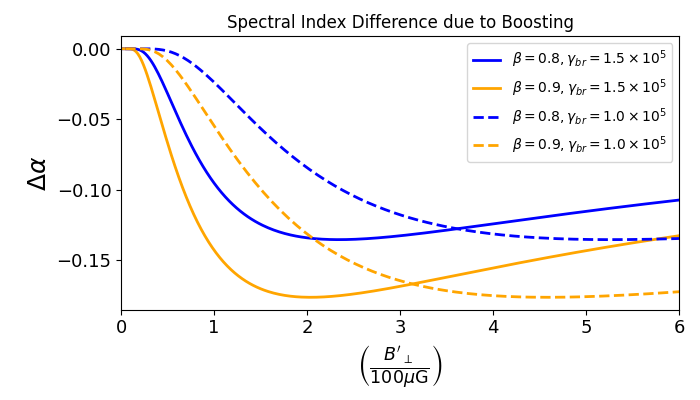}
\caption{\label{index_change_plot} Left panel: Spectral index of synchrotron emission due to pairs having an energy distribution given by a broken power-law with lower energy index $1.35$ and high energy index $2.35$, as a function of the break Lorentz Factor $\gamma_{br}$, comoving magnetic field (orthogonal to the line of sight) $B'_\perp$, and Doppler boosting term $D_{\rm boost}$ (where $\gamma$ is the bulk Lorentz factor of the emitting plasma, and $\beta$ if the component of the bulk velocity along the line of sight. The solid blue line is the analytical value at $\lambda =8.3\mu{\rm m}$; orange dots are values obtained from the ratio of integrated flux in the F560W and F1130W JWST filters band. Right panel: Difference in spectral index between a boosted (toward the observer) and un-boosted emitting region, as a function of the comoving magnetic field for two different values of break Lorentz Factor, and for two values of the bulk velocity of the boosted region, at the effective wavelength of $8.3\mu{\rm m}$.}
\end{figure*}

\subsection{Properties of the Wisps} \label{wisps}

Highly structured continuum-emission features concentrated
in the central regions of the Crab Nebula have long been
understood to be associated with synchrotron radiation.
Optical observations show that these features -- ring-like ``wisps''
along with other more compact ``knot'' structures -- show dynamical behavior
on timescales of weeks to months \citep{scargle69}, and simultaneous
observations with HST and \textit{Chandra} reveal corresponding structures
and variations -- both in position and brightness -- in both the
optical and X-ray bands \citep{hester96,weisskopf00}, as well as in the radio \citep{bietenholz01}. These structures are understood as synchrotron
radiation from the electron-positron wind as it enters the nebula
from the termination shock, where the momentum flux of the wind is
balanced by the pressure within the nebula. The emission is highly
polarized, and the magnetic field orientation follows the structure
of the wisps. Theoretical models \citep[e.g.,][]{delzanna06} explain these wisps and knots as Doppler-boosted regions where the plasma bulk velocity points toward the observer. Wisps are typically located at intermediate latitudes, close to the equator, while the knot traces a polar outflow.

Fig.~\ref{wispfigure} reveals these complex structures in the inner nebula as
observed with JWST. The upper left panel is from the NIRCam observation
using the F480M filter and provides the highest-resolution image in the figure.
Emission is seen from the pulsar itself along with a knot very close
to the pulsar, labeled in the top-right panel of Fig.~\ref{wispfigure}. 
The leading edges of the wisps in the northwest are Doppler-brightened
due to the high-velocity equatorial flow and the geometry of the system,
with the pulsar spin axis tilted into the plane of the sky by $\sim$ 27$^{\circ}$ along a projected orientation roughly 126$^{\circ}$ degrees north of
east \citep{ng04}. 

The top right and bottom left panels in Fig.~\ref{wispfigure} show MIRI observations of the central region obtained on the same day, taken through the F560W and F1130W filters, $\sim$4 months after the NIRCam image.  While the resolution of these images is lower, changes in the wisp structure are evident over the ~4-month period between these
observations and the NIRCam observation. This is consistent with
observations at other wavelengths that show wisp motions corresponding
to velocities of $\sim 0.8-0.9 c$ \citep{Schweizer_Bucciantini+13a}, typical values downstream of a shock in a magnetized wind with the ratio of magnetic to kinetic energy of $\sigma \simeq 1 $. 

The bottom right panel is a spectral index map produced from the F560W
and F1130W maps (see \S\ref{method}). 
The uncertainties on the relative spectral index values between different positions in the PWN are dominated by variations in the residual background emission and are significantly lower than those on the absolute values. The relative 1-$\sigma$ uncertainty on the spectral index in individual pixels ranges from 0.0048 in the bright torus region to 0.024 in the outer PWN. This allows us to search for small variations in the spectral index across the observed structures in the torus region. While the more diffuse regions
of the nebula in Fig.~\ref{wispfigure} have indices of $\sim 0.35$, there are significant
variations for multiple structures. The region corresponding to the
bright NW wisp shows a considerably flatter spectrum, with $\alpha
= 0.27$, while that for the knot region is steeper, with $\alpha =
0.42$.

\begin{figure*}
\center
\includegraphics[height=1.75in]{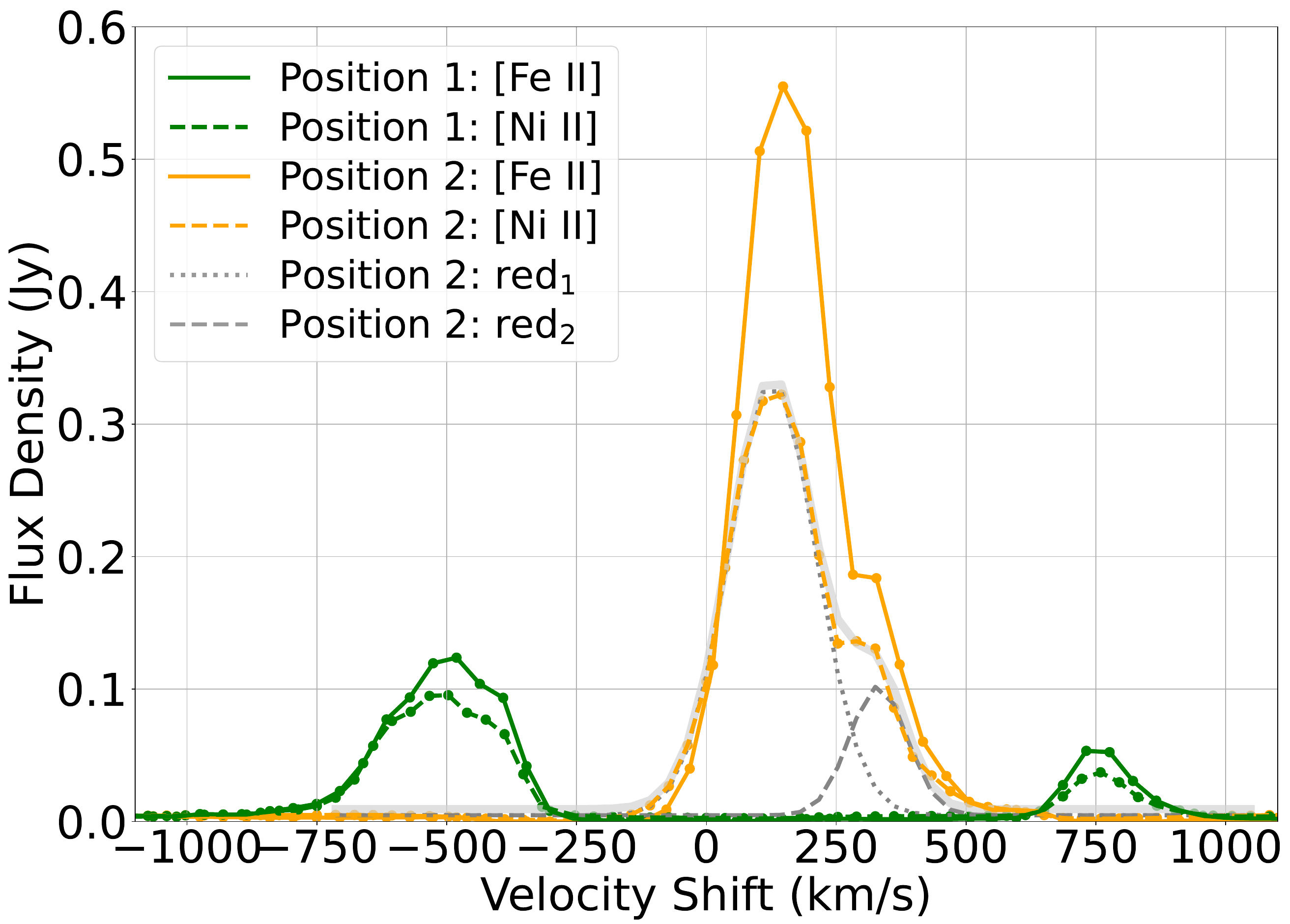}
\includegraphics[height=1.75in]{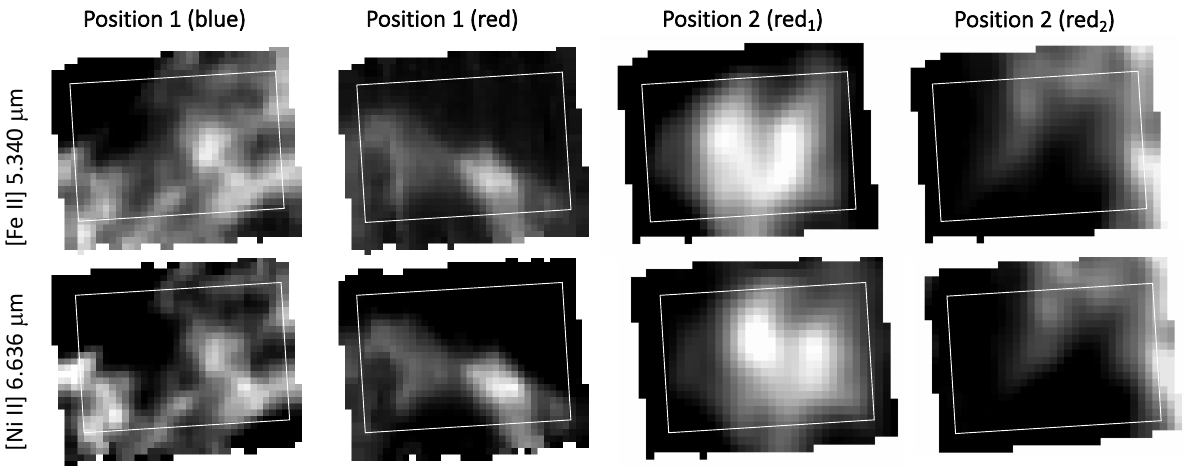}
\caption{\label{mrslineprofiles} Left: The plot shows the [\ion{Fe}{2}]~5.340~\micron\ (solid lines) and [\ion{Ni}{2}]~6.636~\micron\ (dashed lines) line profiles as a function of the velocity shift with respect to the rest wavelength. The line profiles from Position 1 and Position 2 are shown in the green and orange colors, respectively. The solid gray curve shows the best-fit two-component Gaussian model to the [\ion{Ni}{2}]~6.636~\micron\ at Position 2, with the dotted and dashed gray curves showing the individual model components. The locations of the MRS pointings are shown in Fig.~\ref{coverage}. Right: This figure shows line emission maps for the [\ion{Fe}{2}]~5.340~\micron\ and [\ion{Ni}{2}]~6.636~\micron\ lines extracted from the MIRI MRS data. The maps are produced by integrating the MRS cubes over the wavelength range that spans the detected line. While Position 2 shows two distinct red-shifted components (see left panel), Position 1 contains distinct blue and red-shifted components with different spatial distributions. The white rectangles represent the spectral extraction regions used in the analysis.}
\end{figure*}

\begin{deluxetable*}{cccccccccc}
  \tablecaption{Iron and Nickel Emission Lines Used in the Analysis \label{nife_table} }
\tablehead{& & \multicolumn{2}{c}{Position 1 (blue)} & \multicolumn{2}{c}{Position 1 (red)} & \multicolumn{2}{c}{Position 2 (red$_1$)} & \multicolumn{2}{c}{Position 2 (red$_2$)} }
\startdata
Line ($\mu$m) &  $C_{ex}$  & Flux & $n_iV$ & Flux & $n_iV$ & Flux  & $n_iV$  & Flux & $n_iV$ \\  
 &&&&&&&&&\cr
[Fe II] 5.3403& 5.03e-09& 61.3 $\pm$ 1.5& 4.48e48& 14.6 $\pm$ 1.1 & 1.07e48& 33 $\pm$ 4 & 2.41e48& 214 $\pm$ 5 & 1.56e49\cr
[Fe II] 5.6739& 2.19e-11& \nodata & \nodata & \nodata & \nodata & 0.056 $\pm$ 0.027 & 9.98e47& 0.48 $\pm$ 0.03 & 8.46e48\cr
[Fe II] 6.7213& 4.49e-10& 3.76 $\pm$ 0.14 & 3.87e48& 0.96 $\pm$ 0.10 & 9.88e47 & 1.16 $\pm$ 0.25 & 1.19e48& 14.2 $\pm$ 0.2 & 1.47e49\cr
[Fe II] 17.936& 7.86e-09& 30.4 $\pm$ 1.2 & 4.77e48& 4.9 $\pm$ 1.0 & 7.69e47& 25 $\pm$ 4 & 3.92e4 8& 64 $\pm$ 4 & 1.00e49\cr
[Fe II] 24.519& 1.91e-09& 7.2 $\pm$ 0.5 & 6.36e48& 0.66 $\pm$ 0.13 & 5.83e47 & 6.0 $\pm$ 0.9 & 5.30e48 & 16.2 $\pm$ 0.6 & 1.43e49\cr
[Fe II] 25.988& 3.95e-08 & \nodata & \nodata & \nodata & \nodata & 119 $\pm$ 11 & 5.38e48 & 127 $\pm$ 7 & 5.75e48\cr
[Fe III] 22.9258& 4.98e-08& 25.3 $\pm$ 0.7 & 1.06e48& 5.9 $\pm$ 0.5  & 2.47e47& 8.5 $\pm$ 1.0 & 3.55e47& 28.3 $\pm$ 1.1 & 1.18e48\cr
[Ni II] 6.636& 2.42e-08& 39.4 $\pm$ 0.9 & 7.43e47& 7.3 $\pm$ 0.6 & 1.38e47& 20.4 $\pm$ 1.3 & 3.85e47& 104 $\pm$ 2 & 1.97e48\cr
[Ni III] 11.002& 5.87e-09& 1.32 $\pm$ 0.04 & 1.70e47& 0.26 $\pm$ 0.03 & 3.40e46& 0.54 $\pm$ 0.07 & 6.96e46& 1.21 $\pm$ 0.08 & 1.56e47\cr
&&&&&&&&&
\enddata
\tablecomments{Fluxes at earth are in units of $10^{-15}\:erg\:cm^{-2}\:s^{-1}$ and are converted to $n_iV$ by Flux$\times 2.1728e63/h\nu/C_{ex}/n_e$ to derive the total number of ions in the field of view.}
\end{deluxetable*}

\begin{deluxetable*}{cccccccccccccc}
  \tablecaption{Derived Nickel-to-Iron Ratios \label{nife_table2} }
  \tablehead{ & & \multicolumn{12}{c}{Derived Ratios} \\
  & & \multicolumn{3}{c}{Position 1 (blue)} & \multicolumn{3}{c}{Position 1 (red)} & \multicolumn{3}{c}{Position 2 (red$_1$)} & \multicolumn{3}{c}{Position 2 (red$_2$)}} 
  \startdata
Line Ratio &  $C_{ion. bal.}$  & Ion  & Abund. & $\times$Solar & Ion & Abund. & $\times$Solar & Ion  & Abund. & $\times$solar & Ion & Abund. & $\times$Solar \cr \hline
 &&&&&&&&&\cr
[Ni II]/[Fe II] & 1.61& 0.152& 0.245 & 4.6 & 0.162 & 0.261 & 4.9 &  0.147&0.237&4.5 &0.172& 0.277 & 5.2\cr
[Ni III]/[Fe III] &1.13 & 0.160& 0.18 & 3.4 &  0.138& 0.156& 2.9 &  0.196& 0.221& 4.2 & 0.132& 0.149 & 2.8\cr
				  \enddata
\tablecomments{The first column for each filament position gives the ion ratio, the second the element mass abundance ratio that is corrected for the ionization balance, and the third column gives the Ni/Fe mass abundance ratio relative to the solar ratio. For reference, the solar value is Ni/Fe = 0.053.}
\end{deluxetable*}

The cooling timescale for electrons emitting in the MIRI band ($\lambda
\sim 5 \mu{\rm m}$) is $t_{syn} \approx 4.1 B_{100}^{3/2}$~kyr. Thus, even if we assume a magnetic field of $B\sim 400~\mu$G, as expected downstream of the termination shock in a highly magnetized ($\sigma \simeq 1$) wind, the cooling timescale is much too long (given the typical timescales of the wisps) to
explain the spectral index changes indicated in Fig.~\ref{wispfigure}. 
However, the broadband
spectrum of the nebula indicates a break in the IR that is presumably
associated with a break in the electron injection spectrum. The frequency
of the observed synchrotron emission is
\begin{equation}
\nu_{obs} = D_{boost}\nu_{syn} \approx 1.4 D\gamma_e^2 B_{100} {\rm\ Hz} 
\end{equation}
where $\nu_{syn}$ is the synchrotron frequency in the rest frame,
$\gamma_e$ is the Lorentz factor of the electrons, $B_{100}$ is
the magnetic field strength in units of $100~\mu{\rm G}$, and 
$D_{boost} = [\Gamma(1-\beta \cos\theta)]^{-1}$ is the Doppler boost factor associated with the pairs flow
in the nebula ($\theta$  being the angle w.r.t. the line of sight). The observed frequency thus probes different parts
of the electron spectrum depending upon the factor $D_{boost} B_{100}$.
For example, for a uniform field strength, regions with a high Doppler factor
probe lower energy electrons than regions with unboosted emission. 

The flat spectrum of the NW wisp, which is clearly Doppler-brightened,
presumably probes lower electron energies than for adjacent regions,
thus sampling the electron spectrum below the break, while regions
with steeper spectra probe electrons where the spectrum is steepening. The measured spectral indices of these features may be the first direct evidence that the spectral curvature in the injection spectrum of emitting pairs is tied to the acceleration process at the termination shock.

Global spectral energy distribution modeling of the Crab Nebula \citep{Bucciantini_Arons+11a} indicates that emission comes from an
underlying electron spectrum with a broken power law index distribution. The radio emission arises from particles with  $N(\gamma_e) \propto \gamma_e^{-1.5}$,  while optical emission implies
$N(\gamma_e) \propto \gamma_e^{-2.35}$ (with $\gamma_e$ being the Lorentz factor). Fig.~\ref{index_change_plot} shows the spectral index across the 5.6--11.3~\micron\ wavelength range that is expected from synchrotron emission produced by a broken-power-law particle distribution. The values of the spectral index $\alpha$ were derived using the analytical results of \citet{gleeson74}. The orange points represent the index value calculated from the intensities in the F560W to F1130W  filters, while the solid line represents the index values at an effective wavelength $\lambda_{\rm eff} = 8.3 \mu{\rm m}$, where $\gamma_{br}$ is the break Lorentz factor in the electron injection spectrum.

Assuming that break energy $m_e c^2 \gamma_{br}$ is given by the ratio of the pulsar voltage and the pair multiplicity $\kappa$, then $\gamma_{br} \simeq 7.5 \times 10^{10} / \kappa$. Reasonable values of $\kappa \simeq 5\times 10^5$,  $\gamma_{br} = 1.5 \times 10^5$ and $B \simeq 200~\mu$G, and observed wisp
motions of $\sim 0.9 c$ \citep{Schweizer_Bucciantini+13a}, yield $\Delta \alpha = 0.17$ between
boosted and unboosted regions, fully consistent with the variations
seen in Fig.~\ref{wispfigure}.

Emission from the knot region is much steeper than for the NW wisp but is
also Doppler-brightened. While the degeneracy between the effects of the
Doppler factor, the magnetic field strength, and $\gamma_{br}$ makes it impossible to say
with certainty, this is suggestive either of a lower magnetic field from the 
emission region of the knot and the large wisp to the southeast of the pulsar. This is consistent with pictures of the
knot emission arising from high-latitude regions of the termination
shock surface where the field is expected to be lower \citep{delzanna06}.  Another possibility is that different acceleration mechanisms are at play that produce different values of $\gamma_{br}$. For example, in the low latitude regions where the wisps are thought to arise, the pulsar wind is expected to be striped and reconnection at the shock to play a major role in the acceleration. On the other hand, in the polar region associated with the knot, the pulsar wind should be un-striped and reconnection much less efficient \citep{lyubarsky03,Sironi_2011}.

\begin{figure*}
\center
\includegraphics[width=0.35\textwidth]{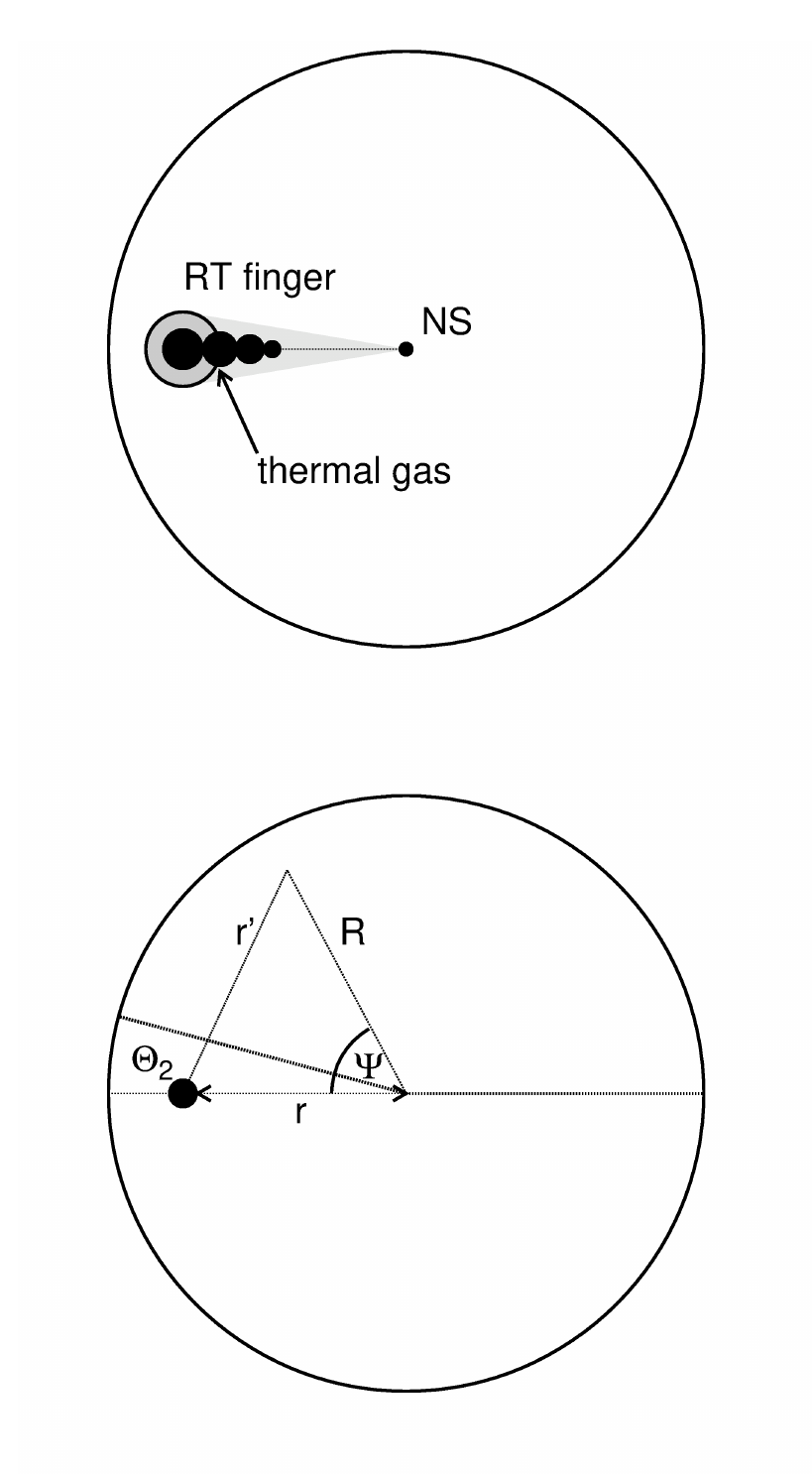}\includegraphics[width=0.55\textwidth]{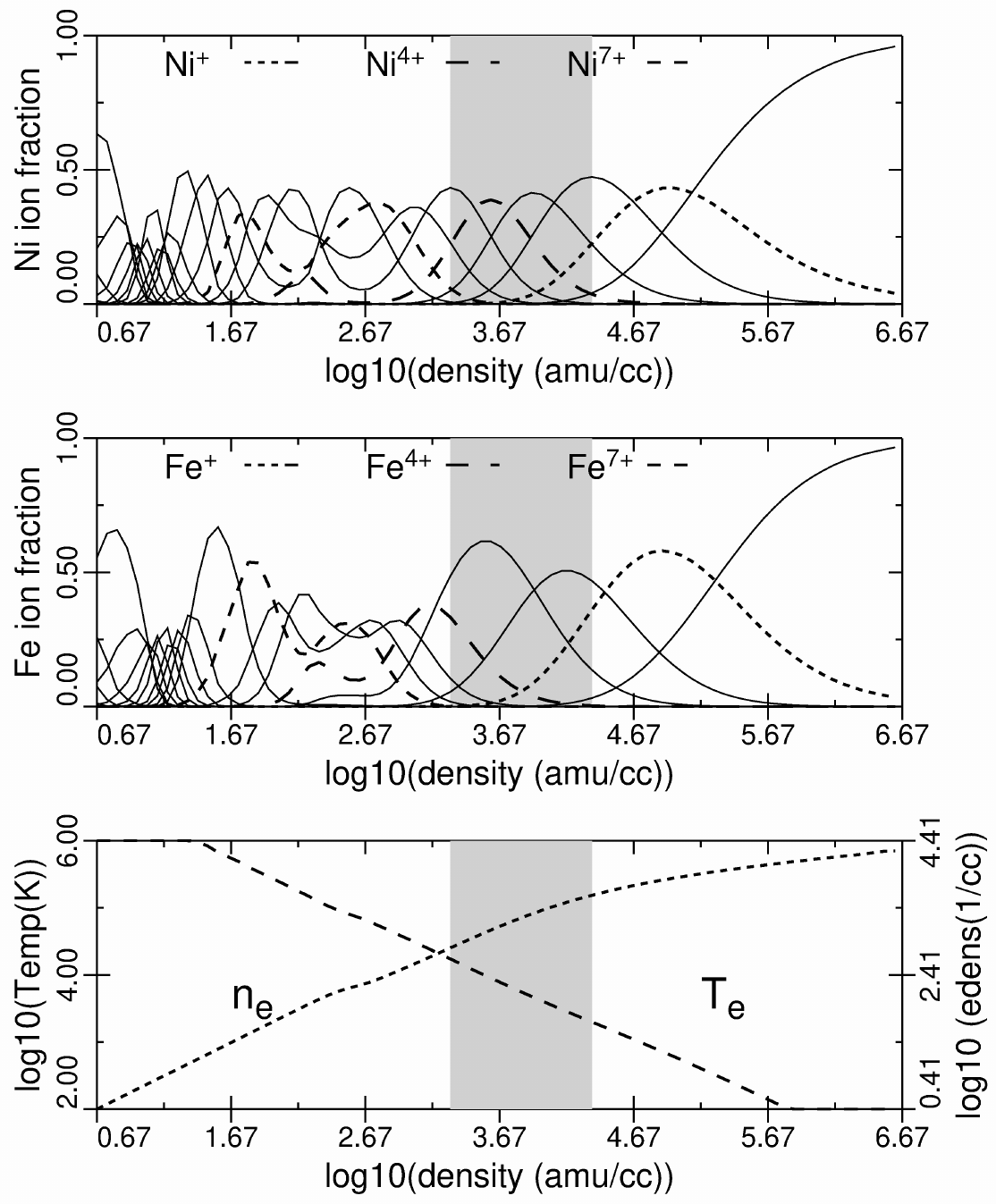}
\caption{\label{Ibalance} Left: Schematic diagrams (not to scale) of assumed PWN geometry. The thermal gas in a Rayleigh-Taylor finger resides in a cone cut out of the otherwise assumed spherical pulsar wind nebula. Synchrotron emission from the nebula irradiates thermal gas in the cone. The lower panel shows the geometry for the calculation. Right: Ionization balance for Ni (top), Fe (middle), and variation of electron density and temperature (bottom) used in the Ni/Fe abundance ratio determination.}
\end{figure*}

\begin{figure*}
\center
\includegraphics[width=0.48\textwidth]{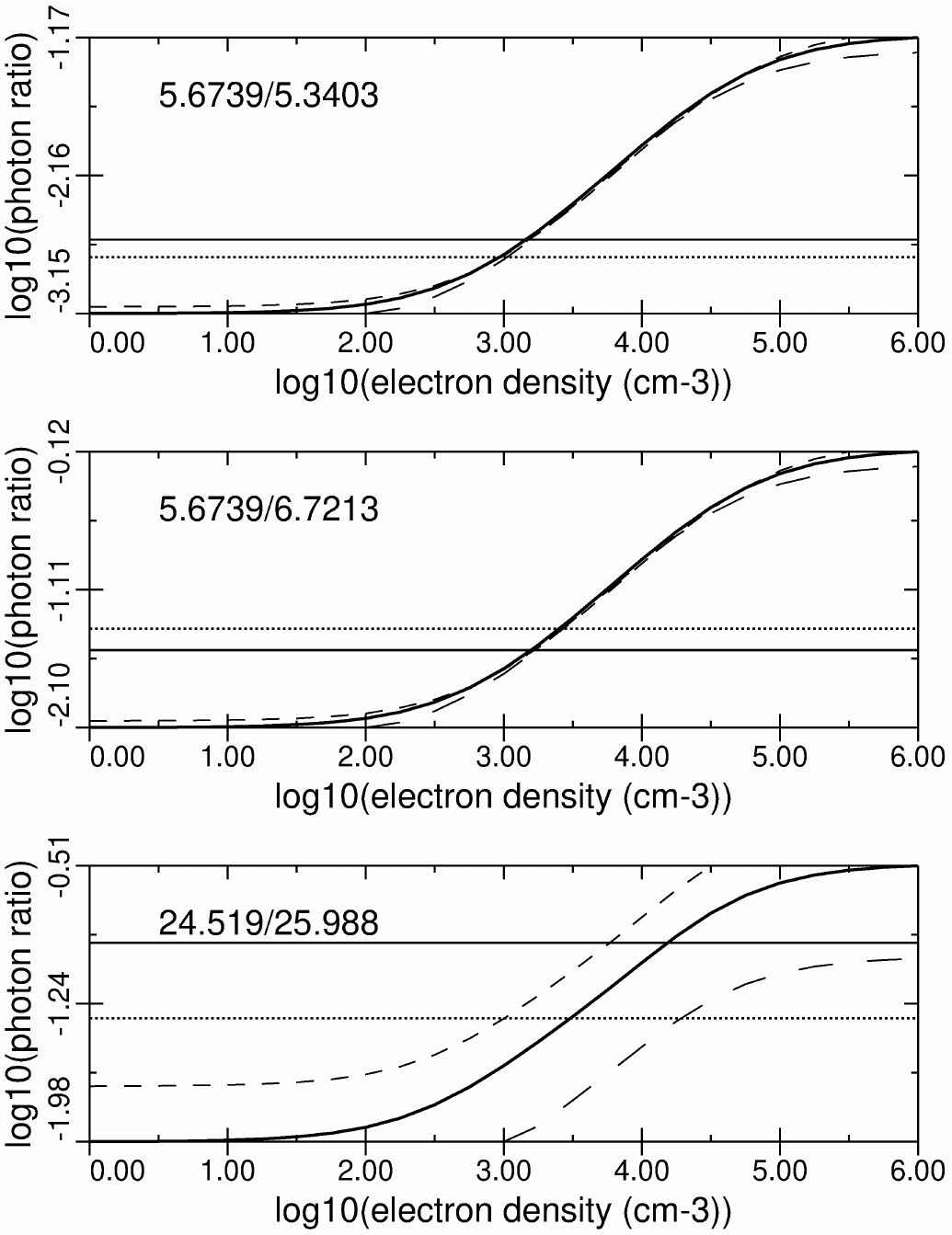}
\includegraphics[width=0.48\textwidth]{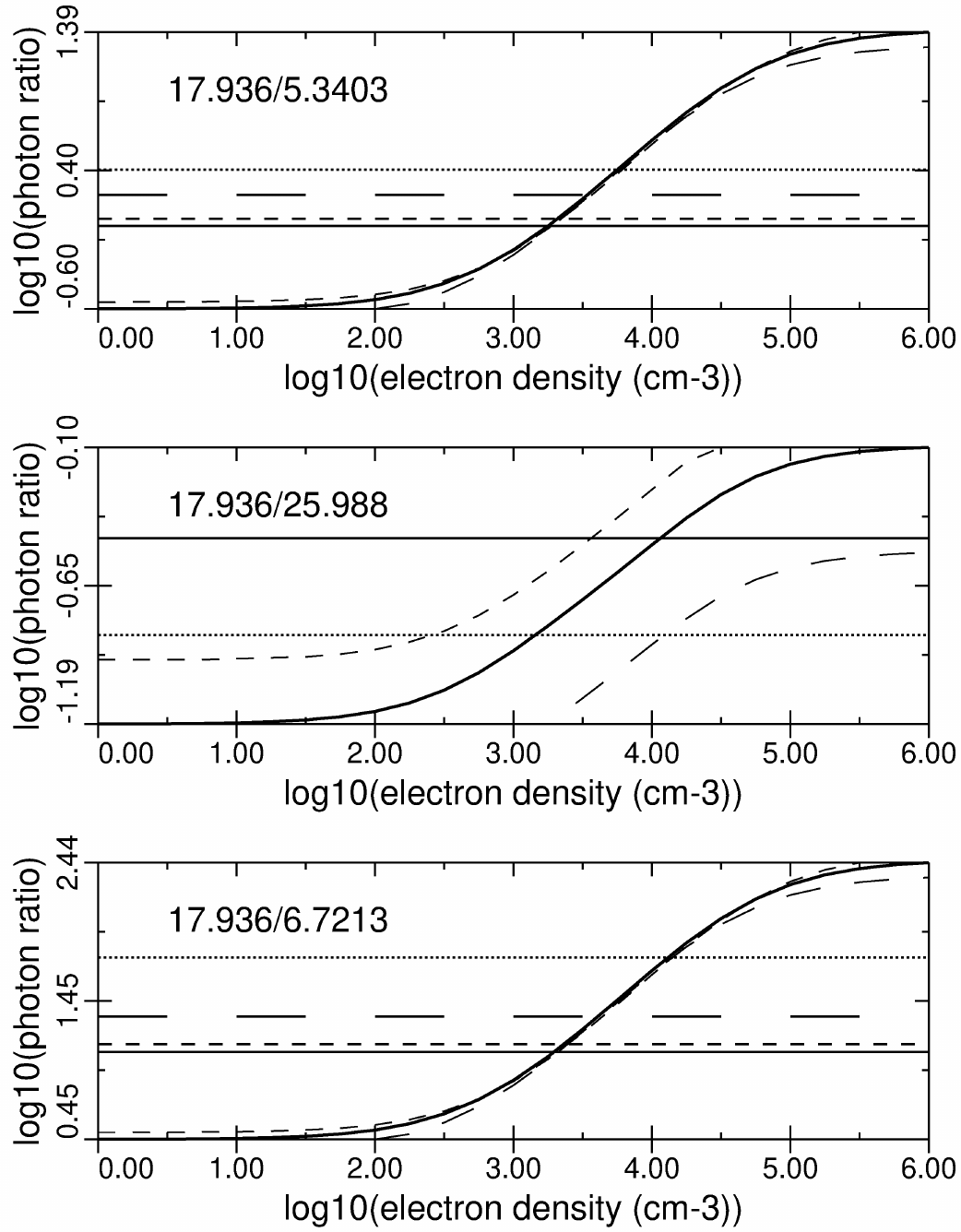}
\caption{\label{densities} Density diagnostic line ratios in \ion{[Fe}{2}] at 
temperature $T=2400$ K (solid), $T=1350$ K (long dash) and $T=4270$ K (short dash). Long and short dashed horizontal lines come from the red and blue velocity components of filament 1, and the dotted and solid horizontal lines come from the red1 and red2 velocity components of filament 2, respectively (see Table~\ref{nife_table}). All ratios indicate electron density in the range $\simeq 1000 - 3000$ cm$^{-3}$ for $T=2400$ K.}
\end{figure*}

\section{Nickel-to-Iron Abundance Ratios} \label{ratios}

The existing estimated Ni/Fe ratios in select regions of the Crab Nebula range from $\sim$ 6 times solar using near-IR lines to 75 times solar using optical lines \citep{henry84, macalpine89, hudgins90}. Reconciling these early near-IR and optical estimates required either higher gas densities than have been generally accepted, or the assumption that the atomic data are inaccurate or incomplete \citep{hudgins90}.
The inconsistencies may also in part be due to the complex ionization structure across the ejecta filaments \citep[e.g.][]{sankrit98} that could not be properly accounted for in lower spatial resolution studies. The spectral and spatial resolution and the sensitivity of the JWST MIRI MRS allow us to measure these abundance ratios much more accurately. As will be seen below, the JWST data not only resolves the gas distribution that we are sampling but also provides many more Fe and Ni lines that make it possible to better constrain the densities and temperature of the emitting gas.

\subsection{Spectral Line Measurements}

The MIRI MRS spectra seen in Fig.~\ref{spectrum} exhibit a wealth of spectral lines arising from the filaments. In this initial analysis, we restrict ourselves to a determination of the Ni/Fe abundance ratios, deferring more complete spectral line identification and analysis to later work.  As explained in \S\ref{mrs}, the MRS spectra in all four channels were extracted from a fixed aperture size shown in Figure~\ref{mrs_apertures}. The left panel of Fig.~\ref{mrslineprofiles} shows the line flux profiles as a function of the velocity shift for the bright [\ion{Fe}{2}] 5.340~\micron\ and [\ion{Ni}{2}] 6.636~\micron\ lines. It can be seen from the plot that the line profiles of the Fe and Ni lines are the same, confirming that these lines arise from the same velocity components. 

At Position 1, two broad velocity components are evident and are well fitted by individual Gaussian components, one blue-shifted by an average velocity of $\rm \sim500\:km\:s^{-1}$ and the other red-shifted to $\rm \sim750\:km\:s^{-1}$. The corresponding FWHM values of these two components are $\rm \sim260\:km\:s^{-1}$ and $\rm \sim140\:km\:s^{-1}$, respectively. At Position 2, almost all emission is red-shifted, spanning a velocity range from 0--500~$\rm km\:s^{-1}$. The line profiles at Position 2 are well-fitted by two broad, blended Gaussian components with centroid (FWHM) values of approximately $\rm 130\:km\:s^{-1}$ ($\rm 200\:km\:s^{-1}$) and $\rm 330\:km\:s^{-1}$ ($\rm 130\:km\:s^{-1}$). The spatial distributions for each of the two velocity components at both positions are shown in the right panel of Fig.~\ref{mrslineprofiles}, with the extraction aperture shown as the white rectangle. It can be seen that the Fe and Ni lines have practically identical spatial distributions in addition to the same velocity profiles, confirming that both lines arise from the same location. The other Fe and Ni lines exhibit the same properties and their emission was fitted in the same way. The list of lines used in our photoionization models and their fitted flux values are listed in Table~\ref{nife_table}.

\subsection{Photoionization Model Results} \label{photoresults}

We model the ionization balance following in part the approach in \citet{laming20}. We calculate the photoionization-recombination equilibrium for thermal gas residing in a Rayleigh-Taylor finger, constrained to have the same pressure as the surrounding PWN, taken to be $1.9\times 10^{-9}$ dyne cm$^{-2}$\citep{porth14b,fraschetti17}, which includes the contribution from electrons and positrons and the assumed magnetic field of 140 $\mu$G.

The schematic geometry is shown in Fig.~\ref{Ibalance}. The PWN is taken to be a sphere of radius 1 pc, with a cone cut out of it where the thermal Rayleigh-Taylor finger gas resides. Referencing the lower left panel, we calculate the intensity of photoionizing radiation, $J$, in the Rayleigh-Taylor finger as 
\begin{equation}
    J=\frac{\epsilon}{4\pi}\int _{R_{min}}^{R_{max}}\int _{cos\theta _1}^{\cos\theta _2}\frac{2\pi d\left(\cos\psi\right) R^2dR}{r^2+R^2-2rR\cos\psi},
\end{equation}
where $r^{\prime 2}=r^2+R^2-2rR\cos\psi $, and in our case $\cos\theta _1 = -1$ and $\cos\theta _2$ is given by the cone opening angle. The synchrotron emission per unit volume is $\epsilon$, and the PWN inner and outer radii are $R_{min}$ and $R_{max}$ respectively. Equation 1 becomes
\begin{equation}
    J=\frac{\epsilon}{4r}\int _{R_{min}}^{R_{max}}\ln\left(\frac{r^2+R^2-2Rr\cos\theta _1}{r^2+R^2-2Rr\cos\theta _2}\right)RdR.
\end{equation}
We put $x=R-r\cos\theta$ so that 
\begin{eqnarray}
    & & \int R\ln\left(r^2+R^2-2Rr\cos\theta\right)dR=\cr
    & & \int\left(x+r\cos\theta\right)\ln\left(x^2+r^2\sin ^2\theta\right)dx
\end{eqnarray}
and with the help of \citet{gradshteyn65} equations 2.733.1 and 2.733.2 we arrive at 
\begin{eqnarray}
    J&=&\frac{\epsilon}{4r}\{f\left(R_{max},\theta _1\right)- f\left(R_{max},\theta _2\right)-f\left(R_{min},\theta_1\right)\cr &+&f\left(R_{min},\theta _2\right)\},
\end{eqnarray}
where 
\begin{eqnarray}
    f\left(R,\theta\right)&=&\frac{1}{2}\left(R^2-r^2\cos 2\theta\right)\ln\left(R^2+r^2-2rR\cos\theta\right)\cr &-&\frac{1}{2}\left(R+r\cos\theta\right)^2 +2r^2\cos ^2\theta \cr &+&2r^2\sin\theta\cos\theta\arctan\left(\frac{R/r}{\sin\theta}-\cot\theta\right).
\end{eqnarray}
It is useful to check that when $\theta _1\rightarrow 180^{\circ}$ and $\theta _2\rightarrow 0^{\circ}$ equations A4 and A5 of \citet{laming20} are recovered.

The PWN luminosity of $1.8\times 10^{37}$ erg s$^{-1}$ in the energy range 300 eV - 10 keV \citep{mori04,fraschetti17} and radius of 1 pc (assumed $>> R_{min}$) give $\epsilon = 1.46\times 10^{-19}$ erg cm$^{-3}$s$^{-1}$. We calculate photoionization rates by using equations 4 and 5 and integrating over a power-law spectrum with a photon index $\Gamma = 2.1$ \citep{mori04}. Collisional atomic rates are taken from \cite{mazzotta98}, with updates to dielectronic recombination as given in \citet{laming20}. Photoionization cross sections are taken from the tabulation of \citet{verner96}, with rates calculated by integrating over the PWN spectrum. Using an abundance set for model V or VI in \citet{owen15}, we iterate the calculation until the input electron density matches that at output. The sample ionization balance for Ni and Fe is given in Fig.~\ref{Ibalance}, and reflects a plasma at radius 0.9 pc in the PWN, in a cone where $\cos\theta = 0.99$, i.e. a filament radius of order 1\% of the PWN radius, matching the geometry modeled in  \citet{sankrit98}.

In Fig.~\ref{Ibalance}, the plasma particle density and temperature are coupled so that the gas pressure equals that of the PWN, given above as $1.9\times 10^{-9}$ dyne cm$^{-2}$. At high densities and low temperatures on the right-hand side of the right-hand panel of Fig.~\ref{Ibalance}, neutral atoms are found, with the equilibrium charge state increasing as the density decreases and temperature increases. The gray-shaded region shows the approximate parameter space from which the emission we analyze arises. The high-density/low-temperature side is dictated by \ion{[Fe}{2}] and \ion{[Ni}{2}], to be discussed in more detail below, and the low-density/high-temperature side comes from the densities and temperatures associated with \ion{[O}{3}] as observed by \citet{fesen82}.

These limits are corroborated by an analysis of electron density diagnostic line ratios in \ion{[Fe}{2}]. We calculate these using energy levels and A-coefficients from \citet{deb11} and excitation rates from \citet{bautista15}, including the lowest lying 52 levels in our model. Fig.~\ref{densities} gives six line intensity ratios from \ion{[Fe}{2}] lines that are sensitive to the electron density, plotted for electron temperatures of 1350 K (long dash) 2400 K (thick solid), and 4270 K (short dash), together with the observed ratios from Positions 1 and 2, using the line fluxes in Table~\ref{nife_table}. At 1350 K and 4270 K, the 24.519/25.988 and 17.936/25.988 ratios agree markedly less well with the other four, moving to higher/lower density as the temperature is decreased/increased. We use a temperature of 2400 K and density 3162 cm$^{-3}$ to evaluate the \ion{[Fe}{2}] and \ion{[Ni}{2}] line intensities in Table~\ref{nife_table} to give the abundance analysis in Table~\ref{nife_table2}. For \ion{[Fe}{3}] and \ion{[Ni}{3}] we take the same temperature and density, which is where these ions are maximized in Fig.~\ref{Ibalance} (right panel). 

In addition to the \ion{[Fe}{2}] atomic data outlined above, for \ion{[Fe}{3}] we use \citet{deb09} and \citet{zhang95}, for \ion{[Ni}{2}] we use \citet{cassidy10} and \citet{cassidy16}, while \ion{[Ni}{3}] is taken from \citet{bautista01}. Abundance ratios individually for Ni$^+$/Fe$^+$ and
Ni$^{++}$/Fe$^{++}$ are given in Table~\ref{nife_table2}, with each case corrected by the ionization balance to give a total element abundance ratio Ni/Fe.
We find Ni/Fe number abundance ratio in the range of 0.156 to 0.277, corresponding to 2.8 -- 5.2 times the solar ratio \citep[now taken to be 0.053;][]{scott15}, with the \ion{[Ni}{2}]/\ion{[Fe}{2}] generally registering higher Ni/Fe. The overlap between the density/temperature ranges where \ion{[Ni}{2}] and \ion{[Fe}{2}] lines are emitted is better than it is for \ion{[Ni}{3}] and \ion{[Fe}{3}]. \ion{[Ni}{3}] comes from slightly lower temperature/higher density than does \ion{[Fe}{3}], and so a larger volume of plasma at the higher temperature would bias the Ni/Fe ratio downwards. Nevertheless, even the higher-end JWST-derived Ni/Fe abundance ratio of $\sim$~5 times solar is significantly lower than the previously reported ratios based on optical data. 

We compare our results with those from optical data of \citet{macalpine89}, who originally found the Ni/Fe abundance ratio to be of order 50 times the solar ratio. We have reprocessed their reported \ion{[Ni}{2}] 0.7378/\ion{[Fe}{2}] 0.8617 flux ratios with our updated atomic physics models, and find Ni/Fe = 0.63, 0.50, 0.57 for their positions a, c, and d.
These amount to 9.4 - 11.9 times the solar ratio, the main difference being more modern and more complete atomic physics, and our accounting for the ionization balance. \citet{hudgins90} measure the \ion{[Ni}{2}] 1.192/\ion {[Fe}{2}] 1.257 intensity ratio to infer the Ni/Fe abundance. Assuming an electron density $< 10^3$ cm$^{-3}$, they give Ni/Fe about 0.3, or around 6 times the solar ratio. However, the \ion{[Fe}{2}] density diagnostic line ratio presented in their paper, the 1.294/1.257 ratio, indicates a density of $5\times 10^3$ cm$^{-3}$. Re-evaluating this with the newer \ion{[Fe}{2}] 
model revises the density upwards to around $1\times 10^4$ cm$^{-3}$. At these densities, the inferred Ni/Fe abundance ratio becomes 30-60 times solar, because the \ion{[Ni}{2}] 1.192 line is density sensitive. \citet{hudgins90} cast doubt on the reliability of this density diagnostic, and we would expect this higher excitation line to be formed at a slightly higher temperature and lower density than the mid-IR lines. 

To have \ion{[Fe}{2}] appearing at a density of $10^3$~cm$^{-3}$, the ambient PWN pressure and/or the flux of ionizing radiation have to be decreased. The first does not seem plausible given the position within the nebula from where the \citet{hudgins90} spectrum was taken, but a reduction in ionizing flux could come about due to local extinction due to dust in the filaments. We find that a reduction by a factor of 0.1 moves the densities at which \ion{[Fe}{2}] and \ion{[Ni}{2}] form to lower values by about a factor of 0.5, so that the mid-IR lines observed by JWST are predicted to form at about 1500 cm$^{-3}$, and the near-IR lines observed by \citet{hudgins90} at closer to 1000 cm$^{-3}$. At these lower densities, the JWST mid-IR lines give Ni-to-Fe abundance ratios in the range 2.3--4.6 times the solar ratio (but with the higher end derived from [\ion{Ni}{2}]/[\ion{Fe}{2}] being favored), the optical lines of
\citet{macalpine89} give 4.6--5.8, and the near-IR analysis of \citet{hudgins90} gives 8.2 times the solar ratio, bringing all of the estimates to a roughly consistent range. This flux reduction corresponds to 2.5 magnitudes in the UV, which is consistent with V-band extinctions in the filaments reported by \citet{grenman17} and \citet{delooze19}. Uncertainties on such abundance values have both observational and theoretical sources and are difficult to evaluate. We calculate the standard deviation of the number of  Fe$^+$ ions in the field of view of each spectrum, taken from the 4-6 lines of \ion{[Fe}{2}] observed, to find uncertainties of order 20 - 50\% for our JWST analysis. Similar uncertainties for the NIR and optical results are plausible.

We note here that if the Ni/Fe ratios vary spatially across the Crab's filaments, it is possible that the value we measure here may not be representative of the ejecta as a whole. However, based on the currently available data, there is no evidence that the Ni/Fe abundance ratios vary by more than $\sim$~50\% across the Crab's filaments. While the JWST data probe only two select filament positions, we were able to isolate four distinct ejecta velocity ranges that all have roughly the same Ni/Fe abundance ratios. The other locations across the filaments that were observed in the near-IR and optical wavelengths have Ni/Fe values that fall within a similar range (see above). The work of \citet{sibley16} did indicate that the fractional Ni abundances relative to solar do vary across the filaments (see their Figure~3), but it does not necessarily follow that the Ni/Fe ratios vary substantially. Additional spectroscopic observations covering a range of filament positions will be required to confirm a lack of significant variation in Ni/Fe across the filaments.

In Sect.~\ref{NiFerat}, we discuss the implications of the Ni/Fe ratio derived above.

\section{The Nature of the Supernova Explosion} \label{explosion}

The Crab Nebula and SN~1054 have long been known to be anomalous compared to what are usually considered ``normal" CCSNe, and for decades \citep{nomoto82} the Crab has been widely considered as the best-observed candidate for an ECSN explosion.  This is based on a number of peculiar observed properties of the Crab: in addition to claims of an elevated Ni/Fe abundance ratio discussed above, these include its very low explosion kinetic energy of less than 10$^{50}$ erg, lack of any obvious $\alpha$-element enrichment, moderately low total mass in the filaments, low Fe abundance in the filaments and low late-time luminosity from historical data (both of which imply a low $^{56}$Ni yield), location of almost 200 pc out of the Galactic plane (implying an environment consistent with a relatively low mass progenitor), and other factors.  However, a variety of theoretical work has continued to explore the end fates of stars that occupy the remarkable transitional range of initial mass (roughly 8-11 $M_{\odot}$) between those that die as planetary nebulae with white dwarfs and those that produce higher-energy CCSNe --- ECSNe are not the only types of explosions that occupy this range.  

In the discussion below, we will refer to a few of these repeatedly, so it is convenient to define some acronyms that serve as shorthand for these explosive ends.  We quote nominal single-star effective initial mass ranges for each, although these will obviously differ slightly from one model to the next depending on adopted physics in stellar evolution and explosion models.

{\it ECSN}:  This is the electron capture SN discussed above, resulting from the collapse of a degenerate ONeMg core, with a stellar initial mass of roughly 8-9 $M_{\odot}$, and a low explosion kinetic energy of $\sim$10$^{50}$ erg.

{\it LMCCSN}:  This refers to low-mass Fe-core-collapse SNe that occur at the lowest end of the Fe core-collapse SN initial mass range immediately above ECSNe, at perhaps 9-10.5 $M_{\odot}$.  These are also expected to produce explosions with low kinetic energy that in many ways resemble expectations for ECSN explosions.

{\it SDSN}:  This stands for Si deflagration SN, a special case of LMCCSN, discussed in detail by \citet{wh15}.  These progenitors have degenerate O and Si cores and ignite O and Si burning off center in degenerate flashes as they approach Fe core collapse.  \citet{wh15} find that for initial masses of 9.8-10.3 $M_{\odot}$, the Si deflagration flashes are robust and strong enough to eject much of the H envelope shortly before the Fe core collapses. 
Because of the pre-SN mass ejection, these events may show evidence of strong CSM interaction.

{\it CCSN}: We will use this shorthand for Fe-core-collapse SNe that produce an explosion kinetic energy of $\gtrsim$~0.3~$\times$~10$^{51}$ erg \citep[e.g.][]{sukhbold16, burrows24}.  These should occur above an initial mass of roughly 10.5-11 $M_{\odot}$, although the mass transition is not sudden and may differ from one set of models to the next.

The first three above (ECSN, LMCCSN, SDSN) are all expected to have relatively low explosion kinetic energy of roughly 10$^{50}$ erg when the core collapses.
Other explosions besides SDSNe may also produce strong pre-SN mass loss in degenerate flashes, super-AGB pulses, or through wave-driven mass loss.  Overall, there is much overlap in properties between some of these different models, which may make it very difficult to tell the difference observationally between a true ECSN and the low-energy explosions from lower-mass Fe core collapse.  Below we discuss a few of these factors that are particularly relevant to the Crab Nebula, including the Ni/Fe abundance ratio, explosion asymmetry and the neutron star kick, and possible CSM interaction.

\subsection{Ni/Fe ratio} \label{NiFerat}

\begin{figure*}
\center
\includegraphics[width=0.8\textwidth]{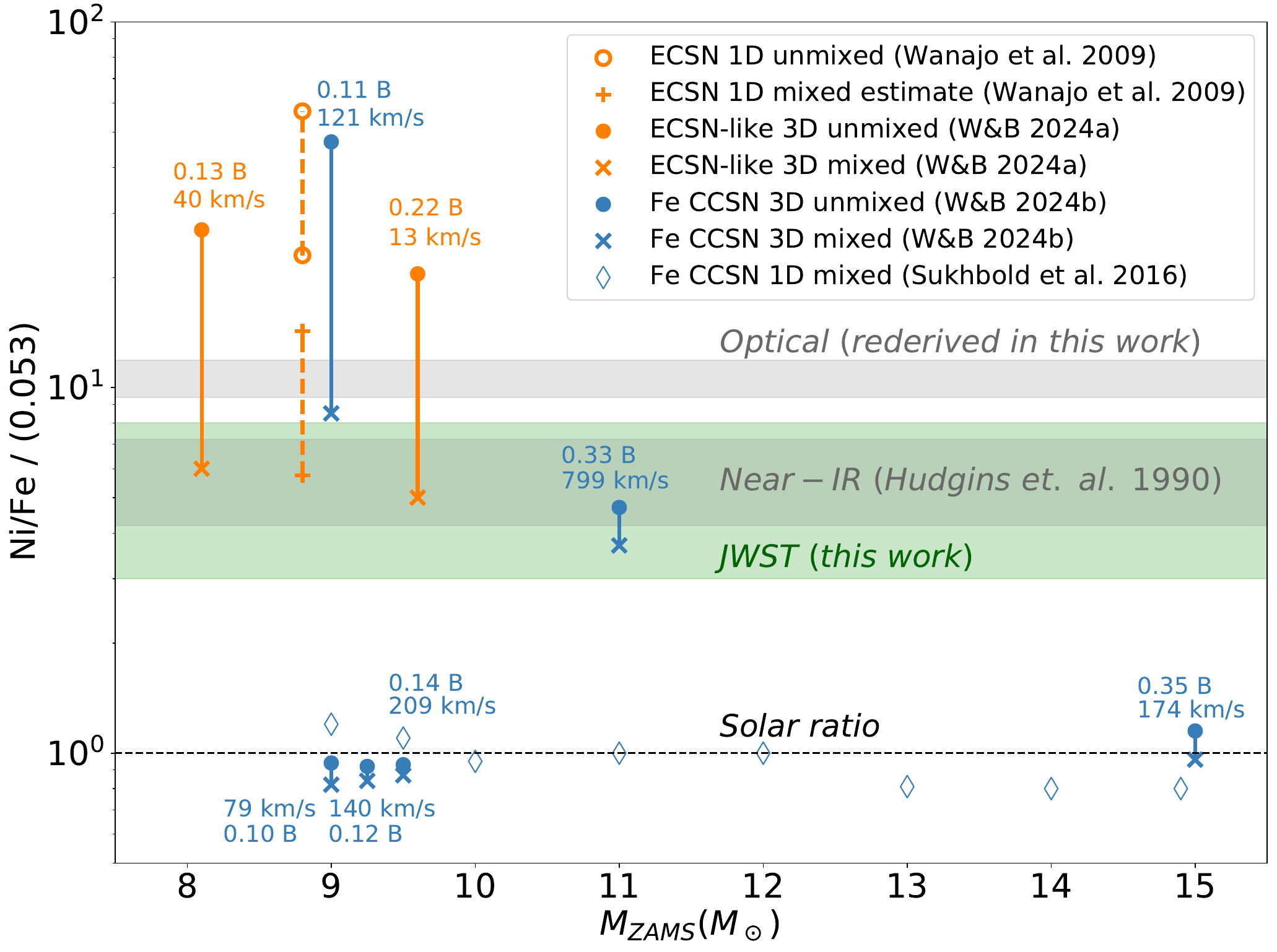}
\caption{\label{ratio_plot} The plot shows a comparison between the measured Ni/Fe ratios for the Crab Nebula and the predictions from some of the available explosion nucleosynthesis models. The green band represents the JWST-measured Ni/Fe ratios (relative to the solar ratio of 0.053 from \cite{scott15}), including the estimated uncertainty of 50\%. The near-IR measurements from \citet{hudgins90} and the optical ratios rederived in this work are shown as the gray bands. We note that the range of optical values does not include uncertainties in the measurements, but represents a range of values measured at different positions in the Crab Nebula. These optical values are likely to be lower if dust in the filaments is accounted for (see \S\ref{ratios} for details)). 
The yields of 3-D models of \citet{wang24b} and \citet{wang24} are shown as orange and blue solid vertical lines, respectively. The vertical lines reflect the variation in the expected ratio due to mixing with the stellar envelope (the dots are Ni/Fe values assuming no mixing with the envelope and the crosses are values that assume the ejecta are mixed with the stellar envelope). The explosion energy (in units of Bethes) and resulting neutron star kick velocity are also labeled for each model \citep{burrows24,wang24b} and can be compared to the $\sim$~160~$\rm km\:s^{-1}$ kick velocity of the Crab's pulsar \citep{kaplan08,hester08}. Nucleosynthetic yields from the 1-D Fe-core-collapse models of \citet{sukhbold16} are shown as blue diamonds for comparison. 
The orange dashed line connecting open orange circles represents the range of Ni/Fe ratios for the 1-D ECSN models of \citet{wanajo09} that assume no mixing with the envelope. The orange dashed line connecting the orange pluses shows how these values might shift if we use a dilution factor of $\sim$~4 (similar to what is found for the ECSN-like models). For more details, please see \S\ref{NiFerat}.}
\end{figure*}

A major observational goal of our study was to measure the Ni/Fe abundance ratios in the Crab Nebula filaments and to compare them to expectations from models of ECSNe and more typical Fe-core-collapse SNe.  Previous studies, especially those using optical [Ni~{\sc ii}] emission lines, had estimated high Ni/Fe abundance ratios in the Crab's filaments of 50--75 times the solar value \citep{hudgins90}. Theoretical predictions for nucleosynthetic yields of ECSNe have produced such high Ni/Fe abundance ratios \citep[e.g.][]{wanajo09}, seemingly favoring an ECSN model for the Crab.  

We investigated the Ni/Fe ratio in detail in Sect.~\ref{ratios}, the main results of which were that: (1) The Ni/Fe abundance ratio in the Crab filaments estimated from our {\it JWST} spectra suggests a modestly elevated Ni/Fe ratio that is 3--8 times the solar Ni/Fe ratio (assuming a 50\% uncertainty).  (2) We reanalyzed previous measurements including optical emission line strengths that led to very high estimates of the Ni/Fe ratio, and we found instead that with modern atomic data and reasonable ionization parameters, these observed optical line strengths are consistent with similarly modest Ni/Fe ratios. Using updated atomic data, the optical data give Ni/Fe ratios of 9--12, but as shown in \S\ref{ratios}, accounting for dust in the filaments would lower these values. Overall, we suggest that observations of the Crab filaments provide a consistent estimate of the Ni/Fe abundance ratio that is on the order of five times the solar value, albeit with some mild differences in this ratio from one filament to the next.

It is important to note that the expected or observed Ni/Fe ratio depends on the degree to which the products of explosive nucleosynthesis have been mixed with the ejecta and the unprocessed stellar envelope.   One must be careful when comparing predicted nucleosynthetic yields for a specific layer or a specific burning region, a mixed emitting region, or the total yields averaged over all the ejecta.   The filaments in the Crab most likely represent synthesized material that has been significantly mixed with the unprocessed outer envelope; the filaments contain several $M_{\odot}$ of gas and show hydrogen emission \citep{sankrit98}. Therefore, the Ni/Fe ratio produced by explosive nucleosynthesis in SN 1054  will be significantly diluted (and the resulting Ni/Fe ratio reduced) as it is mixed with the Ni and Fe in the unprocessed envelope that has a solar Ni/Fe ratio.  Some theoretical model predictions for the Ni/Fe ratio refer to the pure results of the nucleosynthesis, and some refer to the total yield where these are mixed with the envelope. It is very important to distinguish these two when comparing observations to theory, and we refer to these as ``unmixed'' and ``mixed'' below. The relevant numbers for comparison with the Crab filaments will always be the ``mixed'' ratio.

Very few studies report nucleosynthetic yields of stable Ni and Fe in ECSNe. \citet{wanajo09} examined the nucleosynthesis of ECSNe based on 1-dimensional (1-D) models of an O-Mg-Ne core progenitor with an initial mass of 8.8~$M_{\odot}$ \citep{nomoto84,kitaura06}, and found Ni/Fe abundance ratios of 24--57 times the solar value.  These values are, however, unmixed.   A more recent study of 3-dimensional (3-D) Fe-core-collapse explosions of zero and low-metallicity 8.1 and 9.6~$M_{\odot}$ progenitors with tenuous mantles (that might at solar metallicity have O-Ne cores and explode as ECSNe), yielded Ni/Fe abundance ratios of $\sim$20 times solar \citep{wang24b}, though it is still unclear how similar the values would be for solar metallicity progenitors that explode as ECSNe. In all of the above models, the predicted ratio would be significantly lower if a stellar envelope is mixed in with the ejecta. For example, artificially assuming that a solar metallicity envelope is mixed with the ejecta of the ``ECSN-like" models of \citet{wang24b} reduces the Ni/Fe value to 5--6 times solar, consistent with what we estimate for the Crab Nebula (A. Burrows, private communication). While \citet{wanajo09} only provided the unmixed Ni/Fe values for their models, including the envelope would likely reduce their values in a similar way. Figure~\ref{ratio_plot} shows the mixed and unmixed Ni/Fe predictions from these models, including the estimated mixed values for the \citet{wanajo09} models that assume the same dilution factor as found for the \citet{wang24b} models. It can be seen from the figure that the available ECSN yields in the mixed case are consistent with the observed Ni/Fe values in the Crab.

The link between SN explosive nucleosynthesis and the Ni/Fe ratio in Fe-core-collapse SNe was investigated recently by \citet{jerkstrand15}, following the inference of a high Ni/Fe ratio in the Type IIP SN 2012ec \citep{Jerkstrand2015_2012ec}. The yield of stable nickel is dominated by the isotopes of $^{58}$Ni and $^{60}$Ni, having neutron excesses of 0.034 and 0.033, respectively. The yield of Fe is dominated by the synthesis of $^{56}$Ni (which then decays to $^{56}$Fe through $^{56}$Co), with an initial neutron excess of zero. The Ni/Fe ratio is therefore a diagnostic of the neutron richness of the layer that experiences explosive silicon burning. A solar ratio of Ni-to-Fe (0.053), corresponds to a neutron excess of $1.8 \times 10^{-3}$, close to the value found in the oxygen shell at solar metallicity. The next shell is the silicon shell, which has a neutron excess of a few times the solar value \citep{thielemann90,Woosley07,jerkstrand15}. The normal CCSNe that present Ni/Fe ratios of 3--5 may therefore be interpreted as stars in which the dominant Fe and Ni emission lines arise in condensations dominated by material that was explosively burnt and ejected from the deep-lying silicon-shell material. There are theoretical indications that this may happen more easily for lower-mass stars \citep{woosley95} and for asymmetric explosions \citep{nagataki97}. While the more recent 1-D explosion nucleosynthesis models of \citet{sukhbold16} do not show this enhancement, a particularly asymmetric 3-D model of an 11~M$_{\odot}$ progenitor \citep{burrows24,wang24} does indicate a Ni/Fe enhancement of $\sim$4 times solar (envelope included), though the resulting neutron star kick velocity of $\sim$700~$\rm km\:s^{-1}$ is too high compared to the 160~$\rm km\:s^{-1}$ velocity of the Crab's pulsar (see Figure~\ref{ratio_plot}).

The situation becomes more complicated if significant Fe and/or Ni contributions come from the innermost layers affected by neutrino weak interactions, thus changing the neutron excess of the progenitor matter. These layers are too small in mass ($\lesssim 10^{-3}\ M_\odot$) to affect the total Ni and Fe yields for intermediate and high-mass progenitors (i.e. CCSNe from 11-20 $M_{\odot}$ progenitors), but may become relevant for low mass progenitors with lower yields of Ni and Fe that produce LMCCSNe. The recent work by \citet{wang24} analyzing the nucleosynthesis of 3-D Fe-core-collapse SN simulations \citep{wang23,burrows24} found that the 9~$M_{\odot}$ progenitor model with some imposed initial velocity perturbations produced more neutron-rich material leading to a final Ni/Fe abundance ratio of $\sim$ 8 times solar. This value reflects the ratio for the entire ejecta, including the stellar envelope, so it is appropriate to compare it with the observed ratio in the Crab's filaments that might contain mixed-in outer ejecta layers or unprocessed CSM. The fact that two different 9~M$_{\odot}$ Fe-core-collapse models of \citet{wang24} result in very different Ni/Fe ratios (see Figure~\ref{ratio_plot}), with the model with imposed initial velocity perturbations producing higher Ni/Fe, suggests that a wide range of Ni/Fe ratios is possible for models in this mass range of progenitors and that these models are consistent with our reported Ni/Fe ratio for the Crab.

The observationally-derived Ni/Fe ratios for the Crab Nebula and the predictions from the models discussed above are summarized in Figure~\ref{ratio_plot}. It can be seen that the modestly elevated Ni/Fe ratio in the Crab is consistent with the ``mixed'' Ni/Fe ratios for either a low-mass Fe-core-collapse SN (LMCCSN) or an ECSN. These SNe also provide low enough explosion energies (listed in units of Bethes (B) in Figure~\ref{ratio_plot}) to be consistent with the Crab \citep{burrows24,wang24b}. The pulsar kick velocities resulting from the 3-D models are also included in the plot. As discussed in \S\ref{kick}, a stronger line of evidence for an Fe-core-collapse explosion for the Crab may be the higher asymmetry of these explosions that can reproduce the measured kick velocity of the Crab's pulsar (see Figure~\ref{ratio_plot} and \S\ref{kick}).
Other recent evidence includes the honeycomb-like structure revealed in a 3-D reconstruction of the Crab's ejecta \citep{martin21} that may be produced by expanding plumes of radioactive $^{56}$Ni-rich ejecta seen in Fe-core-collapse SN simulations \citep{stockinger20}. Some recent efforts focus on the analysis of emission lines from lighter elements, that may have significant differences between ECSNe and LMCCSNe \citep{jerkstrand2018,hosseinzadeh2018,hiramatsu21}, which may be a useful avenue for further constraining the explosion that produced the Crab Nebula.

\subsection{Asymmetry and pulsar kick} \label{kick}

Simulations of ECSNe and Fe-core-collapse SN models predict different ejecta distributions in the resulting SNR. Fe-core-collapse models predict higher asymmetries in the ejecta in the critical seconds after the explosion which can impart pulsar kick velocities exceeding the 160~${\rm km\:s^{-1}}$ velocity of the Crab's pulsar. The asymmetry in the post-shock ejecta leads to an acceleration of the pulsar in the opposite direction from the heaviest ejecta species, including Fe \citep{wongwathanarat13, janka17}, an effect now observed in multiple SNRs \citep{bhalerao19, hollandashford17, katsuda18b}. 

Recent 3-D SN explosion simulations investigating neutron star kicks through the gravitational tugboat mechanism revealed a potential challenge to interpreting the Crab Nebula as an ECSN. The steep density gradient in the O-Ne-Mg core of ECSNe progenitors leads to explosions that are much more symmetric than Fe-core-collapse SNe, and thus, ECSNe are expected to produce low neutron star kick velocities of only a few to tens of ${\rm km\:s^{-1}}$ \citep{gessner18,burrows24b}. This is much lower than the 160 ${\rm km\:s^{-1}}$ measured for the Crab's pulsar \citep{kaplan08,hester08}. So it seems that the Crab's pulsar received a kick that is perhaps too fast for an ECSN, but too slow for a higher mass CCSN.  Indeed, simulations of $\sim$~9~$M_{\odot}$ Fe-core-collapse SNe can achieve more asymmetric explosions with SN energies and kick velocities in the range of what is observed for the Crab Nebula \citep[e.g.][]{stockinger20,burrows24}, as shown in Figure~\ref{ratio_plot}.  The observed motion of the Crab's pulsar may therefore favor a LMCCSN, an Fe-core-collapse SN with a low progenitor mass. In order to still be consistent with an ECSN, the observed pulsar velocity would require an alternative kick mechanism, such as anisotropic neutrino emission or a breakup of a binary system.

\subsection{CSM interaction} \label{csm}

If SN~1054 was not a true ECSN, but was instead a low-energy Fe core collapse event arising from the lowest initial masses to have this fate (LMCCSN), it is worth discussing how it might have produced such a high luminosity. These low-mass Fe-core-collapse SNe are expected, like ECSNe, to produce relatively low energy explosions (kinetic energy of $\sim$10$^{50}$ erg), they may be equally consistent with the observed Ni/Fe ratio, and they may be more consistent with the observed motion of the pulsar.  With a progenitor resembling a super-AGB star or red supergiant, such explosions would presumably yield relatively low-luminosity SNe II-P.  

Indeed, both ECSNe and low-mass Fe-core-collapse events have been invoked to explain several relatively low-luminosity SNe II-P, including events similar to SN~2005cs.  This is, however, at odds with the fact that SN~1054 was extremely bright, being visible during the daytime for a few weeks with an estimated absolute visual magnitude of $-$18 mag (3 mag brighter than SN~2005cs and other faint SNe II-P, even though the Crab's kinetic energy is even lower than estimates for this class of extragalactic SN). \citet{omand24} explore the possibility that the luminosity of SN~1054 could be explained by a pulsar-driven supernova model, similar to those used for superluminous SNe, but this scenario still leaves the question of the very low observed kinetic energy. \citet{smith13crab} proposed that the low kinetic energy, high peak luminosity, thin filaments, and several other properties of SN~1054 and the Crab could be reconciled if the event was a low-energy explosion that was brightened significantly by strong and immediate interaction with CSM, likely resembling the sub-class of events called SNe IIn-P \citep{mauerhan13,smith13crab}.  In this scenario, the strong CSM interaction occurring during the main SN event would cause much of the ejecta and CSM to be swept up into a very thin shell, which would cool radiatively into a very thin and dense layer and may even form dust in the post-shock layers.  This cold massive shell would later be fragmented by the expansion of the PWN, eventually producing the Crab filaments we see today.  This model, however, invokes the ejection of a massive CSM shell shortly before the final Fe core collapse.

In this context, the phenomena discussed by \citet{wh15} may be of interest.  They found that stars with initial masses just above the ECSN range (corresponding to initial masses of 9-10.3 $M_{\odot}$ in their models) suffer off-center degenerate Si-burning flashes.   In some cases, these Si flashes may be powerful enough to cause the ejection of much of the H envelope shortly before Fe core collapse, leading to a bright CSM-interaction-powered SN.  Above, we referred to these events as SDSNe, with ``SD" referring to the Si deflagration that precedes Fe core collapse.  \citet{wh15} estimate that this type of pre-SN mass ejection happens robustly in the initial mass range 9.8-10.3 $M_{\odot}$ in their models, although similar but less powerful flashes may occur over a wider mass range.   Such a scenario may provide a viable alternative to an ECSN to explain the origin of the Crab Nebula, and makes specific predictions about the timing and energy of pre-SN mass-loss events.  This scenario can be tested directly if light echoes from SN~1054 can be found because spectra of these echoes could confirm the prediction of a Type IIn-like spectrum \citep{smith13crab}, probably similar to other SN IIn-P events.

\section{Conclusions} \label{conclusion}

We presented the analysis of JWST infrared observations of the Crab Nebula composed of MIRI and NIRCam imaging mosaics that sample iron, sulfur, dust, and the PWN's synchrotron emission, as well as smaller field-of-view imaging that samples argon, neon, and oxygen emission. The observations also include two MIRI MRS pointings centered on the Crab's inner Rayleigh-Taylor filaments, and they reveal spectra rich in emission lines, including many lines from iron and nickel.
The unsurpassed sensitivity and spatial resolution of the JWST data reveal the Crab Nebula in unprecedented detail and have allowed us to study the large and small-scale properties of the PWN's synchrotron emission, to isolate emission from dust grains in the ejecta filaments, and determine nickel-to-iron abundance ratios in the ejecta to compare them with explosion and nucleosynthesis models of Fe core collapse and electron-capture SNe. Below we summarize our main findings and conclusions.

\begin{enumerate}

\item All emission lines from the ejecta filaments display the familiar filamentary morphology, but with different spatial distributions that likely reflect a combination of ionization effects and varying elemental composition across the filaments. We used a differencing technique with the longer-wavelength MIRI images to produce a high-resolution map of dust emission across the Crab Nebula. We found that the emission from the mid-infrared-emitting warm grains is concentrated in the bright, highest-density inner filaments. A comparison of the dust emission map from JWST with a \textit{Herschel} image sampling emission from cooler grains shows that the outermost filaments along the Crab's long axis have relatively warmer dust, possibly due to less shielding from radiation in the outer filaments. The emission from cooler grains dominates at the dense tips of the Rayleigh-Taylor fingers, reflecting either enhanced shielding or a grain size distribution that is weighted towards larger grains.

\item The comparison of the synchrotron-dominated JWST images with the dust map that traces the densest filaments of ejecta also leads to interesting observations. Various indentations or ``bays" are evident in the synchrotron-dominated JWST images and are seen to coincide with the location of the densest ejecta filaments, as traced by the dust emission map. The PWN appears to be confined by multiple prominent filaments distributed in a wide band oriented along the torus, and less confined along the Crab's long axis where thinner filaments and smaller PWN indentations are seen. The PWN extends well past the densest filaments along both axes, a morphology consistent with simulations of synchrotron ``blowout" that occurs when the PWN energy exceeds the SN explosion energy \citep{blondin17} or when it encounters a thin shell produced by the interaction of the SN ejecta with significant CSM material \citep{og71, smith13crab}.

\item A synchrotron spectral index map produced from the F560W and F1130W MIRI images shows index values ranging from 0.35 $\pm$ 0.05 in the inner torus to 0.80 $\pm$ 0.07 in the outer PWN. Flux densities extracted from a filament-free region of all NIRCam and MIRI images are well-fitted by a power law model with a spectral index of 0.47 $\pm$ 0.10. With the current MIRI calibration uncertainties, we were not able to confirm a break in the global synchrotron spectrum at JWST wavelengths. However, due to the very small uncertainties in the relative synchrotron spectral index across the PWN, we measured index changes across various small-scale structures in the PWN's torus region. The region corresponding to the bright NW wisp shows a considerably flatter spectrum ($\alpha= 0.27$) than the more diffuse regions
of the torus ($\sim 0.35$). This change in the index can be explained by Doppler boosting of emission from particles with a broken power-law distribution. In this scenario, the flat spectrum of the Doppler-brightened NW wisp probes lower electron energies than for adjacent regions, thus sampling the electron spectrum below the break. The measured index change may be the first direct evidence that a break in the injected particle spectrum is tied to the acceleration process at the termination shock. The knot region and a large wisp southeast of the pulsar have a steeper spectrum ($\alpha = 0.42$) than the surroundings, suggesting either a lower magnetic field in these regions or a different acceleration mechanism.

\item We detected several Ni and Fe lines in the MRS spectra of the ejecta filaments and used photoionization models to determine the Ni/Fe abundance ratios in multiple ejecta velocity components. We find consistent Ni/Fe abundance ratios that are $\sim$ 5 times higher than the solar ratio, with an estimated uncertainty of 20--50\%. Our re-analysis of Ni/Fe values from optical and near-IR data (which were previously interpreted as indicating much higher Ni/Fe abundance) using updated atomic data and allowing for local dust extinction shows that the revised values are roughly consistent with the JWST results. We compared the measured Ni/Fe abundance ratios with nucleosynthetic yields from Fe core collapse and ECSN models available in the literature. We conclude that the modestly elevated Ni/Fe ratios in the Crab Nebula are consistent with either a low-mass Fe-core-collapse explosion or an ECSN. The kick velocity of the Crab's pulsar more strongly favors a low-mass Fe-core-collapse SN that can explode with equally low energy but a higher asymmetry than an ECSN. We noted that some models for low-energy Fe CCSNe at the lowest range of initial masses may provide a suitable mechanism for eruptive pre-SN mass loss that might lead to strong CSM interaction that could explain the high luminosity of SN~1054.

\end{enumerate}

Future studies combining this JWST dataset with additional observations can include an exploration of the small-scale structure of the Rayleigh-Taylor filaments and mixing of ejecta. The photoionization models can be expanded to include the full set of lines detected with MRS to determine other elemental abundances that can be compared to explosion nucleosynthesis models. A more detailed study of the small-scale spatial and temporal variations in the synchrotron emission can be conducted with the NIRCam and MIRI images and compared with other multi-wavelength data. Additional spectroscopic observations sampling other locations within the ejecta filaments will be important for determining if the Ni/Fe abundance ratios vary across the filaments. Finally, a limiting factor in constraining the Crab Nebula's origin is in the uncertainty of theoretical predictions, so refining and expanding theoretical explosion and nucleosynthesis models for progenitors at the low end of the mass range (8--10~$M_{\odot}$) will be crucial for more definitively characterizing the progenitor and explosion that produced the Crab Nebula.

\begin{acknowledgements}

This work is based on observations made with the NASA/ESA/CSA James Webb Space Telescope. The data were obtained from the Mikulski Archive for Space Telescopes at the Space Telescope Science Institute, which is operated by the Association of Universities for Research in Astronomy, Inc., under NASA contract NAS 5-03127 for JWST. These observations are associated with program \#1714. The support for the program was provided by the NASA grant JWST-GO-01714.

T.T. acknowledges support from the NASA grant JWST-GO-01714.001, NSF grant AST-2205314, and the NASA ADAP award 80NSSC23K1130. PJK acknowledges support from the Science Foundation Ireland/Irish Research Council Pathway program under Grant Number 21/PATH-S/9360.
D.M. acknowledges NSF support from grants PHY-2209451 and AST-2206532. 
J.M.L. acknowledges support from the NASA grant JWST-GO-01714.011 and from basic research funds of the Office of Naval Research.
M.M. acknowledges support in part from ADAP program grant No. 80NSSC22K0486, from the NSF grant AST-2206657, and from the HST GO program HST-GO-16656. We also thank Adam Burrows and Tianshu Wang for the productive discussion about their 3D explosions and nucleosynthesis models.
I.D.L. acknowledges funding from the Belgian Science Policy Office (BELSPO) through the PRODEX project “JWST/MIRI Science exploitation” (C4000142239) and from the European Research Council (ERC) under the European Union’s Horizon 2020 research and innovation programme DustOrigin (ERC-2019-StG-851622).

\end{acknowledgements}

\bibliography{Astro_BIB_new}{}

\begin{thebibliography}{}
\expandafter\ifx\csname natexlab\endcsname\relax\def\natexlab#1{#1}\fi
\providecommand{\url}[1]{\href{#1}{#1}}
\providecommand{\dodoi}[1]{doi:~\href{http://doi.org/#1}{\nolinkurl{#1}}}
\providecommand{\doeprint}[1]{\href{http://ascl.net/#1}{\nolinkurl{http://ascl.net/#1}}}
\providecommand{\doarXiv}[1]{\href{https://arxiv.org/abs/#1}{\nolinkurl{https://arxiv.org/abs/#1}}}

\bibitem[{{Argyriou} {et~al.}(2023){Argyriou}, {Glasse}, {Law}, {Labiano}, {{\'A}lvarez-M{\'a}rquez}, {Patapis}, {Kavanagh}, {Gasman}, {Mueller}, {Larson}, {Vandenbussche}, {Glauser}, {Royer}, {Dicken}, {Harkett}, {Sargent}, {Engesser}, {Jones}, {Kendrew}, {Noriega-Crespo}, {Brandl}, {Rieke}, {Wright}, {Lee}, \& {Wells}}]{argyriou23}
{Argyriou}, I., {Glasse}, A., {Law}, D.~R., {et~al.} 2023, \aap, 675, A111, \dodoi{10.1051/0004-6361/202346489}

\bibitem[{{Barlow} {et~al.}(2013){Barlow}, {Swinyard}, {Owen}, {Cernicharo}, {Gomez}, {Ivison}, {Krause}, {Lim}, {Matsuura}, {Miller}, {Olofsson}, \& {Polehampton}}]{barlow13}
{Barlow}, M.~J., {Swinyard}, B.~M., {Owen}, P.~J., {et~al.} 2013, Science, 342, 1343, \dodoi{10.1126/science.1243582}

\bibitem[{{Bautista}(2001)}]{bautista01}
{Bautista}, M.~A. 2001, \aap, 365, 268, \dodoi{10.1051/0004-6361:20000032}

\bibitem[{{Bautista} {et~al.}(2015){Bautista}, {Fivet}, {Ballance}, {Quinet}, {Ferland}, {Mendoza}, \& {Kallman}}]{bautista15}
{Bautista}, M.~A., {Fivet}, V., {Ballance}, C., {et~al.} 2015, \apj, 808, 174, \dodoi{10.1088/0004-637X/808/2/174}

\bibitem[{{Begelman} \& {Li}(1992)}]{begelman92}
{Begelman}, M.~C., \& {Li}, Z.-Y. 1992, \apj, 397, 187, \dodoi{10.1086/171778}

\bibitem[{{Bhalerao} {et~al.}(2019){Bhalerao}, {Park}, {Schenck}, {Post}, \& {Hughes}}]{bhalerao19}
{Bhalerao}, J., {Park}, S., {Schenck}, A., {Post}, S., \& {Hughes}, J.~P. 2019, \apj, 872, 31, \dodoi{10.3847/1538-4357/aafafd}

\bibitem[{{Bietenholz} {et~al.}(2001){Bietenholz}, {Frail}, \& {Hester}}]{bietenholz01}
{Bietenholz}, M.~F., {Frail}, D.~A., \& {Hester}, J.~J. 2001, \apj, 560, 254, \dodoi{10.1086/322244}

\bibitem[{{Bietenholz} {et~al.}(1991){Bietenholz}, {Kronberg}, {Hogg}, \& {Wilson}}]{bietenholz91}
{Bietenholz}, M.~F., {Kronberg}, P.~P., {Hogg}, D.~E., \& {Wilson}, A.~S. 1991, \apjl, 373, L59, \dodoi{10.1086/186051}

\bibitem[{{Bietenholz} \& {Nugent}(2015)}]{bietenholz15}
{Bietenholz}, M.~F., \& {Nugent}, R.~L. 2015, \mnras, 454, 2416, \dodoi{10.1093/mnras/stv2112}

\bibitem[{{Blair} {et~al.}(1997){Blair}, {Davidson}, {Fesen}, {Uomoto}, {MacAlpine}, \& {Henry}}]{blair97}
{Blair}, W.~P., {Davidson}, K., {Fesen}, R.~A., {et~al.} 1997, \apjs, 109, 473, \dodoi{10.1086/312986}

\bibitem[{{Blondin} \& {Chevalier}(2017)}]{blondin17}
{Blondin}, J.~M., \& {Chevalier}, R.~A. 2017, \apj, 845, 139, \dodoi{10.3847/1538-4357/aa8267}

\bibitem[{{Boucaud} {et~al.}(2016){Boucaud}, {Bocchio}, {Abergel}, {Orieux}, {Dole}, \& {Hadj-Youcef}}]{boucaud16}
{Boucaud}, A., {Bocchio}, M., {Abergel}, A., {et~al.} 2016, \aap, 596, A63, \dodoi{10.1051/0004-6361/201629080}

\bibitem[{{Bucciantini} {et~al.}(2011){Bucciantini}, {Arons}, \& {Amato}}]{Bucciantini_Arons+11a}
{Bucciantini}, N., {Arons}, J., \& {Amato}, E. 2011, \mnras, 410, 381, \dodoi{10.1111/j.1365-2966.2010.17449.x}

\bibitem[{{Burrows} {et~al.}(2024{\natexlab{a}}){Burrows}, {Wang}, \& {Vartanyan}}]{burrows24}
{Burrows}, A., {Wang}, T., \& {Vartanyan}, D. 2024{\natexlab{a}}, arXiv e-prints, arXiv:2401.06840, \dodoi{10.48550/arXiv.2401.06840}

\bibitem[{{Burrows} {et~al.}(2024{\natexlab{b}}){Burrows}, {Wang}, {Vartanyan}, \& {Coleman}}]{burrows24b}
{Burrows}, A., {Wang}, T., {Vartanyan}, D., \& {Coleman}, M. S.~B. 2024{\natexlab{b}}, \apj, 963, 63, \dodoi{10.3847/1538-4357/ad2353}

\bibitem[{{Cassidy} {et~al.}(2016){Cassidy}, {Hibbert}, \& {Ramsbottom}}]{cassidy16}
{Cassidy}, C.~M., {Hibbert}, A., \& {Ramsbottom}, C.~A. 2016, \aap, 587, A107, \dodoi{10.1051/0004-6361/201527669}

\bibitem[{{Cassidy} {et~al.}(2010){Cassidy}, {Ramsbottom}, {Scott}, \& {Burke}}]{cassidy10}
{Cassidy}, C.~M., {Ramsbottom}, C.~A., {Scott}, P.~M., \& {Burke}, P.~G. 2010, \aap, 513, A55, \dodoi{10.1051/0004-6361/200913571}

\bibitem[{{Chastenet} {et~al.}(2022){Chastenet}, {De Looze}, {Hensley}, {Vandenbroucke}, {Barlow}, {Rho}, {Ravi}, {Gomez}, {Kirchschlager}, {Mac{\'\i}as-P{\'e}rez}, {Matsuura}, {Pattle}, {Ponthieu}, {Priestley}, {Rela{\~n}o}, {Ritacco}, \& {Wesson}}]{chastnet22}
{Chastenet}, J., {De Looze}, I., {Hensley}, B.~S., {et~al.} 2022, \mnras, 516, 4229, \dodoi{10.1093/mnras/stac2413}

\bibitem[{{Chevalier}(1977)}]{chevalier77}
{Chevalier}, R.~A. 1977, \araa, 15, 175, \dodoi{10.1146/annurev.aa.15.090177.001135}

\bibitem[{{Das} {et~al.}(2020){Das}, {Sil}, {Bhat}, {Gorai}, {Chakrabarti}, \& {Caselli}}]{das19}
{Das}, A., {Sil}, M., {Bhat}, B., {et~al.} 2020, \apj, 902, 131, \dodoi{10.3847/1538-4357/abb5fe}

\bibitem[{{Davidson} \& {Fesen}(1985)}]{davidson85}
{Davidson}, K., \& {Fesen}, R.~A. 1985, \araa, 23, 119, \dodoi{10.1146/annurev.aa.23.090185.001003}

\bibitem[{{De Looze} {et~al.}(2019){De Looze}, {Barlow}, {Bandiera}, {Bevan}, {Bietenholz}, {Chawner}, {Gomez}, {Matsuura}, {Priestley}, \& {Wesson}}]{delooze19}
{De Looze}, I., {Barlow}, M.~J., {Bandiera}, R., {et~al.} 2019, \mnras, 488, 164, \dodoi{10.1093/mnras/stz1533}

\bibitem[{{Deb} \& {Hibbert}(2009)}]{deb09}
{Deb}, N.~C., \& {Hibbert}, A. 2009, J. Phys. B., 42, 065003, \dodoi{10.1088/0953-4075/42/6/065003}

\bibitem[{{Deb} \& {Hibbert}(2011)}]{deb11}
---. 2011, \aap, 536, A74, \dodoi{10.1051/0004-6361/201118059}

\bibitem[{{Del Zanna} {et~al.}(2006){Del Zanna}, {Volpi}, {Amato}, \& {Bucciantini}}]{delzanna06}
{Del Zanna}, L., {Volpi}, D., {Amato}, E., \& {Bucciantini}, N. 2006, \aap, 453, 621, \dodoi{10.1051/0004-6361:20064858}

\bibitem[{{Dicken} {et~al.}(2022){Dicken}, {Rieke}, {Ressler}, {Morrison}, {Garcia Marin}, {Argyriou}, {Gordon}, {Regan}, {Cossou}, {Gaspar}, {Glasse}, {Guillard}, {Labiano}, \& {Wright}}]{dicken22}
{Dicken}, D., {Rieke}, G., {Ressler}, M., {et~al.} 2022, in Society of Photo-Optical Instrumentation Engineers (SPIE) Conference Series, Vol. 12180, Space Telescopes and Instrumentation 2022: Optical, Infrared, and Millimeter Wave, ed. L.~E. {Coyle}, S.~{Matsuura}, \& M.~D. {Perrin}, 121802R, \dodoi{10.1117/12.2630027}

\bibitem[{{Dicken} {et~al.}(2024){Dicken}, {Garc{\'\i}a Mar{\'\i}n}, {Shivaei}, {Guillard}, {Libralato}, {Glasse}, {Gordon}, {Cossou}, {Kavanagh}, {Temim}, {Flagey}, {Klaassen}, {Rieke}, {Wright}, {Alberts}, {Azzollini}, {{\'A}lvarez-M{\'a}rquez}, {Bouchet}, {Bright}, {Cracraft}, {Coulais}, {Hunor Detre}, {Engesser}, {Fox}, {Gaspar}, {Gastaud}, {Glauser}, {Hines}, {Kendrew}, {Labiano}, {Lagage}, {Lee}, {Law}, {Morrison}, {Noriega-Crespo}, {Jones}, {Patapis}, {Scheithauer}, {Sloan}, \& {Tamaz}}]{dicken24}
{Dicken}, D., {Garc{\'\i}a Mar{\'\i}n}, M., {Shivaei}, I., {et~al.} 2024, arXiv e-prints, arXiv:2403.16686.
\newblock \doarXiv{2403.16686}

\bibitem[{{Fesen} \& {Blair}(1990)}]{fesen90}
{Fesen}, R., \& {Blair}, W.~P. 1990, \apjl, 351, L45, \dodoi{10.1086/185676}

\bibitem[{{Fesen} {et~al.}(2008){Fesen}, {Rudie}, {Hurford}, \& {Soto}}]{fesen08}
{Fesen}, R., {Rudie}, G., {Hurford}, A., \& {Soto}, A. 2008, \apjs, 174, 379, \dodoi{10.1086/522781}

\bibitem[{{Fesen} \& {Kirshner}(1982)}]{fesen82}
{Fesen}, R.~A., \& {Kirshner}, R.~P. 1982, \apj, 258, 1, \dodoi{10.1086/160043}

\bibitem[{{Fesen} {et~al.}(1992){Fesen}, {Martin}, \& {Shull}}]{fesen92}
{Fesen}, R.~A., {Martin}, C.~L., \& {Shull}, J.~M. 1992, \apj, 399, 599, \dodoi{10.1086/171951}

\bibitem[{{Frail} {et~al.}(1995){Frail}, {Kassim}, {Cornwell}, \& {Goss}}]{frail95}
{Frail}, D.~A., {Kassim}, N.~E., {Cornwell}, T.~J., \& {Goss}, W.~M. 1995, \apjl, 454, L129, \dodoi{10.1086/309794}

\bibitem[{{Fraschetti} \& {Pohl}(2017)}]{fraschetti17}
{Fraschetti}, F., \& {Pohl}, M. 2017, \mnras, 471, 4856, \dodoi{10.1093/mnras/stx1833}

\bibitem[{{Gardner} {et~al.}(2023){Gardner}, {Mather}, {Abbott}, {Abell}, {Abernathy}, {Abney}, {Abraham}, {Abraham}, {Abul-Huda}, {Acton}, {Adams}, {Adams}, {Adler}, {Adriaensen}, {Aguilar}, {Ahmed}, {Ahmed}, {Ahmed}, {Albat}, {Albert}, {Alberts}, {Aldridge}, {Allen}, {Allen}, {Altenburg}, {Altunc}, {Alvarez}, {{\'A}lvarez-M{\'a}rquez}, {Alves de Oliveira}, {Ambrose}, {Anandakrishnan}, {Andersen}, {Anderson}, {Anderson}, {Anderson}, {Anderson}, {Aprea}, {Archer}, {Arenberg}, {Argyriou}, {Arribas}, {Artigau}, {Arvai}, {Atcheson}, {Atkinson}, {Averbukh}, {Aymergen}, {Bacinski}, {Baggett}, {Bagnasco}, {Baker}, {Balzano}, {Banks}, {Baran}, {Barker}, {Barrett}, {Barringer}, {Barto}, {Bast}, {Baudoz}, {Baum}, {Beatty}, {Beaulieu}, {Bechtold}, {Beck}, {Beddard}, {Beichman}, {Bellagama}, {Bely}, {Berger}, {Bergeron}, {Bernier}, {Bertch}, {Beskow}, {Betz}, {Biagetti}, {Birkmann}, {Bjorklund}, {Blackwood}, {Blazek}, {Blossfeld}, {Bluth}, {Boccaletti}, {Boegner}, {Bohlin}, {Boia}, {B{\"o}ker}, {Bonaventura}, {Bond},
  {Bosley}, {Boucarut}, {Bouchet}, {Bouwman}, {Bower}, {Bowers}, {Bowers}, {Boyce}, {Boyer}, {Boyer}, {Boyer}, {Boyer}, {Bradley}, {Brady}, {Brandl}, {Brannen}, {Breda}, {Bremmer}, {Brennan}, {Bresnahan}, {Bright}, {Broiles}, {Bromenschenkel}, {Brooks}, {Brooks}, {Brown}, {Brown}, {Brown}, {Bruce}, {Bryson}, {Bujanda}, {Bullock}, {Bunker}, {Bureo}, {Burt}, {Bush}, {Bushouse}, {Bussman}, {Cabaud}, {Cale}, {Calhoon}, {Calvani}, {Canipe}, {Caputo}, {Cara}, {Carey}, {Case}, {Cesari}, {Cetorelli}, {Chance}, {Chandler}, {Chaney}, {Chapman}, {Charlot}, {Chayer}, {Cheezum}, {Chen}, {Chen}, {Cherinka}, {Chichester}, {Chilton}, {Chittiraibalan}, {Clampin}, {Clark}, {Clark}, {Clark}, {Claybrooks}, {Cleveland}, {Cohen}, {Cohen}, {Col{\'o}n}, {Coleman}, {Colina}, {Comber}, {Comeau}, {Comer}, {Conde Reis}, {Connolly}, {Conroy}, {Contos}, {Contreras}, {Cook}, {Cooper}, {Cooper}, {Correia}, {Correnti}, {Cossou}, {Costanza}, {Coulais}, {Cox}, {Coyle}, {Cracraft}, {Crew}, {Curtis}, {Cusveller}, {Da Costa Maciel}, {Dailey},
  {Daugeron}, {Davidson}, {Davies}, {Davis}, {Davis}, {Day}, {de Chambure}, {de Jong}, {De Marchi}, {Dean}, {Decker}, {Delisa}, {Dell}, {Dellagatta}, {Dembinska}, {Demosthenes}, {Dencheva}, {Deneu}, {DePriest}, {Deschenes}, {Dethienne}, {Detre}, {Diaz}, {Dicken}, {DiFelice}, {Dillman}, {Disharoon}, {Dixon}, {Doggett}, {Dominguez}, {Donaldson}, {Doria-Warner}, {Santos}, {Doty}, {Douglas}, {Doyon}, {Dressler}, {Driggers}, {Driggers}, {Dunn}, {DuPrie}, {Dupuis}, {Durning}, {Dutta}, {Earl}, {Eccleston}, {Ecobichon}, {Egami}, {Ehrenwinkler}, {Eisenhamer}, {Eisenhower}, {Eisenstein}, {El Hamel}, {Elie}, {Elliott}, {Elliott}, {Engesser}, {Espinoza}, {Etienne}, {Etxaluze}, {Evans}, {Fabreguettes}, {Falcolini}, {Falini}, {Fatig}, {Feeney}, {Feinberg}, {Fels}, {Ferdous}, {Ferguson}, {Ferrarese}, {Ferreira}, {Ferruit}, {Ferry}, {Filippazzo}, {Firre}, {Fix}, {Flagey}, {Flanagan}, {Fleming}, {Florian}, {Flynn}, {Foiadelli}, {Fontaine}, {Fontanella}, {Forshay}, {Fortner}, {Fox}, {Framarini}, {Francisco}, {Franck}, {Franx},
  {Franz}, {Friedman}, {Friend}, {Frost}, {Fu}, {Fullerton}, {Gaillard}, {Galkin}, {Gallagher}, {Galyer}, {Garc{\'\i}a Mar{\'\i}n}, {Gardner}, {Garland}, {Garrett}, {Gasman}, {G{\'a}sp{\'a}r}, {Gastaud}, {Gaudreau}, {Gauthier}, {Geers}, {Geithner}, {Gennaro}, {Gerber}, {Gereau}, {Giampaoli}, {Giardino}, {Gibbons}, {Gilbert}, {Gilman}, {Girard}, {Giuliano}, {Gkountis}, {Glasse}, {Glassmire}, {Glauser}, {Glazer}, {Goldberg}, {Golimowski}, {Gonzaga}, {Gordon}, {Gordon}, {Goudfrooij}, {Gough}, {Graham}, {Grau}, {Green}, {Greene}, {Greene}, {Greenfield}, {Greenhouse}, {Greve}, {Greville}, {Grimaldi}, {Groe}, {Groebner}, {Grumm}, {Grundy}, {G{\"u}del}, {Guillard}, {Guldalian}, {Gunn}, {Gurule}, {Gutman}, {Guy}, {Guyot}, {Hack}, {Haderlein}, {Hagan}, {Hagedorn}, {Hainline}, {Haley}, {Hami}, {Hamilton}, {Hammann}, {Hammel}, {Hanley}, {Hansen}, {Hardy}, {Harnisch}, {Harr}, {Harris}, {Hart}, {Hartig}, {Hasan}, {Hashim}, {Hashimoto}, {Haskins}, {Hawkins}, {Hayden}, {Hayden}, {Healy}, {Hecht}, {Heeg}, {Hejal}, {Helm},
  {Hengemihle}, {Henning}, {Henry}, {Henry}, {Henshaw}, {Hernandez}, {Herrington}, {Heske}, {Hesman}, {Hickey}, {Hilbert}, {Hines}, {Hinz}, {Hirsch}, {Hitcho}, {Hodapp}, {Hodge}, {Hoffman}, {Holfeltz}, {Holler}, {Hoppa}, {Horner}, {Howard}, {Howard}, {Huber}, {Hunkeler}, {Hunter}, {Hunter}, {Hurd}, {Hurst}, {Hutchings}, {Hylan}, {Ignat}, {Illingworth}, {Irish}, {Isaacs}, {Jackson}, {Jaffe}, {Jahic}, {Jahromi}, {Jakobsen}, {James}, {James}, {James}, {Jamieson}, {Jandra}, {Jayawardhana}, {Jedrzejewski}, {Jeffers}, {Jensen}, {Joanne}, {Johns}, {Johnson}, {Johnson}, {Johnson}, {Johnson}, {Johnson}, {Johnson}, {Johnstone}, {Jollet}, {Jones}, {Jones}, {Jones}, {Jones}, {Jones}, {Jordan}, {Jordan}, {Jue}, {Jurkowski}, {Justis}, {Justtanont}, {Kaleida}, {Kalirai}, {Kalmanson}, {Kaltenegger}, {Kammerer}, {Kan}, {Kanarek}, {Kao}, {Karakla}, {Karl}, {Kassin}, {Kauffman}, {Kavanagh}, {Kelley}, {Kelly}, {Kendrew}, {Kennedy}, {Kenny}, {Keski-Kuha}, {Keyes}, {Khan}, {Kidwell}, {Kimble}, {King}, {King}, {Kinzel}, {Kirk},
  {Kirkpatrick}, {Klaassen}, {Klingemann}, {Klintworth}, {Knapp}, {Knight}, {Knollenberg}, {Knutsen}, {Koehler}, {Koekemoer}, {Kofler}, {Kontson}, {Kovacs}, {Kozhurina-Platais}, {Krause}, {Kriss}, {Krist}, {Kristoffersen}, {Krogel}, {Krueger}, {Kulp}, {Kumari}, {Kwan}, {Kyprianou}, {Labador}, {Labiano}, {Lafreni{\`e}re}, {Lagage}, {Laidler}, {Laine}, {Laird}, {Lajoie}, {Lallo}, {Lam}, {LaMassa}, {Lambros}, {Lampenfield}, {Lander}, {Langston}, {Larson}, {Larson}, {LaVerghetta}, {Law}, {Lawrence}, {Lee}, {Lee}, {Lee}, {Leisenring}, {Leveille}, {Levenson}, {Levi}, {Levine}, {Lewis}, {Lewis}, {Lewis}, {Libralato}, {Lidon}, {Liebrecht}, {Lightsey}, {Lilly}, {Lim}, {Lim}, {Ling}, {Link}, {Link}, {Lipinski}, {Liu}, {Lo}, {Lobmeyer}, {Logue}, {Long}, {Long}, {Long}, {Long}, {L{\'o}pez-Caniego}, {Lotz}, {Love-Pruitt}, {Lubskiy}, {Luers}, {Luetgens}, {Luevano}, {Lui}, {Lund}, {Lundquist}, {Lunine}, {L{\"u}tzgendorf}, {Lynch}, {MacDonald}, {MacDonald}, {Macias}, {Macklis}, {Maghami}, {Maharaja}, {Maiolino},
  {Makrygiannis}, {Malla}, {Malumuth}, {Manjavacas}, {Marini}, {Marrione}, {Marston}, {Martel}, {Martin}, {Martin}, {Martinez}, {Maschmann}, {Masci}, {Masetti}, {Maszkiewicz}, {Matthews}, {Matuskey}, {McBrayer}, {McCarthy}, {McCaughrean}, {McClare}, {McClare}, {McCloskey}, {McClurg}, {McCoy}, {McElwain}, {McGregor}, {McGuffey}, {McKay}, {McKenzie}, {McLean}, {McMaster}, {McNeil}, {De Meester}, {Mehalick}, {Meixner}, {Mel{\'e}ndez}, {Menzel}, {Menzel}, {Merz}, {Mesterharm}, {Meyer}, {Meyett}, {Meza}, {Midwinter}, {Milam}, {Miller}, {Miller}, {Miskey}, {Misselt}, {Mitchell}, {Mohan}, {Montoya}, {Moran}, {Morishita}, {Moro-Mart{\'\i}n}, {Morrison}, {Morrison}, {Morse}, {Moschos}, {Moseley}, {Mosier}, {Mosner}, {Mountain}, {Muckenthaler}, {Mueller}, {Mueller}, {Muhiem}, {M{\"u}hlmann}, {Mullally}, {Mullen}, {Munger}, {Murphy}, {Murray}, {Muzerolle}, {Mycroft}, {Myers}, {Myers}, {Myers}, {Myers}, {Myrick}, {Nagle}, {Nayak}, {Naylor}, {Neff}, {Nelan}, {Nella}, {Nguyen}, {Nguyen}, {Nickson}, {Nidhiry}, {Niedner},
  {Nieto-Santisteban}, {Nikolov}, {Nishisaka}, {Noriega-Crespo}, {Nota}, {O'Mara}, {Oboryshko}, {O'Brien}, {Ochs}, {Offenberg}, {Ogle}, {Ohl}, {Olmsted}, {Osborne}, {O'Shaughnessy}, {{\"O}stlin}, {O'Sullivan}, {Otor}, {Ottens}, {Ouellette}, {Outlaw}, {Owens}, {Pacifici}, {Page}, {Paranilam}, {Park}, {Parrish}, {Paschal}, {Patapis}, {Patel}, {Patrick}, {Pattishall}, {Paul}, {Paul}, {Pauly}, {Pavlovsky}, {Pe{\~n}a-Guerrero}, {Pedder}, {Peek}, {Pelham}, {Penanen}, {Perriello}, {Perrin}, {Perrine}, {Perrygo}, {Peslier}, {Petach}, {Peterson}, {Pfarr}, {Pierson}, {Pietraszkiewicz}, {Pilchen}, {Pipher}, {Pirzkal}, {Pitman}, {Player}, {Plesha}, {Plitzke}, {Pohner}, {Poletis}, {Pollizzi}, {Polster}, {Pontius}, {Pontoppidan}, {Porges}, {Potter}, {Prescott}, {Proffitt}, {Pueyo}, {Quispe Neira}, {Radich}, {Rager}, {Rameau}, {Ramey}, {Ramos Alarcon}, {Rampini}, {Rapp}, {Rashford}, {Rauscher}, {Ravindranath}, {Rawle}, {Rawlings}, {Ray}, {Regan}, {Rehm}, {Rehm}, {Reid}, {Reis}, {Renk}, {Reoch}, {Ressler}, {Rest},
  {Reynolds}, {Richon}, {Richon}, {Ridgaway}, {Riedel}, {Rieke}, {Rieke}, {Rifelli}, {Rigby}, {Riggs}, {Ringel}, {Ritchie}, {Rix}, {Robberto}, {Robinson}, {Robinson}, {Robinson}, {Rock}, {Rodriguez}, {Rodr{\'\i}guez del Pino}, {Roellig}, {Rohrbach}, {Roman}, {Romelfanger}, {Romo}, {Rosales}, {Rose}, {Roteliuk}, {Roth}, {Rothwell}, {Rouzaud}, {Rowe}, {Rowlands}, {Roy}, {Royer}, {Rui}, {Rumler}, {Rumpl}, {Russ}, {Ryan}, {Ryan}, {Saad}, {Sabata}, {Sabatino}, {Sabbi}, {Sabelhaus}, {Sabia}, {Sahu}, {Saif}, {Salvignol}, {Samara-Ratna}, {Samuelson}, {Sanders}, {Sappington}, {Sargent}, {Sauer}, {Savadkin}, {Sawicki}, {Schappell}, {Scheffer}, {Scheithauer}, {Scherer}, {Schiff}, {Schlawin}, {Schmeitzky}, {Schmitz}, {Schmude}, {Schneider}, {Schreiber}, {Schroeven-Deceuninck}, {Schultz}, {Schwab}, {Schwartz}, {Scoccimarro}, {Scott}, {Scott}, {Seaton}, {Seely}, {Seery}, {Seidleck}, {Sembach}, {Shanahan}, {Shaughnessy}, {Shaw}, {Shay}, {Sheehan}, {Sheth}, {Shih}, {Shivaei}, {Siegel}, {Sienkiewicz}, {Simmons}, {Simon},
  {Sirianni}, {Sivaramakrishnan}, {Slade}, {Sloan}, {Slocum}, {Slowinski}, {Smith}, {Smith}, {Smith}, {Smith}, {Smith}, {Smith}, {Smolik}, {Soderblom}, {Sohn}, {Sokol}, {Sonneborn}, {Sontag}, {Sooy}, {Soummer}, {Southwood}, {Spain}, {Sparmo}, {Speer}, {Spencer}, {Sprofera}, {Stallcup}, {Stanley}, {Stansberry}, {Stark}, {Starr}, {Stassi}, {Steck}, {Steeley}, {Stephens}, {Stephenson}, {Stewart}, {Stiavelli}, {}, {Strada}, {Straughn}, {Streetman}, {Strickland}, {Strobele}, {Stuhlinger}, {Stys}, {Such}, {Sukhatme}, {Sullivan}, {Sullivan}, {Sumner}, {Sun}, {Sunnquist}, {Swade}, {Swam}, {Swenton}, {Swoish}, {Tam Litten}, {Tamas}, {Tao}, {Taylor}, {Taylor}, {te Plate}, {Van Tea}, {Teague}, {Telfer}, {Temim}, {Texter}, {Thatte}, {Thompson}, {Thompson}, {Thomson}, {Thronson}, {Tierney}, {Tikkanen}, {Tinnin}, {Tippet}, {Todd}, {Tran}, {Trauger}, {Trejo}, {Vinh Truong}, {Tsukamoto}, {Tufail}, {Tumlinson}, {Tustain}, {Tyra}, {Ubeda}, {Underwood}, {Uzzo}, {Vaclavik}, {Valenduc}, {Valenti}, {Van Campen}, {van de Wetering},
  {Van Der Marel}, {van Haarlem}, {Vandenbussche}, {van Dishoeck}, {Vanterpool}, {Vernoy}, {Vila Costas}, {Volk}, {Voorzaat}, {Voyton}, {Vydra}, {Waddy}, {Waelkens}, {Wahlgren}, {Walker}, {Wander}, {Warfield}, {Warner}, {Wasiak}, {Wasiak}, {Wehner}, {Weiler}, {Weilert}, {Weiss}, {Wells}, {Welty}, {Wheate}, {Wheeler}, {White}, {Whitehouse}, {Whiteleather}, {Whitman}, {Williams}, {Willmer}, {Willott}, {Willoughby}, {Wilson}, {Wilson}, {Wilson}, {Windhorst}, {Wislowski}, {Wolfe}, {Wolfe}, {Wolff}, {Wondel}, {Woo}, {Woods}, {Worden}, {Workman}, {Wright}, {Wu}, {Wu}, {Wun}, {Wymer}, {Yadetie}, {Yan}, {Yang}, {Yates}, {Yeager}, {Yerger}, {Young}, {Young}, {Yu}, {Yu}, {Zak}, {Zeidler}, {Zepp}, {Zhou}, {Zincke}, {Zonak}, \& {Zondag}}]{gardner23}
{Gardner}, J.~P., {Mather}, J.~C., {Abbott}, R., {et~al.} 2023, \pasp, 135, 068001, \dodoi{10.1088/1538-3873/acd1b5}

\bibitem[{{Gessner} \& {Janka}(2018)}]{gessner18}
{Gessner}, A., \& {Janka}, H.-T. 2018, \apj, 865, 61, \dodoi{10.3847/1538-4357/aadbae}

\bibitem[{{Glaccum} {et~al.}(1982){Glaccum}, {Harper}, {Loewenstein}, {Pernic}, \& {Low}}]{glaccum82}
{Glaccum}, W., {Harper}, D.~A., {Loewenstein}, R.~F., {Pernic}, R., \& {Low}, F.~J. 1982, in Bulletin of the American Astronomical Society, Vol.~14, 612

\bibitem[{{Gleeson} {et~al.}(1974){Gleeson}, {Legg}, \& {Westfold}}]{gleeson74}
{Gleeson}, L.~J., {Legg}, M.~P.~C., \& {Westfold}, K.~C. 1974, \mnras, 168, 379, \dodoi{10.1093/mnras/168.2.379}

\bibitem[{{Gomez} {et~al.}(2012){Gomez}, {Krause}, {Barlow}, {Swinyard}, {Owen}, {Clark}, {Matsuura}, {Gomez}, {Rho}, {Besel}, {Bouwman}, {Gear}, {Henning}, {Ivison}, {Polehampton}, \& {Sibthorpe}}]{gomez12a}
{Gomez}, H.~L., {Krause}, O., {Barlow}, M.~J., {et~al.} 2012, \apj, 760, 96, \dodoi{10.1088/0004-637X/760/1/96}

\bibitem[{{Gordon}(2023)}]{gordon23a}
{Gordon}, K. 2023, {karllark/dust\_extinction: OneRelationForAllWaves}, v1.2,  Zenodo, \dodoi{10.5281/zenodo.7799360}

\bibitem[{{Gordon} {et~al.}(2023){Gordon}, {Clayton}, {Decleir}, {Fitzpatrick}, {Massa}, {Misselt}, \& {Tollerud}}]{gordon23}
{Gordon}, K.~D., {Clayton}, G.~C., {Decleir}, M., {et~al.} 2023, \apj, 950, 86, \dodoi{10.3847/1538-4357/accb59}

\bibitem[{{Gradshteyn} \& {Ryzhik}(1965)}]{gradshteyn65}
{Gradshteyn}, I.~S., \& {Ryzhik}, I.~M. 1965, {Table of Integrals, Series and Products}

\bibitem[{{Grenman} {et~al.}(2017){Grenman}, {Gahm}, \& {Elfgren}}]{grenman17}
{Grenman}, T., {Gahm}, G.~F., \& {Elfgren}, E. 2017, \aap, 599, A110, \dodoi{10.1051/0004-6361/201629693}

\bibitem[{{Henry}(1984)}]{henry84}
{Henry}, R.~B.~C. 1984, \apj, 281, 644, \dodoi{10.1086/162140}

\bibitem[{{Hester}(2008)}]{hester08}
{Hester}, J.~J. 2008, \araa, 46, 127, \dodoi{10.1146/annurev.astro.45.051806.110608}

\bibitem[{{Hester} {et~al.}(1995){Hester}, {Scowen}, {Sankrit}, {Burrows}, {Gallagher}, {Holtzman}, {Watson}, {Trauger}, {Ballester}, {Casertano}, {Clarke}, {Crisp}, {Evans}, {Griffiths}, {Hoessel}, {Krist}, {Lynds}, {Mould}, {O'Neil}, {Stapelfeldt}, \& {Westphal}}]{hester95}
{Hester}, J.~J., {Scowen}, P.~A., {Sankrit}, R., {et~al.} 1995, \apj, 448, 240, \dodoi{10.1086/175956}

\bibitem[{{Hester} {et~al.}(1996){Hester}, {Stone}, {Scowen}, {Jun}, {Gallagher}, {Norman}, {Ballester}, {Burrows}, {Casertano}, {Clarke}, {Crisp}, {Griffiths}, {Hoessel}, {Holtzman}, {Krist}, {Mould}, {Sankrit}, {Stapelfeldt}, {Trauger}, {Watson}, \& {Westphal}}]{hester96}
{Hester}, J.~J., {Stone}, J.~M., {Scowen}, P.~A., {et~al.} 1996, \apj, 456, 225, \dodoi{10.1086/176643}

\bibitem[{{Hiramatsu} {et~al.}(2021){Hiramatsu}, {Howell}, {Van Dyk}, {Goldberg}, {Maeda}, {Moriya}, {Tominaga}, {Nomoto}, {Hosseinzadeh}, {Arcavi}, {McCully}, {Burke}, {Bostroem}, {Valenti}, {Dong}, {Brown}, {Andrews}, {Bilinski}, {Williams}, {Smith}, {Smith}, {Sand}, {Anand}, {Xu}, {Filippenko}, {Bersten}, {Folatelli}, {Kelly}, {Noguchi}, \& {Itagaki}}]{hiramatsu21}
{Hiramatsu}, D., {Howell}, D.~A., {Van Dyk}, S.~D., {et~al.} 2021, Nature Astronomy, 5, 903, \dodoi{10.1038/s41550-021-01384-2}

\bibitem[{{Hitomi Collaboration} {et~al.}(2018){Hitomi Collaboration}, {Aharonian}, {Akamatsu}, {Akimoto}, {Allen}, {Angelini}, {Audard}, {Awaki}, {Axelsson}, {Bamba}, {Bautz}, {Blandford}, {Brenneman}, {Brown}, {Bulbul}, {Cackett}, {Chernyakova}, {Chiao}, {Coppi}, {Costantini}, {de Plaa}, {de Vries}, {den Herder}, {Done}, {Dotani}, {Ebisawa}, {Eckart}, {Enoto}, {Ezoe}, {Fabian}, {Ferrigno}, {Foster}, {Fujimoto}, {Fukazawa}, {Furuzawa}, {Galeazzi}, {Gallo}, {Gandhi}, {Giustini}, {Goldwurm}, {Gu}, {Guainazzi}, {Haba}, {Hagino}, {Hamaguchi}, {Harrus}, {Hatsukade}, {Hayashi}, {Hayashi}, {Hayashida}, {Hiraga}, {Hornschemeier}, {Hoshino}, {Hughes}, {Ichinohe}, {Iizuka}, {Inoue}, {Inoue}, {Ishida}, {Ishikawa}, {Ishisaki}, {Kaastra}, {Kallman}, {Kamae}, {Kataoka}, {Katsuda}, {Kawai}, {Kelley}, {Kilbourne}, {Kitaguchi}, {Kitamoto}, {Kitayama}, {Kohmura}, {Kokubun}, {Koyama}, {Koyama}, {Kretschmar}, {Krimm}, {Kubota}, {Kunieda}, {Laurent}, {Lee}, {Leutenegger}, {Limousin}, {Loewenstein}, {Long}, {Lumb}, {Madejski},
  {Maeda}, {Maier}, {Makishima}, {Markevitch}, {Matsumoto}, {Matsushita}, {McCammon}, {McNamara}, {Mehdipour}, {Miller}, {Miller}, {Mineshige}, {Mitsuda}, {Mitsuishi}, {Miyazawa}, {Mizuno}, {Mori}, {Mori}, {Mukai}, {Murakami}, {Mushotzky}, {Nakagawa}, {Nakajima}, {Nakamori}, {Nakashima}, {Nakazawa}, {Nobukawa}, {Nobukawa}, {Noda}, {Odaka}, {Ohashi}, {Ohno}, {Okajima}, {Ota}, {Ozaki}, {Paerels}, {Paltani}, {Petre}, {Pinto}, {Porter}, {Pottschmidt}, {Reynolds}, {Safi-Harb}, {Saito}, {Sakai}, {Sasaki}, {Sato}, {Sato}, {Sato}, {Sato}, {Sawada}, {Schartel}, {Serlemtsos}, {Seta}, {Shidatsu}, {Simionescu}, {Smith}, {Soong}, {Stawarz}, {Sugawara}, {Sugita}, {Szymkowiak}, {Tajima}, {Takahashi}, {Takahashi}, {Takeda}, {Takei}, {Tamagawa}, {Tamura}, {Tanaka}, {Tanaka}, {Tanaka}, {Tashiro}, {Tawara}, {Terada}, {Terashima}, {Tombesi}, {Tomida}, {Tsuboi}, {Tsujimoto}, {Tsunemi}, {Tsuru}, {Uchida}, {Uchiyama}, {Uchiyama}, {Ueda}, {Ueda}, {Uno}, {Urry}, {Ursino}, {Watanabe}, {Werner}, {Wilkins}, {Williams}, {Yamada},
  {Yamaguchi}, {Yamaoka}, {Yamasaki}, {Yamauchi}, {Yamauchi}, {Yaqoob}, {Yatsu}, {Yonetoku}, {Zhuravleva}, {Zoghbi}, {Tominaga}, \& {Moriya}}]{hitomi18}
{Hitomi Collaboration}, {Aharonian}, F., {Akamatsu}, H., {et~al.} 2018, \pasj, 70, 14, \dodoi{10.1093/pasj/psx072}

\bibitem[{{Holland-Ashford} {et~al.}(2017){Holland-Ashford}, {Lopez}, {Auchettl}, {Temim}, \& {Ramirez-Ruiz}}]{hollandashford17}
{Holland-Ashford}, T., {Lopez}, L.~A., {Auchettl}, K., {Temim}, T., \& {Ramirez-Ruiz}, E. 2017, \apj, 844, 84, \dodoi{10.3847/1538-4357/aa7a5c}

\bibitem[{{Hosseinzadeh} {et~al.}(2018){Hosseinzadeh}, {Valenti}, {McCully}, {Howell}, {Arcavi}, {Jerkstrand}, {Guevel}, {Tartaglia}, {Rui}, {Mo}, {Wang}, {Huang}, {Song}, {Zhang}, \& {Itagaki}}]{hosseinzadeh2018}
{Hosseinzadeh}, G., {Valenti}, S., {McCully}, C., {et~al.} 2018, \apj, 861, 63, \dodoi{10.3847/1538-4357/aac5f6}

\bibitem[{{Hudgins} {et~al.}(1990){Hudgins}, {Herter}, \& {Joyce}}]{hudgins90}
{Hudgins}, D., {Herter}, T., \& {Joyce}, R.~J. 1990, \apjl, 354, L57, \dodoi{10.1086/185722}

\bibitem[{{Janka}(2017)}]{janka17}
{Janka}, H.-T. 2017, \apj, 837, 84, \dodoi{10.3847/1538-4357/aa618e}

\bibitem[{{Janka} {et~al.}(2008){Janka}, {M{\"u}ller}, {Kitaura}, \& {Buras}}]{janka08}
{Janka}, H.~T., {M{\"u}ller}, B., {Kitaura}, F.~S., \& {Buras}, R. 2008, \aap, 485, 199, \dodoi{10.1051/0004-6361:20079334}

\bibitem[{{Jerkstrand} {et~al.}(2018){Jerkstrand}, {Ertl}, {Janka}, {M{\"u}ller}, {Sukhbold}, \& {Woosley}}]{jerkstrand2018}
{Jerkstrand}, A., {Ertl}, T., {Janka}, H.~T., {et~al.} 2018, \mnras, 475, 277, \dodoi{10.1093/mnras/stx2877}

\bibitem[{{Jerkstrand} {et~al.}(2015{\natexlab{a}}){Jerkstrand}, {Timmes}, {Magkotsios}, {Sim}, {Fransson}, {Spyromilio}, {M{\"u}ller}, {Heger}, {Sollerman}, \& {Smartt}}]{jerkstrand15}
{Jerkstrand}, A., {Timmes}, F.~X., {Magkotsios}, G., {et~al.} 2015{\natexlab{a}}, \apj, 807, 110, \dodoi{10.1088/0004-637X/807/1/110}

\bibitem[{{Jerkstrand} {et~al.}(2015{\natexlab{b}}){Jerkstrand}, {Smartt}, {Sollerman}, {Inserra}, {Fraser}, {Spyromilio}, {Fransson}, {Chen}, {Barbarino}, {Dall'Ora}, {Botticella}, {Della Valle}, {Gal-Yam}, {Valenti}, {Maguire}, {Mazzali}, \& {Tomasella}}]{Jerkstrand2015_2012ec}
{Jerkstrand}, A., {Smartt}, S.~J., {Sollerman}, J., {et~al.} 2015{\natexlab{b}}, \mnras, 448, 2482, \dodoi{10.1093/mnras/stv087}

\bibitem[{{Kaplan} {et~al.}(2008){Kaplan}, {Chatterjee}, {Gaensler}, \& {Anderson}}]{kaplan08}
{Kaplan}, D.~L., {Chatterjee}, S., {Gaensler}, B.~M., \& {Anderson}, J. 2008, \apj, 677, 1201, \dodoi{10.1086/529026}

\bibitem[{{Katsuda} {et~al.}(2018){Katsuda}, {Morii}, {Janka}, {Wongwathanarat}, {Nakamura}, {Kotake}, {Mori}, {M{\"u}ller}, {Takiwaki}, {Tanaka}, {Tominaga}, \& {Tsunemi}}]{katsuda18b}
{Katsuda}, S., {Morii}, M., {Janka}, H.-T., {et~al.} 2018, \apj, 856, 18, \dodoi{10.3847/1538-4357/aab092}

\bibitem[{{Kitaura} {et~al.}(2006){Kitaura}, {Janka}, \& {Hillebrandt}}]{kitaura06}
{Kitaura}, F.~S., {Janka}, H.~T., \& {Hillebrandt}, W. 2006, \aap, 450, 345, \dodoi{10.1051/0004-6361:20054703}

\bibitem[{{Laming} \& {Temim}(2020)}]{laming20}
{Laming}, J.~M., \& {Temim}, T. 2020, \apj, 904, 115, \dodoi{10.3847/1538-4357/abc1e5}

\bibitem[{{Law} {et~al.}(2023){Law}, {E. Morrison}, {Argyriou}, {Patapis}, {{\'A}lvarez-M{\'a}rquez}, {Labiano}, \& {Vandenbussche}}]{Law2023}
{Law}, D.~R., {E. Morrison}, J., {Argyriou}, I., {et~al.} 2023, \aj, 166, 45, \dodoi{10.3847/1538-3881/acdddc}

\bibitem[{{Li} \& {Begelman}(1992)}]{li92}
{Li}, Z.-Y., \& {Begelman}, M.~C. 1992, \apj, 400, 186, \dodoi{10.1086/171985}

\bibitem[{{Loh} {et~al.}(2012){Loh}, {Baldwin}, {Ferland}, {Curtis}, {Richardson}, {Fabian}, \& {Salom{\'e}}}]{loh12}
{Loh}, E.~D., {Baldwin}, J.~A., {Ferland}, G.~J., {et~al.} 2012, \mnras, 421, 789, \dodoi{10.1111/j.1365-2966.2011.20353.x}

\bibitem[{{Loll} {et~al.}(2013){Loll}, {Desch}, {Scowen}, \& {Foy}}]{loll13}
{Loll}, A.~M., {Desch}, S.~J., {Scowen}, P.~A., \& {Foy}, J.~P. 2013, \apj, 765, 152, \dodoi{10.1088/0004-637X/765/2/152}

\bibitem[{{Lyubarsky}(2003)}]{lyubarsky03}
{Lyubarsky}, Y.~E. 2003, \mnras, 345, 153, \dodoi{10.1046/j.1365-8711.2003.06927.x}

\bibitem[{{Lyutikov} {et~al.}(2019){Lyutikov}, {Temim}, {Komissarov}, {Slane}, {Sironi}, \& {Comisso}}]{lyutikov19}
{Lyutikov}, M., {Temim}, T., {Komissarov}, S., {et~al.} 2019, \mnras, 489, 2403, \dodoi{10.1093/mnras/stz2023}

\bibitem[{{MacAlpine} {et~al.}(1989){MacAlpine}, {McGaugh}, {Mazzarella}, \& {Uomoto}}]{macalpine89}
{MacAlpine}, G.~M., {McGaugh}, S.~S., {Mazzarella}, J.~M., \& {Uomoto}, A. 1989, \apj, 342, 364, \dodoi{10.1086/167598}

\bibitem[{{MacAlpine} \& {Satterfield}(2008)}]{macalpine08}
{MacAlpine}, G.~M., \& {Satterfield}, T.~J. 2008, \aj, 136, 2152, \dodoi{10.1088/0004-6256/136/5/2152}

\bibitem[{{Martin} {et~al.}(2021){Martin}, {Milisavljevic}, \& {Drissen}}]{martin21}
{Martin}, T., {Milisavljevic}, D., \& {Drissen}, L. 2021, \mnras, 502, 1864, \dodoi{10.1093/mnras/staa4046}

\bibitem[{{Mauerhan} {et~al.}(2013){Mauerhan}, {Smith}, {Silverman}, {Filippenko}, {Morgan}, {Cenko}, {Ganeshalingam}, {Clubb}, {Bloom}, {Matheson}, \& {Milne}}]{mauerhan13}
{Mauerhan}, J.~C., {Smith}, N., {Silverman}, J.~M., {et~al.} 2013, \mnras, 431, 2599, \dodoi{10.1093/mnras/stt360}

\bibitem[{{Mayall}(1939)}]{mayall39}
{Mayall}, N.~U. 1939, Leaflet of the Astronomical Society of the Pacific, 3, 145

\bibitem[{{Mazzotta} {et~al.}(1998){Mazzotta}, {Mazzitelli}, {Colafrancesco}, \& {Vittorio}}]{mazzotta98}
{Mazzotta}, P., {Mazzitelli}, G., {Colafrancesco}, S., \& {Vittorio}, N. 1998, \aaps, 133, 403, \dodoi{10.1051/aas:1998330}

\bibitem[{{Michel} {et~al.}(1991){Michel}, {Scowen}, {Dufour}, \& {Hester}}]{michel91}
{Michel}, F.~C., {Scowen}, P.~A., {Dufour}, R.~J., \& {Hester}, J.~J. 1991, \apj, 368, 463, \dodoi{10.1086/169710}

\bibitem[{{Miyaji} {et~al.}(1980){Miyaji}, {Nomoto}, {Yokoi}, \& {Sugimoto}}]{miyaji80}
{Miyaji}, S., {Nomoto}, K., {Yokoi}, K., \& {Sugimoto}, D. 1980, \pasj, 32, 303

\bibitem[{{Mori} {et~al.}(2004){Mori}, {Burrows}, {Hester}, {Pavlov}, {Shibata}, \& {Tsunemi}}]{mori04}
{Mori}, K., {Burrows}, D.~N., {Hester}, J.~J., {et~al.} 2004, \apj, 609, 186, \dodoi{10.1086/421011}

\bibitem[{{Nagataki} {et~al.}(1997){Nagataki}, {Hashimoto}, {Sato}, \& {Yamada}}]{nagataki97}
{Nagataki}, S., {Hashimoto}, M.-a., {Sato}, K., \& {Yamada}, S. 1997, \apj, 486, 1026, \dodoi{10.1086/304565}

\bibitem[{{Ng} \& {Romani}(2004)}]{ng04}
{Ng}, C.~Y., \& {Romani}, R.~W. 2004, \apj, 601, 479, \dodoi{10.1086/380486}

\bibitem[{{Nomoto}(1987)}]{nomoto87}
{Nomoto}, K. 1987, \apj, 322, 206, \dodoi{10.1086/165716}

\bibitem[{{Nomoto} {et~al.}(1982){Nomoto}, {Sparks}, {Fesen}, {Gull}, {Miyaji}, \& {Sugimoto}}]{nomoto82}
{Nomoto}, K., {Sparks}, W.~M., {Fesen}, R.~A., {et~al.} 1982, \nat, 299, 803, \dodoi{10.1038/299803a0}

\bibitem[{{Nomoto} {et~al.}(1984){Nomoto}, {Thielemann}, \& {Wheeler}}]{nomoto84}
{Nomoto}, K., {Thielemann}, F.-K., \& {Wheeler}, J.~C. 1984, \apjl, 279, L23, \dodoi{10.1086/184247}

\bibitem[{{Olmi} {et~al.}(2016){Olmi}, {Del Zanna}, {Amato}, {Bucciantini}, \& {Mignone}}]{olmi16}
{Olmi}, B., {Del Zanna}, L., {Amato}, E., {Bucciantini}, N., \& {Mignone}, A. 2016, Journal of Plasma Physics, 82, 635820601, \dodoi{10.1017/S0022377816000957}

\bibitem[{{Omand} {et~al.}(2024){Omand}, {Sarin}, \& {Temim}}]{omand24}
{Omand}, C. M.~B., {Sarin}, N., \& {Temim}, T. 2024, arXiv e-prints, arXiv:2404.19017.
\newblock \doarXiv{2404.19017}

\bibitem[{{Ostriker} \& {Gunn}(1969)}]{Ostriker69}
{Ostriker}, J.~P., \& {Gunn}, J.~E. 1969, \apj, 157, 1395, \dodoi{10.1086/150160}

\bibitem[{{Ostriker} \& {Gunn}(1971)}]{og71}
---. 1971, \apjl, 164, L95, \dodoi{10.1086/180699}

\bibitem[{{Owen} \& {Barlow}(2015)}]{owen15}
{Owen}, P.~J., \& {Barlow}, M.~J. 2015, \apj, 801, 141, \dodoi{10.1088/0004-637X/801/2/141}

\bibitem[{{Porth} {et~al.}(2014{\natexlab{a}}){Porth}, {Komissarov}, \& {Keppens}}]{porth14a}
{Porth}, O., {Komissarov}, S.~S., \& {Keppens}, R. 2014{\natexlab{a}}, \mnras, 438, 278, \dodoi{10.1093/mnras/stt2176}

\bibitem[{{Porth} {et~al.}(2014{\natexlab{b}}){Porth}, {Komissarov}, \& {Keppens}}]{porth14b}
---. 2014{\natexlab{b}}, \mnras, 443, 547, \dodoi{10.1093/mnras/stu1082}

\bibitem[{{Priestley} {et~al.}(2019){Priestley}, {Barlow}, \& {De Looze}}]{priestley19}
{Priestley}, F.~D., {Barlow}, M.~J., \& {De Looze}, I. 2019, \mnras, 485, 440, \dodoi{10.1093/mnras/stz414}

\bibitem[{{Priestley} {et~al.}(2017){Priestley}, {Barlow}, \& {Viti}}]{priestley17}
{Priestley}, F.~D., {Barlow}, M.~J., \& {Viti}, S. 2017, \mnras, 472, 4444, \dodoi{10.1093/mnras/stx2327}

\bibitem[{{Rest} {et~al.}(2023){Rest}, {Pierel}, {Correnti}, {Canipe}, {Hilbert}, {Engesser}, {Sunnquist}, \& {Fox}}]{rest23}
{Rest}, A., {Pierel}, J., {Correnti}, M., {et~al.} 2023, {arminrest/jhat: The JWST HST Alignment Tool (JHAT)}, v2,  Zenodo, \dodoi{10.5281/zenodo.7892935}

\bibitem[{{Reynolds} \& {Aller}(1988)}]{reynolds88}
{Reynolds}, S.~P., \& {Aller}, H.~D. 1988, \apj, 327, 845, \dodoi{10.1086/166242}

\bibitem[{{Rieke} {et~al.}(2023){Rieke}, {Kelly}, {Misselt}, {Stansberry}, {Boyer}, {Beatty}, {Egami}, {Florian}, {Greene}, {Hainline}, {Leisenring}, {Roellig}, {Schlawin}, {Sun}, {Tinnin}, {Williams}, {Willmer}, {Wilson}, {Clark}, {Rohrbach}, {Brooks}, {Canipe}, {Correnti}, {DiFelice}, {Gennaro}, {Girard}, {Hartig}, {Hilbert}, {Koekemoer}, {Nikolov}, {Pirzkal}, {Rest}, {Robberto}, {Sunnquist}, {Telfer}, {Wu}, {Ferry}, {Lewis}, {Baum}, {Beichman}, {Doyon}, {Dressler}, {Eisenstein}, {Ferrarese}, {Hodapp}, {Horner}, {Jaffe}, {Johnstone}, {Krist}, {Martin}, {McCarthy}, {Meyer}, {Rieke}, {Trauger}, \& {Young}}]{rieke23}
{Rieke}, M.~J., {Kelly}, D.~M., {Misselt}, K., {et~al.} 2023, \pasp, 135, 028001, \dodoi{10.1088/1538-3873/acac53}

\bibitem[{{Rigby} {et~al.}(2023){Rigby}, {Perrin}, {McElwain}, {Kimble}, {Friedman}, {Lallo}, {Doyon}, {Feinberg}, {Ferruit}, {Glasse}, {Rieke}, {Rieke}, {Wright}, {Willott}, {Colon}, {Milam}, {Neff}, {Stark}, {Valenti}, {Abell}, {Abney}, {Abul-Huda}, {Acton}, {Adams}, {Adler}, {Aguilar}, {Ahmed}, {Albert}, {Alberts}, {Aldridge}, {Allen}, {Altenburg}, {{\'A}lvarez-M{\'a}rquez}, {Alves de Oliveira}, {Andersen}, {Anderson}, {Anderson}, {Argyriou}, {Armstrong}, {Arribas}, {Artigau}, {Arvai}, {Atkinson}, {Bacon}, {Bair}, {Banks}, {Barrientes}, {Barringer}, {Bartosik}, {Bast}, {Baudoz}, {Beatty}, {Bechtold}, {Beck}, {Bergeron}, {Bergkoetter}, {Bhatawdekar}, {Birkmann}, {Blazek}, {Blome}, {Boccaletti}, {B{\"o}ker}, {Boia}, {Bonaventura}, {Bond}, {Bosley}, {Boucarut}, {Bourque}, {Bouwman}, {Bower}, {Bowers}, {Boyer}, {Bradley}, {Brady}, {Braun}, {Breda}, {Bresnahan}, {Bright}, {Britt}, {Bromenschenkel}, {Brooks}, {Brooks}, {Brown}, {Brown}, {Brown}, {Bunker}, {Burger}, {Bushouse}, {Cale}, {Cameron}, {Cameron},
  {Canipe}, {Caplinger}, {Caputo}, {Cara}, {Carey}, {Carniani}, {Carrasquilla}, {Carruthers}, {Case}, {Catherine}, {Chance}, {Chapman}, {Charlot}, {Charlow}, {Chayer}, {Chen}, {Cherinka}, {Chichester}, {Chilton}, {Chonis}, {Clampin}, {Clark}, {Clark}, {Coe}, {Coleman}, {Comber}, {Comeau}, {Connolly}, {Cooper}, {Cooper}, {Coppock}, {Correnti}, {Cossou}, {Coulais}, {Coyle}, {Cracraft}, {Curti}, {Cuturic}, {Davis}, {Davis}, {Dean}, {DeLisa}, {deMeester}, {Dencheva}, {Dencheva}, {DePasquale}, {Deschenes}, {Hunor Detre}, {Diaz}, {Dicken}, {DiFelice}, {Dillman}, {Dixon}, {Doggett}, {Donaldson}, {Douglas}, {DuPrie}, {Dupuis}, {Durning}, {Easmin}, {Eck}, {Edeani}, {Egami}, {Ehrenwinkler}, {Eisenhamer}, {Eisenhower}, {Elie}, {Elliott}, {Elliott}, {Ellis}, {Engesser}, {Espinoza}, {Etienne}, {Etxaluze}, {Falini}, {Feeney}, {Ferry}, {Filippazzo}, {Fincham}, {Fix}, {Flagey}, {Florian}, {Flynn}, {Fontanella}, {Ford}, {Forshay}, {Fox}, {Franz}, {Fu}, {Fullerton}, {Galkin}, {Galyer}, {Garc{\'\i}a Mar{\'\i}n}, {Gardner},
  {Gardner}, {Garland}, {Garrett}, {Gasman}, {Gaspar}, {Gaudreau}, {Gauthier}, {Geers}, {Geithner}, {Gennaro}, {Giardino}, {Girard}, {Giuliano}, {Glassmire}, {Glauser}, {Glazer}, {Godfrey}, {Golimowski}, {Gollnitz}, {Gong}, {Gonzaga}, {Gordon}, {Gordon}, {Goudfrooij}, {Greene}, {Greenhouse}, {Grimaldi}, {Groebner}, {Grundy}, {Guillard}, {Gutman}, {Ha}, {Haderlein}, {Hagedorn}, {Hainline}, {Haley}, {Hami}, {Hamilton}, {Hammel}, {Hansen}, {Harkins}, {Harr}, {Hart}, {Hart}, {Hartig}, {Hashimoto}, {Haskins}, {Hathaway}, {Havey}, {Hayden}, {Hecht}, {Heller-Boyer}, {Henriques}, {Henry}, {Hermann}, {Hernandez}, {Hesman}, {Hicks}, {Hilbert}, {Hines}, {Hoffman}, {Holfeltz}, {Holler}, {Hoppa}, {Hott}, {Howard}, {Howard}, {Hunter}, {Hunter}, {Hurst}, {Husemann}, {Hustak}, {Ilinca Ignat}, {Illingworth}, {Irish}, {Jackson}, {Jahromi}, {Jakobsen}, {James}, {James}, {Januszewski}, {Jenkins}, {Jirdeh}, {Johnson}, {Johnson}, {Jones}, {Jones}, {Jones}, {Jones}, {Jordan}, {Jordan}, {Jurczyk}, {Jurling}, {Kaleida}, {Kalmanson},
  {Kammerer}, {Kang}, {Kao}, {Karakla}, {Kavanagh}, {Kelly}, {Kendrew}, {Kennedy}, {Kenny}, {Keski-kuha}, {Keyes}, {Kidwell}, {Kinzel}, {Kirk}, {Kirkpatrick}, {Kirshenblat}, {Klaassen}, {Knapp}, {Knight}, {Knollenberg}, {Koehler}, {Koekemoer}, {Kovacs}, {Kulp}, {Kumari}, {Kyprianou}, {La Massa}, {Labador}, {Labiano}, {Lagage}, {Lajoie}, {Lallo}, {Lam}, {Lamb}, {Lambros}, {Lampenfield}, {Langston}, {Larson}, {Law}, {Lawrence}, {Lee}, {Leisenring}, {Lepo}, {Leveille}, {Levenson}, {Levine}, {Levy}, {Lewis}, {Lewis}, {Libralato}, {Lightsey}, {Link}, {Liu}, {Lo}, {Lockwood}, {Logue}, {Long}, {Long}, {Loomis}, {Lopez-Caniego}, {Lorenzo Alvarez}, {Love-Pruitt}, {Lucy}, {Luetzgendorf}, {Maghami}, {Maiolino}, {Major}, {Malla}, {Malumuth}, {Manjavacas}, {Mannfolk}, {Marrione}, {Marston}, {Martel}, {Maschmann}, {Masci}, {Masciarelli}, {Maszkiewicz}, {Mather}, {McKenzie}, {McLean}, {McMaster}, {Melbourne}, {Mel{\'e}ndez}, {Menzel}, {Merz}, {Meyett}, {Meza}, {Miskey}, {Misselt}, {Moller}, {Morrison}, {Morse}, {Moseley},
  {Mosier}, {Mountain}, {Mueckay}, {Mueller}, {Mullally}, {Murphy}, {Murray}, {Murray}, {Mustelier}, {Muzerolle}, {Mycroft}, {Myers}, {Myrick}, {Nanavati}, {Nance}, {Nayak}, {Naylor}, {Nelan}, {Nickson}, {Nielson}, {Nieto-Santisteban}, {Nikolov}, {Noriega-Crespo}, {O'Shaughnessy}, {O'Sullivan}, {Ochs}, {Ogle}, {Oleszczuk}, {Olmsted}, {Osborne}, {Ottens}, {Owens}, {Pacifici}, {Pagan}, {Page}, {Park}, {Parrish}, {Patapis}, {Paul}, {Pauly}, {Pavlovsky}, {Pedder}, {Peek}, {Pena-Guerrero}, {Penanen}, {Perez}, {Perna}, {Perriello}, {Phillips}, {Pietraszkiewicz}, {Pinaud}, {Pirzkal}, {Pitman}, {Piwowar}, {Platais}, {Player}, {Plesha}, {Pollizi}, {Polster}, {Pontoppidan}, {Porterfield}, {Proffitt}, {Pueyo}, {Pulliam}, {Quirt}, {Quispe Neira}, {Ramos Alarcon}, {Ramsay}, {Rapp}, {Rapp}, {Rauscher}, {Ravindranath}, {Rawle}, {Regan}, {Reichard}, {Reis}, {Ressler}, {Rest}, {Reynolds}, {Rhue}, {Richon}, {Rickman}, {Ridgaway}, {Ritchie}, {Rix}, {Robberto}, {Robinson}, {Robinson}, {Robinson}, {Rock}, {Rodriguez}, {Rodriguez
  Del Pino}, {Roellig}, {Rohrbach}, {Roman}, {Romelfanger}, {Rose}, {Roteliuk}, {Roth}, {Rothwell}, {Rowlands}, {Roy}, {Royer}, {Royle}, {Rui}, {Rumler}, {Runnels}, {Russ}, {Rustamkulov}, {Ryden}, {Ryer}, {Sabata}, {Sabatke}, {Sabbi}, {Samuelson}, {Sapp}, {Sappington}, {Sargent}, {Sauer}, {Scheithauer}, {Schlawin}, {Schlitz}, {Schmitz}, {Schneider}, {Schreiber}, {Schulze}, {Schwab}, {Scott}, {Sembach}, {Shanahan}, {Shaughnessy}, {Shaw}, {Shawger}, {Shay}, {Sheehan}, {Shen}, {Sherman}, {Shiao}, {Shih}, {Shivaei}, {Sienkiewicz}, {Sing}, {Sirianni}, {Sivaramakrishnan}, {Skipper}, {Sloan}, {Slocum}, {Slowinski}, {Smith}, {Smith}, {Smith}, {Smith}, {Snyder}, {Soh}, {Sohn}, {Soto}, {Spencer}, {Stallcup}, {Stansberry}, {Starr}, {Starr}, {Stewart}, {Stiavelli}, {Straughn}, {Strickland}, {Stys}, {Summers}, {Sun}, {Sunnquist}, {Swade}, {Swam}, {Swaters}, {Swoish}, {Taylor}, {Taylor}, {Te Plate}, {Tea}, {Teague}, {Telfer}, {Temim}, {Thatte}, {Thompson}, {Thompson}, {Thomson}, {Tikkanen}, {Tippet}, {Todd}, {Toolan},
  {Tran}, {Trejo}, {Truong}, {Tsukamoto}, {Tustain}, {Tyra}, {Ubeda}, {Underwood}, {Uzzo}, {Van Campen}, {Vandal}, {Vandenbussche}, {Vila}, {Volk}, {Wahlgren}, {Waldman}, {Walker}, {Wander}, {Warfield}, {Warner}, {Wasiak}, {Watkins}, {Weaver}, {Weilert}, {Weiser}, {Weiss}, {Weissman}, {Welty}, {West}, {Wheate}, {Wheatley}, {Wheeler}, {White}, {Whiteaker}, {Whitehouse}, {Whiteleather}, {Whitman}, {Williams}, {Willmer}, {Willoughby}, {Wilson}, {Wirth}, {Wislowski}, {Wolf}, {Wolfe}, {Wolff}, {Workman}, {Wright}, {Wu}, {Wu}, {Wymer}, {Yates}, {Yeager}, {Yeates}, {Yerger}, {Yoon}, {Young}, {Yu}, {Zak}, {Zeidler}, {Zhou}, {Zielinski}, {Zincke}, \& {Zonak}}]{rigby23}
{Rigby}, J., {Perrin}, M., {McElwain}, M., {et~al.} 2023, \pasp, 135, 048001, \dodoi{10.1088/1538-3873/acb293}

\bibitem[{{Sankrit} \& {Hester}(1997)}]{sankrit97}
{Sankrit}, R., \& {Hester}, J.~J. 1997, \apj, 491, 796, \dodoi{10.1086/304967}

\bibitem[{{Sankrit} {et~al.}(1998){Sankrit}, {Hester}, {Scowen}, {Ballester}, {Burrows}, {Clarke}, {Crisp}, {Evans}, {Gallagher}, {Griffiths}, {Hoessel}, {Holtzman}, {Krist}, {Mould}, {Stapelfeldt}, {Trauger}, {Watson}, \& {Westphal}}]{sankrit98}
{Sankrit}, R., {Hester}, J.~J., {Scowen}, P.~A., {et~al.} 1998, \apj, 504, 344, \dodoi{10.1086/306078}

\bibitem[{{Scargle}(1969)}]{scargle69}
{Scargle}, J.~D. 1969, \apj, 156, 401, \dodoi{10.1086/149978}

\bibitem[{{Schweizer} {et~al.}(2013){Schweizer}, {Bucciantini}, {Idec}, {Nilsson}, {Tennant}, {Weisskopf}, \& {Zanin}}]{Schweizer_Bucciantini+13a}
{Schweizer}, T., {Bucciantini}, N., {Idec}, W., {et~al.} 2013, \mnras, 433, 3325, \dodoi{10.1093/mnras/stt995}

\bibitem[{{Scott} {et~al.}(2015){Scott}, {Asplund}, {Grevesse}, {Bergemann}, \& {Sauval}}]{scott15}
{Scott}, P., {Asplund}, M., {Grevesse}, N., {Bergemann}, M., \& {Sauval}, A.~J. 2015, \aap, 573, A26, \dodoi{10.1051/0004-6361/201424110}

\bibitem[{{Seward} {et~al.}(2006){Seward}, {Gorenstein}, \& {Smith}}]{seward06}
{Seward}, F.~D., {Gorenstein}, P., \& {Smith}, R.~K. 2006, \apj, 636, 873, \dodoi{10.1086/498105}

\bibitem[{{Sibley} {et~al.}(2016){Sibley}, {Katz}, {Satterfield}, {Vanderveer}, \& {MacAlpine}}]{sibley16}
{Sibley}, A.~R., {Katz}, A.~M., {Satterfield}, T.~J., {Vanderveer}, S.~J., \& {MacAlpine}, G.~M. 2016, \aj, 152, 93, \dodoi{10.3847/0004-6256/152/4/93}

\bibitem[{Sironi \& Spitkovsky(2011)}]{Sironi_2011}
Sironi, L., \& Spitkovsky, A. 2011, The Astrophysical Journal, 741, 39, \dodoi{10.1088/0004-637X/741/1/39}

\bibitem[{{Slane} {et~al.}(2004){Slane}, {Helfand}, {van der Swaluw}, \& {Murray}}]{slane04}
{Slane}, P., {Helfand}, D.~J., {van der Swaluw}, E., \& {Murray}, S.~S. 2004, \apj, 616, 403, \dodoi{10.1086/424814}

\bibitem[{{Smith}(2003)}]{smith03}
{Smith}, N. 2003, \mnras, 346, 885, \dodoi{10.1111/j.1365-2966.2003.07135.x}

\bibitem[{{Smith}(2013)}]{smith13crab}
---. 2013, \mnras, 434, 102, \dodoi{10.1093/mnras/stt1004}

\bibitem[{{Sollerman} {et~al.}(2000){Sollerman}, {Lundqvist}, {Lindler}, {Chevalier}, {Fransson}, {Gull}, {Pun}, \& {Sonneborn}}]{sollerman00}
{Sollerman}, J., {Lundqvist}, P., {Lindler}, D., {et~al.} 2000, \apj, 537, 861, \dodoi{10.1086/309062}

\bibitem[{{Stockinger} {et~al.}(2020){Stockinger}, {Janka}, {Kresse}, {Melson}, {Ertl}, {Gabler}, {Gessner}, {Wongwathanarat}, {Tolstov}, {Leung}, {Nomoto}, \& {Heger}}]{stockinger20}
{Stockinger}, G., {Janka}, H.~T., {Kresse}, D., {et~al.} 2020, \mnras, 496, 2039, \dodoi{10.1093/mnras/staa1691}

\bibitem[{{Sukhbold} {et~al.}(2016){Sukhbold}, {Ertl}, {Woosley}, {Brown}, \& {Janka}}]{sukhbold16}
{Sukhbold}, T., {Ertl}, T., {Woosley}, S.~E., {Brown}, J.~M., \& {Janka}, H.-T. 2016, \apj, 821, 38, \dodoi{10.3847/0004-637X/821/1/38}

\bibitem[{{Temim} \& {Dwek}(2013)}]{temim13}
{Temim}, T., \& {Dwek}, E. 2013, \apj, 774, 8, \dodoi{10.1088/0004-637X/774/1/8}

\bibitem[{{Temim} {et~al.}(2009){Temim}, {Slane}, {Gaensler}, {Hughes}, \& {Van Der Swaluw}}]{temim09}
{Temim}, T., {Slane}, P., {Gaensler}, B.~M., {Hughes}, J.~P., \& {Van Der Swaluw}, E. 2009, \apj, 691, 895, \dodoi{10.1088/0004-637X/691/2/895}

\bibitem[{{Temim} {et~al.}(2012){Temim}, {Sonneborn}, {Dwek}, {Arendt}, {Gehrz}, {Slane}, \& {Roellig}}]{temim12b}
{Temim}, T., {Sonneborn}, G., {Dwek}, E., {et~al.} 2012, \apj, 753, 72, \dodoi{10.1088/0004-637X/753/1/72}

\bibitem[{{Temim} {et~al.}(2006){Temim}, {Gehrz}, {Woodward}, {Roellig}, {Smith}, {Rudnick}, {Polomski}, {Davidson}, {Yuen}, \& {Onaka}}]{temim06}
{Temim}, T., {Gehrz}, R.~D., {Woodward}, C.~E., {et~al.} 2006, \aj, 132, 1610, \dodoi{10.1086/507076}

\bibitem[{{Thielemann} {et~al.}(1990){Thielemann}, {Hashimoto}, \& {Nomoto}}]{thielemann90}
{Thielemann}, F.-K., {Hashimoto}, M.-A., \& {Nomoto}, K. 1990, \apj, 349, 222, \dodoi{10.1086/168308}

\bibitem[{{Trimble}(1968)}]{trimble68}
{Trimble}, V. 1968, \aj, 73, 535, \dodoi{10.1086/110658}

\bibitem[{{Trimble}(1977)}]{trimble77}
---. 1977, \aplett, 18, 145

\bibitem[{{Uomoto} \& {MacAlpine}(1987)}]{uomoto87}
{Uomoto}, A., \& {MacAlpine}, G.~M. 1987, \aj, 93, 1511, \dodoi{10.1086/114431}

\bibitem[{{Verner} {et~al.}(1996){Verner}, {Ferland}, {Korista}, \& {Yakovlev}}]{verner96}
{Verner}, D.~A., {Ferland}, G.~J., {Korista}, K.~T., \& {Yakovlev}, D.~G. 1996, \apj, 465, 487, \dodoi{10.1086/177435}

\bibitem[{{Veron-Cetty} \& {Woltjer}(1993)}]{veron-cetty93}
{Veron-Cetty}, M.~P., \& {Woltjer}, L. 1993, \aap, 270, 370

\bibitem[{{Wanajo} {et~al.}(2009){Wanajo}, {Nomoto}, {Janka}, {Kitaura}, \& {M{\"u}ller}}]{wanajo09}
{Wanajo}, S., {Nomoto}, K., {Janka}, H.~T., {Kitaura}, F.~S., \& {M{\"u}ller}, B. 2009, \apj, 695, 208, \dodoi{10.1088/0004-637X/695/1/208}

\bibitem[{{Wang} \& {Burrows}(2023)}]{wang23}
{Wang}, T., \& {Burrows}, A. 2023, \apj, 954, 114, \dodoi{10.3847/1538-4357/ace7b2}

\bibitem[{{Wang} \& {Burrows}(2024{\natexlab{a}})}]{wang24b}
---. 2024{\natexlab{a}}, submitted to \apj

\bibitem[{{Wang} \& {Burrows}(2024{\natexlab{b}})}]{wang24}
---. 2024{\natexlab{b}}, \apj, 962, 71, \dodoi{10.3847/1538-4357/ad12b8}

\bibitem[{{Weisskopf} {et~al.}(2000){Weisskopf}, {Hester}, {Tennant}, {Elsner}, {Schulz}, {Marshall}, {Karovska}, {Nichols}, {Swartz}, {Kolodziejczak}, \& {O'Dell}}]{weisskopf00}
{Weisskopf}, M.~C., {Hester}, J.~J., {Tennant}, A.~F., {et~al.} 2000, \apjl, 536, L81, \dodoi{10.1086/312733}

\bibitem[{{Woltjer} \& {Veron-Cetty}(1987)}]{woltjer87}
{Woltjer}, L., \& {Veron-Cetty}, M.~P. 1987, \aap, 172, L7

\bibitem[{{Wongwathanarat} {et~al.}(2013){Wongwathanarat}, {Janka}, \& {M{\"u}ller}}]{wongwathanarat13}
{Wongwathanarat}, A., {Janka}, H.~T., \& {M{\"u}ller}, E. 2013, \aap, 552, A126, \dodoi{10.1051/0004-6361/201220636}

\bibitem[{{Woosley} \& {Heger}(2007)}]{Woosley07}
{Woosley}, S.~E., \& {Heger}, A. 2007, \physrep, 442, 269, \dodoi{10.1016/j.physrep.2007.02.009}

\bibitem[{{Woosley} \& {Heger}(2015)}]{wh15}
---. 2015, \apj, 810, 34, \dodoi{10.1088/0004-637X/810/1/34}

\bibitem[{{Woosley} \& {Weaver}(1995)}]{woosley95}
{Woosley}, S.~E., \& {Weaver}, T.~A. 1995, \apjs, 101, 181, \dodoi{10.1086/192237}

\bibitem[{{Wright} {et~al.}(2023){Wright}, {Rieke}, {Glasse}, {Ressler}, {Garc{\'\i}a Mar{\'\i}n}, {Aguilar}, {Alberts}, {{\'A}lvarez-M{\'a}rquez}, {Argyriou}, {Banks}, {Baudoz}, {Boccaletti}, {Bouchet}, {Bouwman}, {Brandl}, {Breda}, {Bright}, {Cale}, {Colina}, {Cossou}, {Coulais}, {Cracraft}, {De Meester}, {Dicken}, {Engesser}, {Etxaluze}, {Fox}, {Friedman}, {Fu}, {Gasman}, {G{\'a}sp{\'a}r}, {Gastaud}, {Geers}, {Glauser}, {Gordon}, {Greene}, {Greve}, {Grundy}, {G{\"u}del}, {Guillard}, {Haderlein}, {Hashimoto}, {Henning}, {Hines}, {Holler}, {Detre}, {Jahromi}, {James}, {Jones}, {Justtanont}, {Kavanagh}, {Kendrew}, {Klaassen}, {Krause}, {Labiano}, {Lagage}, {Lambros}, {Larson}, {Law}, {Lee}, {Libralato}, {Lorenzo Alverez}, {Meixner}, {Morrison}, {Mueller}, {Murray}, {Mycroft}, {Myers}, {Nayak}, {Naylor}, {Nickson}, {Noriega-Crespo}, {{\"O}stlin}, {O'Sullivan}, {Ottens}, {Patapis}, {Penanen}, {Pietraszkiewicz}, {Ray}, {Regan}, {Roteliuk}, {Royer}, {Samara-Ratna}, {Samuelson}, {Sargent}, {Scheithauer},
  {Schneider}, {Schreiber}, {Shaughnessy}, {Sheehan}, {Shivaei}, {Sloan}, {Tamas}, {Teague}, {Temim}, {Tikkanen}, {Tustain}, {van Dishoeck}, {Vandenbussche}, {Weilert}, {Whitehouse}, \& {Wolff}}]{wright23}
{Wright}, G.~S., {Rieke}, G.~H., {Glasse}, A., {et~al.} 2023, \pasp, 135, 048003, \dodoi{10.1088/1538-3873/acbe66}

\bibitem[{{Yang} \& {Chevalier}(2015)}]{yang15}
{Yang}, H., \& {Chevalier}, R.~A. 2015, \apj, 806, 153, \dodoi{10.1088/0004-637X/806/2/153}

\bibitem[{{Zhang} \& {Pradhan}(1995)}]{zhang95}
{Zhang}, H.~L., \& {Pradhan}, A.~K. 1995, J. Phys. B., 28, 3403, \dodoi{10.1088/0953-4075/28/15/026}

\end{thebibliography}
\bibliographystyle{aasjournal}



\end{document}